%% file: ms.tex
\documentclass[apj,tighten,iop]{emulateapj2}

\input{defs}

\begin{document}
\title{PRIMUS + DEEP2: Clustering of X-ray, Radio and IR-AGN at $\lowercase{\textit{z}}\sim0.7$}
\shorttitle{PRIMUS: AGN Clustering}
\shortauthors{Mendez et al.}
\email{ajmendez@jhu.edu}

\author{Alexander J. Mendez\altaffilmark{1,2},
        Alison L. Coil\altaffilmark{1},
        James Aird\altaffilmark{3},
        Ramin A. Skibba\altaffilmark{1},
        Aleksandar M. Diamond-Stanic\altaffilmark{4},
        John Moustakas\altaffilmark{5},
        Michael R. Blanton\altaffilmark{6},
        Richard J. Cool\altaffilmark{7},
        Daniel J. Eisenstein\altaffilmark{8},
        Kenneth C. Wong\altaffilmark{9},
        Guangtun Zhu\altaffilmark{2}}
\altaffiltext{1}{Center for Astrophysics and Space Sciences, Department of Physics, University of California, 9500 Gilman Dr., La Jolla, San Diego, CA 92093, USA}
\altaffiltext{2}{Department of Physics and Astronomy, Johns Hopkins University, 3400 N. Charles Street, Baltimore, MD 21218, USA}
\altaffiltext{3}{Department of Physics, Durham University, Durham DH1 3LE, UK}
\altaffiltext{4}{Department of Astronomy, University of Wisconsin-Madison, Madison, WI 53706-1582, USA}
\altaffiltext{5}{Department of Physics and Astronomy, Siena College, 515 Loudon Road, Loudonville, NY 12211, USA}
\altaffiltext{6}{Center for Cosmology and Particle Physics, Department of Physics, New York University, 4 Washington Place, New York, NY 10003, USA}
\altaffiltext{7}{MMT Observatory, 1540 E Second Street, University of Arizona, Tucson, AZ 85721, USA}
\altaffiltext{8}{Harvard College Observatory, 60 Garden St., Cambridge, MA 02138, USA}
\altaffiltext{9}{Steward Observatory, The University of Arizona, 933 N. Cherry Ave., Tucson, AZ 85721, USA}

\begin{abstract}
We measure the clustering of X-ray, radio, and mid-IR-selected active galactic
nuclei (AGN) at $0.2 < z < 1.2$ using multi-wavelength imaging and
spectroscopic redshifts from the PRIMUS and DEEP2 redshift surveys, covering 7
separate fields spanning $\sim$10 square degrees. Using the cross-correlation
of AGN with dense galaxy samples, we measure the clustering scale length and
slope, as well as the bias, of AGN selected at different wavelengths. Similar
to previous studies, we find that X-ray and radio AGN are more clustered than
mid-IR-selected AGN. We further compare the clustering of each AGN sample with
matched galaxy samples designed to have the same stellar mass, star formation
rate, and redshift distributions as the AGN host galaxies and find no
significant differences between their clustering properties. The observed
differences in the clustering of AGN selected at different wavelengths can
therefore be explained by the clustering differences of their host populations,
which have different distributions in both stellar mass and star formation
rate. Selection biases inherent in AGN selection, therefore, determine the
clustering of observed AGN samples. We further find no significant difference
between the clustering of obscured and unobscured AGN, using IRAC or WISE
colors or X-ray hardness ratio.
\end{abstract}

% \input{intro}
\input{intro}

\input{data}

\input{samples}

\input{methods}

\input{results}

\input{discussion}

\input{conclusions}

\bibliography{refs}

\end{document}

%% file: defs.tex
\usepackage{natbib}
\usepackage[usenames]{color}
\definecolor{red}{rgb}{0.75,0,0}
\definecolor{green}{rgb}{0,0.5,0}
\definecolor{blue}{rgb}{0,0,0.75}
\usepackage[ colorlinks=true, % false for boxed links
             urlcolor=blue,
             filecolor=blue,
             linkcolor=red,
             citecolor=blue,
             pdftitle={PRIMUS: IR-AGN and X-ray AGN},
             pdfauthor={Alexander Mendez, et al.},
             pagebackref, % Add for citation links
             bookmarks,
             bookmarksopen=true]{hyperref}

\usepackage{xspace}    % space after new command
\usepackage{microtype} % Cleans up spacing.
\usepackage{amsmath}   % Some nice math things
\usepackage{amssymb}

\usepackage[normalem]{ulem}
\newcommand\redline{\bgroup\markoverwith{\textcolor{red}{\rule[0.5ex]{2pt}{0.4pt}}}\ULon}

\usepackage{makecell} % enables multiple colhead lines.
\usepackage{rotating} %[2011.09.05] rotate the deluxetable. \begin{turnpage}
\newcommand{\um}{\ensuremath{\mu\mathrm{m}}\xspace}
\newcommand{\uJy}{\ensuremath{\mu\mathrm{Jy}}\xspace}

\newcommand{\Xray}{{X-ray AGN}\xspace}
\newcommand{\IR}{{IR-AGN}\xspace}
\newcommand{\WISE}{{WISE IR-AGN}\xspace}
\newcommand{\Radio}{{radio AGN}\xspace}
\newcommand{\Stern}{{Stern et al. \IR}\xspace}
\newcommand{\Donley}{{Donley et al. \IR}\xspace}
\newcommand{\Assef}{{Assef et al. \IR}\xspace}
\newcommand{\Mateos}{{Assef et al. \IR}\xspace}

\newcommand{\Spitzer}{\textit{Spitzer}\xspace}
\newcommand{\Chandra}{\textit{Chandra}\xspace}
\newcommand{\XMM}{\textit{XMM-Newton}\xspace}
\newcommand{\ergscm}{\ensuremath{\mathrm{erg\;s^{-1} cm^{-2}}}\xspace}

\newcommand{\ergs}{\ensuremath{\mathrm{erg\;s^{-1}}}}
\newcommand{\wattshz}{\ensuremath{\mathrm{Watts\;Hz^{-1}}}}

\newcommand{\degsq}{\ensuremath{\mathrm{deg}^2}}
\newcommand{\msun}{\mathcal{M}_{\sun}}

\newcommand{\fxx}[2]{\ensuremath{f_{X}\sim{#1}\times{10}^{#2}}\,\ergscm}
\newcommand{\fhard}[3]{$f_{\mathrm{{#1}keV}}\sim{#2}\times{10^{#3}}$~\ergscm}
\newcommand{\LX}{\ensuremath{L_{X}}\xspace}
\newcommand{\Lx}[2]{\ensuremath{\LX\;{#1}\;{10}^{#2}\;\ergs}\xspace}

\renewcommand{\Pr}{\ensuremath{P_\mathrm{1.4Ghz}}\xspace}
\newcommand{\PR}[2]{\ensuremath{P_\mathrm{1.4Ghz} {#1} {10}^{#2}}~\wattshz\xspace}
\newcommand{\mstar}{\ensuremath{\mathcal{M}_*}}

\newcommand{\specific}{\ensuremath{\lambda}\xspace}
\newcommand{\SpecificP}[2]{\ensuremath{\specific\ {#1}\ {#2}}\xspace}

\newcommand{\xisp}{\ensuremath{\xi(r_p,\,\pi)}}
\newcommand{\pimax}{\ensuremath{\pi_\textrm{max}}}
\newcommand{\hMpc}{\ensuremath{\,h^{-1}\, \textrm{Mpc}}}
\newcommand{\wAA}{\ensuremath{w_\textrm{AA}}}
\newcommand{\wAG}{\ensuremath{w_\textrm{AG}}}
\newcommand{\wGG}{\ensuremath{w_\textrm{GG}}}

\newcommand{\iSEDfit}{\texttt{iSEDfit}\xspace}

\newcommand{\maketmp}[1]{\expandafter\MakeUppercase\expandafter{#1}}

\newcommand{\massunit}{\ensuremath{\msun}\xspace}
\newcommand{\uJybeam}{\ensuremath{\uJy\,\mathrm{beam}^{-1}}\xspace}
\newcommand{\mass}[2]{\ensuremath{\mstar\,{#1}\,{10}^{#2}\,\massunit}}
\newcommand{\median}[2]{\ensuremath{\langle{#1}_{#2}\rangle}\xspace}

\newcommand{\mhalo}{\ensuremath{M_\textrm{halo}}\xspace}
\newcommand{\mhalounit}{\ensuremath{\textrm{h}^{-1}\,\msun}\xspace}
\newcommand{\Mhalo}[2]{\ensuremath{\mhalo\,{#1}\,10^{#2}\,\mhalounit}\xspace}
\newcommand{\medianMhalo}[1]{\ensuremath{\langle\mhalo\rangle\,\sim\,10^{#1}\,\mhalounit}\xspace}

\newcommand{\medianMass}[1]{\ensuremath{\langle\mstar\rangle\,\sim\,10^{#1}\,\massunit}\xspace}

\newcommand{\lambdaunit}{\ensuremath{\ergs\,\msun^{-1}}\xspace}
\newcommand{\lambdavalue}[2]{\ensuremath{\specific\,{#1}\,{10}^{#2}\,\lambdaunit}\xspace}

\newcommand{\mJy}{\ensuremath{\mathrm{mJy}}\xspace}
\newcommand{\hKpc}{\ensuremath{\,h^{-1}\, \textrm{kpc}}}
\newcommand{\chandra}{\textit{Chandra}}
\newcommand{\xmm}{\textit{XMM-Newton}}
\newcommand{\mbh}{\ensuremath{\mathcal{M}_\mathrm{BH}}\xspace}
\newcommand{\sfrunit}{\ensuremath{\msun\,\textrm{yr}^{-1}}}

\newcommand{\ssfr}{\ensuremath{\mathrm{sSFR}}\xspace}
\newcommand{\medianLx}[1]{$\median{L}{X}\;{\sim}\;{10}^{#1}$\;\ergs}

\newcommand{\galex}{\emph{GALEX} }

\newcommand{\arcsecsq}{\ensuremath{\mathrm{arcsec}^2}\xspace}

\renewcommand{\wp}{\ensuremath{w_p}\xspace}
\newcommand{\wprp}{\ensuremath{w_p(r_p)}\xspace}
\newcommand{\ssfrunit}{\ensuremath{\mathrm{yr}^{-1}}}

\newcommand{\zsim}[1]{\ensuremath{z\sim{#1}\xspace}}

\newcommand{\medianzsim}[1]{\ensuremath{\langle{z}\rangle\sim{#1}}\xspace}

\newcommand{\xir}{\ensuremath{\xi(r)}\xspace}
\newcommand{\rp}{\ensuremath{r_p}\xspace}
\newcommand{\rpvalue}[2]{\ensuremath{\rp\,{#1}\,{#2}\hMpc}\xspace}
\newcommand{\smallrp}{\rpvalue{<}{1}}
\newcommand{\largerp}{\rpvalue{>}{1}}

\newcommand{\bias}[2]{\ensuremath{b\,=\,{#1}\pm{#2}}\xspace}

\newcommand{\bootes}{Bo{\"o}tes\xspace}

%% file: intro.tex
%!TEX root = ms.tex
\section{Introduction}\label{sec:intro}

It is now well established that most galaxies host a supermassive black hole
(SMBH) \citep[e.g., ][]{KormendyRichstone95, Richstone98, Ferrarese05,
Kormendy13}. However, it is not well understood what physical processes trigger
intense episodes of accretion onto the SMBH, creating an observed active
galactic nucleus (AGN). The broad similarities between the cosmic star
formation history and AGN mass accretion history, both peaking at \zsim{2} and
declining sharply at lower redshift \citep[e.g.,][]{Soltan82, Madau96,
Franceschini99, Ueda03, Zheng09,Serjeant10, Aird15}, and the relatively tight
observed correlation between SMBH mass and mass of the host galaxy bulge
\citep[$M-\sigma$ relationship; e.g.,][]{Magorrian98, Gebhardt00, Tremaine02}
hint at the possibility of a coeval evolution between SMBHs and their host
galaxies.

The vast scale difference between galaxies and SMBHs, coupled with the relative
rarity of the active accretion phase, has made it difficult to determine the
physical mechanism(s) connecting galaxy and AGN growth. Constraining the
triggering and fueling mechanism(s) of AGN is key to uncovering the relevant
physics connecting SMBHs and their host galaxies.

Clustering measurements on scales larger than a typical dark matter halo
($r_p\gtrsim1\hMpc$) estimate the mean dark matter halo mass of AGN hosts,
effectively placing AGN in a cosmological context \citep[e.g.,][]{Mo96,
Sheth99}. On smaller scales ($r_p\lesssim1\hMpc$) clustering measurements
estimate the fraction of AGN that are hosted by satellite galaxies and place
constraints on triggering and fueling from galaxy-galaxy interactions and
mergers. Theoretical models that assume different internal or external AGN
triggering mechanisms predict different large-scale clustering properties of
AGN, as a function of both luminosity and redshift \citep[e.g.,][]{Silk98,
Springel05a, Hopkins06, Hopkins09, Croton09, Booth10, Fanidakis13, Hutsi14}.
However, observational data suggest that only a weak luminosity dependence
exists \citep[e.g.,][]{Coil09, Krumpe10, Cappelluti10, Allevato12,
Koutoulidis13}. The measurement of clustering properties of AGN across a range
of redshifts and luminosities provides strong constraints to theoretical models
of AGN.

With the advent of the \XMM and \Chandra X-ray telescopes, early \Xray
clustering measurements at \zsim{0.5-2} targeted small fields and found that
they reside in massive halos from \Mhalo{\sim}{12-13} \citep{Gilli05,Yang06}.
Later \citet{Coil09} measured the clustering of \Xray sources at \zsim{1} with
higher accuracy by using the cross-correlation of \Xray sources with DEEP2
galaxies and using a larger field. They found that \Xray are more strongly
clustered, similar to elliptical galaxies, which are more clustered than
star-forming galaxies. Generally, \Xray at \zsim{1-2} are more clustered than
optically-identified quasars as the same redshift and reside in relatively
dense environments suggestive of being within group like environments
\citep[e.g.][]{Gilli05, Yang06, Puccetti06, Coil09, Hickox09}.

The NVSS \citep{Condon98} and FIRST \citep{Becker94} wide-area radio surveys
identified large populations of luminous, low accretion rate,
mechanically-driven AGN \citep[e.g.,][]{Sijacki07}. Clustering studies using
these \Radio found them to be strongly clustered, residing in very massive
halos with \Mhalo{>}{13} \citep{Cress96, Magliocchetti04, Best05}.
\citet{Hickox09} studied the connection between AGN selected using X-ray,
radio, and mid-IR techniques by measuring the clustering, host properties, and
AGN properties of sources in the \bootes field. They found that \Xray and
\Radio reside in dark matter halos of mass \Mhalo{\sim}{13} and
\Mhalo{\sim}{13.5}, respectively, while \IR typically reside in lower mass
halos with \Mhalo{<}{12}.

The observed differences in the clustering of \Xray, \Radio, and \IR samples
indicate that it is crucial to test for any obscuration dependence in AGN
clustering. The simplest unified AGN models \citep[e.g., ][]{Antonucci85,
Urry95} would suggest that unobscured (type-1) and obscured (type-2) AGN should
have the same distribution of environments, with differences in the observed
obscuration due only to the orientation of the AGN relative to the observer. It
has been suggested, however, that obscured and unobscured AGN are similar
objects observed at different evolutionary stages of SMBH accretion
\cite[e.g.,][]{Hopkins08, Hickox09}. Most optical and X-ray AGN clustering
studies do not find significant differences between the clustering of obscured
and unobscured AGN \citep[e.g.,][]{Coil09, Gilli09}. \citet{Hickox11} found a
marginal ($\sim2\,\sigma$) increase in the clustering amplitude between
obscured and unobscured \IR selected AGN at \zsim{1.25}, suggesting that
obscured AGN may reside in more massive halos. More recently,
\citet{Dipompeo14} and \citet{Donoso13} found a significantly higher angular
clustering amplitude for obscured compared to unobscured \WISE at \zsim{0.9}.
However, these results measure only the angular projected clustering amplitude,
due to a lack of spectroscopic redshifts in their sample.

Selection biases inherent in AGN identification may also contribute to the
observed clustering signals, in that \Radio are generally found in luminous,
quiescent galaxies, \Xray are found in a mixture of quiescent and star forming
galaxies, and \IR are typically found in star forming galaxies
\citep[e.g.,][]{Hickox09, Aird12, Mendez13, Goulding14}. As quiescent galaxies
are more strongly clustered than star forming galaxies at a given redshift
\citep[e.g.,][]{LeFevre05, Zehavi05, Coil08, Skibba14}, the observed clustering
differences between AGN selected at different wavelengths could be due in part
to differences in their host populations. In order to understand the magnitude
of this effect, one can compare the clustering of AGN selected at different
wavelengths to matched samples of inactive galaxies \citep[e.g.,][]{Wake08,
Mandelbaum09, Coil09, Hickox09}. While \citet{Coil09} found that \Xray are more
clustered than color and magnitude matched galaxy samples, \citet{Hickox09}
found that \IR are less clustered than color and magnitude matched samples.
Interestingly, using weak lensing measurements at \zsim{0.1},
\citet{Mandelbaum09} found that dark matter halos of radio-loud AGN are twice
as massive as control galaxies of the same stellar mass and that \Radio are
more clustered than optically-selected AGN.

In order to address these outstanding issues, here we measure the clustering
properties of \Xray, \Radio, and \IR at $0.2<z<1.2$ using the DEEP2 and PRIMUS
redshift surveys. The wealth of deep multi-wavelength data, combined with
precise spectroscopic redshifts in these multiple fields makes this sample both
larger and deeper than similar previous studies at these redshifts. We use data
from multiple fields, limiting the affect of cosmic variance. We measure the
cross-correlation function of AGN with dense galaxy samples, used to trace the
large scale structure in our fields. This leads to lower statistical errors
than measuring the auto-correlation function of the AGN directly. We
investigate the dependence of clustering with intrinsic AGN properties (e.g.,
X-ray luminosity, specific accretion rate, hardness ratio, and obscuration). We
create galaxy samples that are matched in stellar mass, star formation rate
(SFR), and redshift to the AGN samples identified in each wavelength, to
compare the clustering of AGN with similar inactive galaxies. This limits
potential selection biases in comparing AGN samples selected at different
wavelengths.

The paper is organized as follows. In \S\ref{sec:data} we present the
spectroscopic redshift surveys and multi-wavelength datasets used here. In
\S\ref{sec:sample} we detail the different AGN selection techniques and the AGN
and galaxy samples used. In \S\ref{sec:results} we present the clustering
measurements of the various AGN and matched galaxy samples. We discuss our
results in \S\ref{sec:discussion} and conclude in \S\ref{sec:conclusions}.
Throughout the paper we assume a standard flat $\Lambda$CDM model with
$\Omega_m=0.3$, $\Omega_\Lambda=0.7$, and $H_{0}=72$~km s$^{-1}$~Mpc$^{-1}$.

%% file: data.tex
%!TEX root = ms.tex
\input{table_01_numbers}

\section{Data}\label{sec:data}
Our analysis combines multi-wavelength imaging with spectroscopic redshifts
from the PRIMUS and DEEP2 galaxy redshift surveys, covering eight well-known
extragalactic fields: the CDFS-SWIRE field \citep{Lonsdale03}, the COSMOS field
\citep{Scoville07}, the DEEP2 \citep[DEEP2; ][]{Davis03} 02hr and 23hr fields,
as well as the Extended Groth Strip (EGS), the Elais-S1 (ES1) field
\citep{Oliver00}, and the XMM-Large Scale Structure field \citep[XMM-LSS;
][]{Pierre04}. We describe the X-ray catalogs that we use in
Section~\ref{sec:xraydata}, the radio catalogs in Section~\ref{sec:radiodata},
and the mid-IR catalogs in Section~\ref{sec:iracdata}. In
Section~\ref{sec:primusdata} and Section~\ref{sec:deep2data} we briefly
describe the PRIMUS and DEEP2 redshift surveys, respectively. In
Section~\ref{sec:massdata} we explain the methods used to estimate stellar
masses and SFRs for PRIMUS and DEEP2 sources. In Section~\ref{sec:windowdata}
we provide information on the spatial selection function of the PRIMUS and
DEEP2 surveys that we use for our clustering analysis.

\subsection{X-ray Data}\label{sec:xraydata}
We use existing \Chandra\ and \XMM\ X-ray source catalogs of various depths in
the COSMOS, DEEP2, ES1, EGS, and XMM-LSS fields (see \citet{Aird12} and
\citet{Mendez13} for details). Due to the large positional uncertainty of the
X-ray point sources, we use the likelihood ratio matching technique
\citep[e.g.,][]{Sutherland92, Ciliegi03, Brusa07, Laird09} to identify optical
counterparts to each X-ray source in each field. The likelihood-ratio technique
accounts for both the optical and X-ray positional uncertainties, by
calculating the probability of having a counterpart with a given magnitude
above the probability of a spurious match. We place a lower limit on the
positional uncertainty for the X-ray source location of 0.5\arcsec\ and require
an optical match within 5\arcsec\ in any field. We restrict our sample to
robust optical counterparts with likelihood ratios above $> 0.5$ and choose the
counterpart with the largest likelihood, when there are multiple counterparts.
Table~\ref{table:numbers} lists the area of the X-ray coverage in each field,
as well as the number of X-ray sources with redshifts (see
Section~\ref{sec:primusdata} for details).

In the COSMOS field we use the public \xmm\ X-ray point source catalog
\citep{Cappelluti09, Brusa10}, which covers the entire 2~\degsq\ to a depth of
\fhard{2-10}{3}{-15}. We further use the deeper \chandra\ point source catalog
that has a depth of \fhard{2-10}{8}{-16}\ and covers the central
$\sim$0.9~\degsq\ \citep{Elvis09, Civano12}.

In the ES1 field, we use the \citet{Puccetti06} point source catalog from four
partially overlapping \xmm\ pointings which has a depth of
\fhard{2-10}{2}{-15}\ and covers 0.52~\degsq\ of the PRIMUS area in this field.

We use the public X-ray point source catalog from the deep \Chandra\ Advanced
CCD Imagining Spectrometer (ACIS-I) XDEEP2 survey \citep{Goulding12} for the
EGS and DEEP2-02hr, DEEP2-16hr, and DEEP2-23hr fields. In the EGS the XDEEP2
survey contains 96 \Chandra\ pointings across the field, covering an area of
0.66 \degsq. The typical full band flux limit in the merged observations in
this field is \fxx{2.8}{-16}, though this varies across the field due to the
number of overlapping pointings. The DEEP2-02hr, DEEP2-16hr, and DEEP2-23hr
fields contain 12, 12 and 17 \Chandra\ pointings respectively, with a full-band
flux-limit of \fxx{4.6}{-15} for all fields. In order to match the reported
hard-band flux in the other fields, we convert the reported $2-7$ keV hard
X-ray band flux into an equivalent $2-10$ keV hard X-ray band flux assuming a
$\Gamma=1.9$ power-law.

In the XMM-LSS field we use the final release of the public XMM X-ray catalog
from \citet{Chiappetti12}, which consists of 124 pointings of the XMM-Newton
X-ray telescope which includes the Subaru XMM-Newton Deep Survey \citep[SXDS;
][]{Ueda08}. This catalog contains sources to a hard-band flux limit of
\fxx{1.3}{-15} and \fxx{9.3}{-17} in the shallower XMM-LSS and deeper XMM-SXDS
regions, respectively. We match the X-ray catalogs using the likelihood ratio
matching technique described above.

Following \citet{Aird12} and \citet{Mendez13}, we apply an ``X-ray weight" for
each X-ray detected source based on the ratio of the total number of X-ray
detected sources to the predicted log(N)-log(S) relation of
\citet{Georgakakis08} at a given flux. These X-ray weights correct observed
number densities of X-ray sources to the intrinsic number density and account
for variations in the flux limit across the fields due to vignetting and the
change in sensitivity of the telescope as a function of axis angle.

\subsection{Radio Data}\label{sec:radiodata}
To select radio AGN we use existing deep Very Large Array (VLA) 1.4 GHz
radio data in the COSMOS, EGS, and XMM-LSS fields. In the COSMOS field, we use
the VLA-COSMOS Deep Project \citep{Schinnerer10}, which combines the shallower
data of the VLA-COSMOS Large Project \citep{Schinnerer07} with deeper coverage
in the central degree of the field. The survey provides radio continuum
coverage for $\sim$2,900 sources with $\sim1.5\arcsec$ resolution and a mean
1$\sigma$ sensitivity of 12 \uJybeam in the central square degree and
$\sim2\arcsec$ resolution and sensitivity of 15 \uJybeam in the outer region.
In the EGS, we use the AEGIS20 (\citet{Ivison07}; \citet{Willner12}) VLA radio
catalog which identifies 1,122 sources from six overlapping pointings in the
northern two-thirds of the field. The lower third of the EGS was not imaged due
to the proximity to a bright radio source, \texttt{3C 295}. The data were
obtained from the VLA with a $5\sigma$ sensitivity limit of 50 \uJybeam with
$\sim3.8\arcsec$ resolution. In the XMM-LSS field, we use the $100-\uJy$
catalog \citep{Simpson06} which contains fourteen overlapping pointings. The
radio imaging identifies 505 radio sources and reaches an sensitivity limit of
12 \uJybeam over 0.8 \degsq of the field. In the DEEP2-02hr, DEEP2-16hr, and
DEEP2-23hr fields we additionally include relatively shallow VLA data from the
Faint Images of the Radio Sky at Twenty-one centimeters survey \citep[FIRST;
][]{Becker95}. We use the \texttt{14Mar04} catalog which contains 946,432 radio
sources above the sensitivity limit of $\sim200$ \uJybeam and above the
detection limit of 1 \mJy.

We use the Australian Telescope Large Area Survey (ATLAS) in the CDFS-SWIRE
field \citep{Norris06} and ES1 field \citep{Middelberg07}. ATLAS used the
Australian Telescope Compact Array (ACTA) at 1.4 GHz to survey both fields. The
CDFS-SWIRE data contains 21 pointings with 784 radio galaxies reaching a
$1\sigma$ sensitivity limit of $\sim40$ \uJybeam, while the ES1 data contains
12 pointings with 1276 radio galaxies reaching a $1\sigma$ sensitivity limit of
$\sim30$ \uJybeam. We find no major astrometric offsets between these radio
catalogs and the PRIMUS spectroscopic catalog (described below), such that we
assign radio counterparts to the optical redshift catalog by using
\texttt{SPHEREMATCH} in IDL to identify counterparts within 2\arcsec, 
corresponding to the approximately astrometric uncertainty in the radio 
catalogs.

\subsection{Mid-IR Data}\label{sec:iracdata}
To identify mid-IR AGN, we use existing public \Spitzer IRAC photometry in the
CDFS-SWIRE, COSMOS, EGS, ES1, and XMM-LSS fields. IRAC provides 3.6, 4.5, 5.8,
and 8.0 \um data which we will reference as [3.6], [4.5], [5.8] and [8.0]. In
the CDFS-SWIRE, ES1, and XMM-LSS fields we use existing shallow IRAC imaging
from Data Release 2 (DR2) from the \Spitzer Wide-area Infrared Extragalactic
Survey \citep[SWIRE; ][]{Lonsdale03} (see \citet{Mendez13} for details). We
find no major astrometric offsets between these catalogs and the PRIMUS optical
redshift catalog, and we assign IRAC counterparts to the optical redshift
sources in all of the fields by matching to the closest object within 1\arcsec.
The CDFS ``proper'' field is not included here; instead, we use the larger
CDFS-SWIRE field at slightly lower declination, which was covered by the PRIMUS
survey. In the COSMOS field, we reproduce the SWIRE source detection procedure
from the SWIRE DR2 documentation using the IRAC mosaic images \citep[see][for
details]{Mendez13}. This ensures that we measure robust fluxes and flux
uncertainties using a consistent technique across all of our fields. For the
majority of sources, our flux measurements are similar to those in the S-COSMOS
public catalog, although the public catalog tends to have larger uncertainties
for similar brightness objects from the SWIRE catalogs due to their aggressive
deblending of sources.

In the DEEP2-02hr field we use a four-band detected catalog\footnote{Catalog
from A. Goulding 2013, private communication}. The sample is drawn from
\Spitzer IRAC observations as part of the DEEP2\_CY5A/50660 program (PI: C.
Jones). The IRAC imaging contains 34 pointings in each band covering the
majority of the DEEP-02hr field. In the EGS field, we use the \citet{Barro11}
publicly-available IRAC [3.6] and [4.5] selected catalog. The catalog contains
$\sim76,000$ sources with [3.6] $\le 23.75$. The sample is drawn from \Spitzer
as part of the Guaranteed Time Observations (GTO; PI: G. Fazio) and presented
in \citet{Barmby08} with additional data from the GO program (ID 41023; PI: K.
Nandra). The GTO IRAC imaging comprises 52 pointing of all four IRAC bands over
the central region of the EGS. The additional GO data cover the upper and lower
regions of the EGS, flanking the original strip.

Additionally, we use data from the \textit{Wide-field Infrared Survey Explorer}
\citep[WISE; ][]{Wright10}, which provides 3.4, 4.5, 12, and 22
\um\ photometry (bands W1, W2, W3, and W4, respectively) in all of our fields.
Here we use the public all-sky catalog from March 2012 which has a $5\sigma$
point source sensitivity better than 0.08, 0.11, 1, and 6 \mJy in each of the
bands, respectively. We remove sources with spurious photometric detections and
require sources to have SNR $>$ 3 in the W1 and W2 bands \citep[See ][ for more
details]{Cutri11}. WISE surveyed the sky in an ecliptic polar-orbit, which
increased the number of observations with increasing ecliptic latitude, causing
the median coverage to vary for different fields. See Table~\ref{table:numbers}
for the IRAC and WISE area (where at least W1 and W2 photometry
was required) of each field.

\subsection{PRIMUS Spectroscopic Redshifts}\label{sec:primusdata}
We use spectroscopic redshifts from the PRIMUS redshift survey to define
samples for our clustering analysis. PRIMUS \citep{Coil11, Cool13} is the
largest faint galaxy redshift survey completed to date, covering $\sim
9\,\degsq$ in seven well-studied fields on the sky with multi-wavelength
imaging from the X-ray to the far infrared (IR). The survey obtained
low-resolution ($\lambda/\Delta\lambda \sim 40$) spectra for $\sim 300,000$
objects, targeting $80\%$ of galaxies in these fields with $i < 22$. PRIMUS
used the IMACS instrument \citep{Bigelow03} on the Magellan-I Baade 6.5 m
telescope to observe $\sim 2,500$ objects at once using a slitmask that covered
0.18 $\degsq$. PRIMUS contains a statistically-complete sample of $\sim120,000$
spectroscopic redshifts to $i_{\mathrm{AB}}\sim23.5$. Redshifts are derived by
fitting a large suite of galaxy, broad-line AGN, and stellar spectral templates
to the low-resolution spectra and optical photometry \citep[see][for
details]{Cool13}. Objects are classified as galaxies, broad-line AGN or stars
depending on the best $\chi^2$ template fit. The PRIMUS redshifts are very
precise ($\sigma_z/(1+z) \sim 0.5\%$) and have a low catastrophic outlier rate,
less than $3\%$ ($\Delta{z}/(1+z) \ge 0.03$). Here we use robust
($z_\textrm{quality} \ge 3$, see \citet{Coil11}) PRIMUS redshifts between $0.2
< z < 1.2$ for the fields listed in Table~\ref{table:numbers}. For further
details of the survey design, targeting, and data see \citet{Coil11}; for
details of the data reduction, redshift confidence, and completeness see
\citet{Cool13}.

The PRIMUS survey generally targeted all sources above $i < 22.5$ and
sparse-sampled $22.5 < i < 23$ sources, so that faint galaxy sources at the
flux limit would not dominate the target selection. The sampling rates are well
defined a-priori such that building a statistically complete flux-limited
sample requires the tracking of both the ``sparse sampling'' weight and the
``density dependent'' weight of each object. The magnitude-dependent sparse
sampling weight accounts for the fraction of sources selected at random in the
0.5 mag interval above the targeting limit in each field. The density-dependent
weight accounts for the sources in high density areas on the sky that are
missed due to slit collisions and the finite number of masks observed. In these
regions the observed spectra of adjacent galaxies would overlap on the
detector if all galaxies were targeted \citep[see ][ for more details]{Coil11,
Moustakas13}.

Additionally, here we include a spatially-varying redshift success fraction weight
to account for changes in the observed redshift success rate across a field
(i.e., due to differences in observing conditions for different slitmasks). In
the PRIMUS field we use the \texttt{pixelize} function in
\texttt{Mangle}\footnote{\url{http://space.mit.edu/~molly/mangle/}}. We
estimate the redshift success fraction by taking the ratio of highly confident
sources with $z_{\textrm{quality}} \ge 3$ to all targeted sources in the field
in pixels of size $\sim 36~\arcsecsq$. We use a larger pixel size in the PRIMUS
fields than in the DEEP2 fields (see Section~\ref{sec:deep2data}) to limit
Poisson noise in the shallower PRIMUS data.

% The inclusion of these targeting and completeness weights is important, as they
% correct the observations to more accurately represent the full galaxy
% population. However, the $i\sim23.5$ selection corresponds to a rest-frame
% selection at $\sim$5000\AA \ at $z=0.7$, such that at the highest redshifts we
% are incomplete for fainter, red galaxies. The redshift completeness of PRIMUS
% is only a weak function of the color of galaxies and is a stronger function of
% luminosity \citep{Cool13}. For the purposes of this paper, we compare the
% clustering of AGN samples either with each other or with matched galaxy
% samples, and these weights are applied to all samples.

% from the response
The inclusion of these targeting and completeness weights is important, as they
correct the observations to more accurately represent the full galaxy
population. However, the $i\sim23.5$ selection corresponds to a rest-frame
selection $\sim$5000\AA at $z=0.7$, such that at the highest redshifts we are
incomplete for fainter, red galaxies. The redshift completeness is only a weak
function of the color of galaxies and is a stronger function of luminosity
\citep{Cool13}. For the purposes of this paper, we compare the clustering of
AGN samples either with each other or with matched galaxy samples, and these
weights are applied to all samples.

\subsection{DEEP2 Spectroscopic Redshifts}\label{sec:deep2data}
We also use spectroscopic redshifts from the Deep Extragalactic Evolutionary
Probe \citep[DEEP2;][]{Davis03, Newman12} redshift survey. In the DEEP2-02hr
and DEEP2-23hr fields, PRIMUS did not target the $0.7 <z<1.4$ redshift range
already covered by DEEP2. The combination of PRIMUS redshifts and DEEP2
redshifts in these fields selects galaxies uniformly from $z=0.2$ to $z=1.4$.
The DEEP2 survey provides spectroscopic redshifts in the EGS, the DEEP2-02hr
field, the DEEP2-16hr field, and the DEEP2-23hr field. The DEEP2 survey was
conducted with the DEIMOS spectrograph \citep{Faber03} on the 10m Keck-II
telescope. In the EGS, the DEEP2 survey has measured $\sim17,000$
high-confidence redshifts ($Q \ge 3$, See \citet{Newman12}) to $R_{AB} = 24.1$.
In the DEEP2-02hr, DEEP2-16hr and DEEP2-23hr fields, the survey used a
photometric color selection to target galaxies at $0.7 < z < 1.4$ to $R_{AB} =
24.1$. We use the Data Release 4 (DR4)
catalog\footnote{http://deep.ps.uci.edu/dr4/home.html} and associated window
function from \citep{Newman12}. Here we use redshifts between $0.2 < z < 1.2$
in the EGS and redshifts between $0.7 < z < 1.2$ in the other DEEP2 fields. For
all of the DEEP2 fields we require a redshift with a confidence greater than
95\% ($Q \ge 3$). We use the extended optical photometry from
\citet{Matthews13} which contains additional Canada-France-Hawaii Telescope
Legacy Survey (CFHTLS) $ugriz$ and the Sloan Digital Sky Survey (SDSS) $ugriz$
photometry matched to the redshift catalog. K-corrections, absolute $M_B$
magnitudes, and rest-frame colors are derived from K-correct \citep{Blanton07}
from the optical photometry in these fields. The numbers of sources with the
above redshift quality cuts and with estimated stellar masses are given in
Table~\ref{table:numbers}.

We use those sources that fall within the recoverable spatial selection
function of the DEEP2 survey. For the EGS, this precludes the use of the data
from the northern 25\% of the field, which had shallower BRI photometry and
non-uniform targeting. For the other DEEP2 fields we include all of the
pointings presented in \citet{Newman12}. The spatial redshift success fraction
reflects the probability that a targeted source has a secure
$z_\mathrm{quality} \ge 3$ redshift. For the DEEP2 fields we calculate this in
$\sim 6~\arcsecsq$ pixels. Using the average of six adjacent pixels to match
the $\sim 36~\arcsecsq$ pixels used in PRIMUS does not change the resulting
clustering measurements in these fields.

\begin{figure*}
  \epsscale{1.22}
  \epstrim{0.9in 0.6in 1.1in 1.0in}
  \plotone{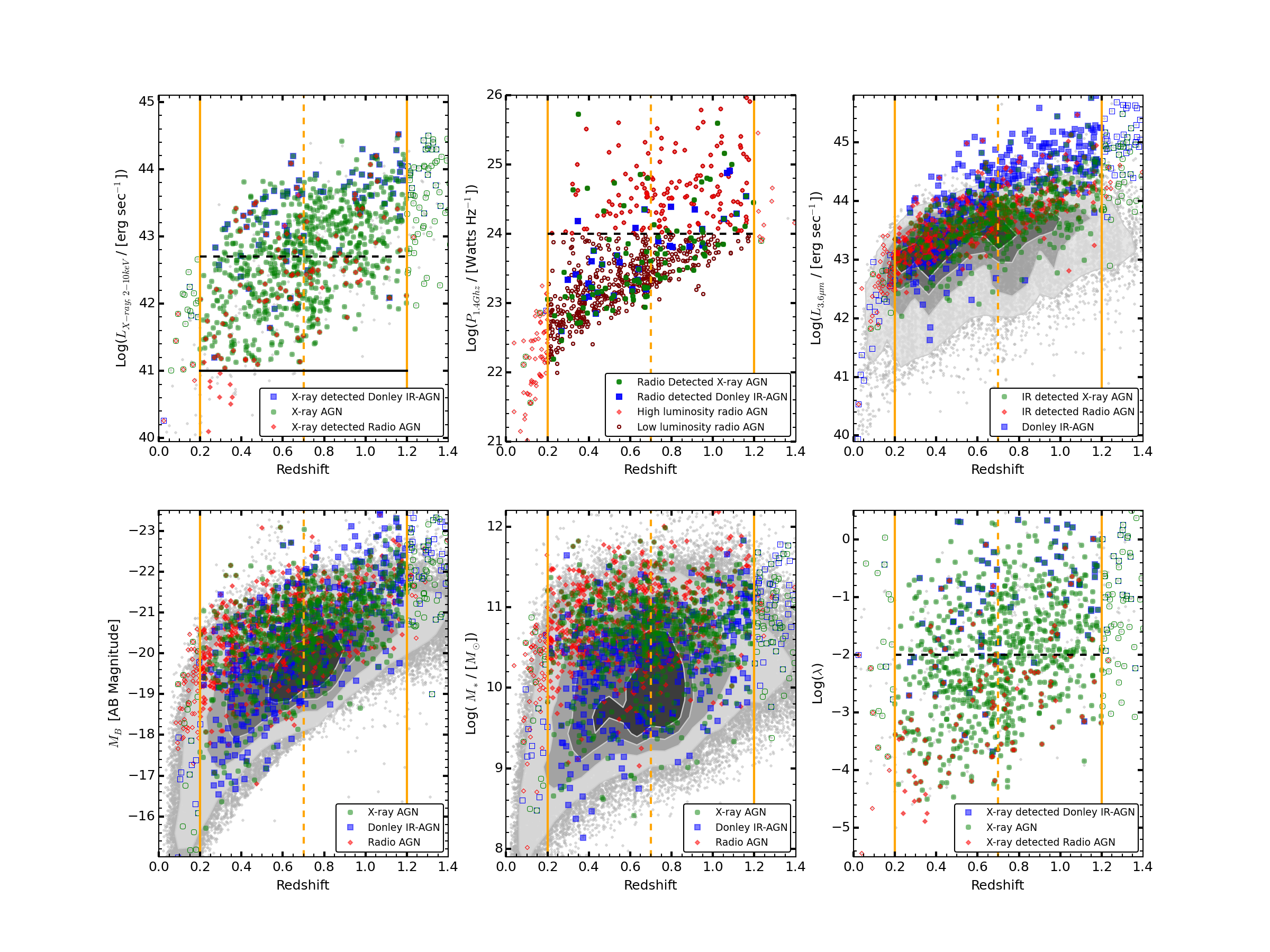}
  \caption{
    The distributions of \Xray, \Radio, and \Donley in \Xray luminosity (upper
    left), radio power (upper center), and 3.6\um\ IR luminosity (upper right),
    as well as host galaxy absolute optical magnitude ($M_B$) (lower left),
    stellar mass (lower center), and X-ray specific accretion rate (lower
    right), all as a function of redshift for $0 < z < 1.4$. \Xray are shown
    with green circles, \Radio with red diamonds, and \Donley with blue
    squares, along with inactive galaxies shown as grey dots with greyscale
    contours containing 30\%, 50\% and 80\% of the full sample. Sources shown
    here have robust spectroscopic redshifts and are not classified as
    broad-line AGN. Solid vertical orange lines show the full redshift range
    used here ($0.2 < z < 1.2$), while the dashed vertical orange line shows
    the redshift used ($z=0.7$) to split the full samples into higher and lower
    redshift samples. In the upper left panel we show a dashed black line for
    the X-ray luminosity cut used to create high and low X-ray luminosity
    samples, while the solid black line shows the lower \Lx{=}{41} luminosity
    cut used for all of the X-ray AGN samples. In upper center panel the dashed
    black line shows the \PR{=}{24} luminosity cut that we use to create high
    and low luminosity radio samples. In the lower right panel we show a dashed
    black line for the specific accretion rate cut used to create high and low
    specific accretion rate samples. \Donley tend to have higher X-ray and IR
    luminosities compared to \Xray and \Radio sources.
}
\label{fig:redshift}
\end{figure*}

\subsection{\iSEDfit Stellar Masses and Star Formation Rates}
\label{sec:massdata}
We estimate stellar masses and SFRs by fitting the
spectral energy distributions (SEDs) of our sources with population synthesis
models using \iSEDfit \citep{Moustakas13}. \iSEDfit is a Bayesian fitting code
that compares the observed photometry for each source to a large Monte Carlo
grid of SED models which span a wide range of stellar population parameters
(e.g. age, metallicity, and star formation history) to estimate the stellar
mass and SFR of a galaxy. We assume a \citet{Chabrier03} initial mass function
from $0.1$ to $100~\mstar$ and use \citet{Bruzual03} stellar population
synthesis models. We assume the following priors to construct the Monte Carlo
grids: uniform stellar metallicity in the range of $0.004 < Z < 0.04$;
\citet{Charlot00} dust attenuation law, with an exponential distribution of
dust, ($0.25 < \gamma < 2.0$); an exponentially declining-$\tau$ ($\phi_{s}(t)
= (\mathcal{M}/\tau) e^{-t/\tau} $) star-formation history (SFH) with $0.01 <
\tau < 5.0$. Stochastic bursts of star formation of varying amplitude,
duration, and onset time are superimposed, allowing for a wide range of
possible star formation histories \citep{Kauffmann03,Salim07}.
While a delayed-$\tau$ model encompasses both a linearly rising ($t/\tau \ll
1$) and an exponentially declining ($t/\tau \gg 1$) SFH history, we find no
significant SFR or stellar mass offsets or trends using different SFH models
for our sources at $z<1.2$, and we therefore choose to use a simpler model of
an exponentially declining SFH. \iSEDfit marginalizes the full posterior
probability distribution of stellar masses and SFRs over all other parameters
and thus encapsulates both the uncertainties in the observations and the model
parameter degeneracies. For each source we take the median stellar mass and SFR
from the full probability distribution functions as the best estimate of the
stellar mass and SFR. The median uncertainties on the log stellar mass and 
SFR are 0.08 dex and 0.2 dex, respectively. In our analysis below we are 
primarily interested in the relative stellar
mass and SFR between sources, such that any overall offsets do not affect 
our results.

We use \iSEDfit stellar masses derived from photometry spanning the UV to the
optical bands. Including the first two IRAC bands ([3.6] and [4.5])
systematically increases the median galaxy sample stellar mass by 0.1 dex. This
is also the case for the X-ray detected sample; however, for the IRAC \Donley
selected sample (details are given below in Section~\ref{sec:irsample}) the
median mass offset is much larger (0.5 dex). As shown in \citet{Mendez13}, this
is due to AGN light contributing to these channels causing the IR-AGN to have
overestimated stellar masses. We therefore do not include the IRAC bands in any
of our stellar mass estimates, such that all stellar masses are derived using
the same photometric bands, minimizing systematic offsets between our samples.
As $\sim82\%$ of
the area covered by PRIMUS has GALEX UV coverage, we include the observed 
FUV and NUV photometry where available to improve the SFR estimates. Including
the GALEX UV bands (compared to just using optical bands alone) slightly
decreases the estimated stellar mass ($\sim0.02\,$ dex) for the galaxy and AGN
samples. We do not estimate stellar masses or SFRs for sources that are deemed
to be broad-line AGN (BLAGN), where their spectra are better matched by BLAGN
templates than by galaxy templates, as their optical photometry will be
dominated by light from the AGN. Table~\ref{table:numbers} lists the total
number of sources with spectroscopic redshifts in each field
($N_\mathrm{galaxy}$) and the number of sources for which we estimate a stellar
mass ($N_\mathrm{mass}$) and SFR.

% replaced from referee report below
% Azadi et al. (2015) investigate how the PRIMUS SFR estimates derived using
% SED fits compare with estimates using 100\micron data for those galaxies
% detected with {\it Herschel}.  They find that a histogram of the SED to
% {\it Herschel}-based SFR differences peaks at zero, though there is a
% non-symmetric tail to higher SFR estimates using the {\it Herschel} data,
% such that the median difference is 0.6 dex.  Most galaxies and AGN are not
% {\it Herschel}-detected, such that the typical difference for the full sample
% could be smaller, as {\it Herschel} detections are biased towards more dusty
% sources.  However, as stated above, in this paper we are concerned only
% with {\it relative} SFRs, such that any overall offsets in the SFR
% estimates will not change our results.

% From Referee Report
\citet{Azadi14} investigate how the PRIMUS SFR estimates derived using SED fits
compare with estimates using 100\micron data for those galaxies detected with
{\it Herschel}. They find that a histogram of the SED to {\it Herschel}-based
SFR differences peaks at zero, though there is a non-symmetric tail to higher
SFR estimates using the Herschel data, such that the median difference is 0.6
dex. Most galaxies and AGN are not {\it Herschel}-detected, such that the
typical difference for the full sample could be smaller, as {\it Herschel}
detections are biased towards more dusty sources. However, in this paper we are
concerned only with {\it relative} SFRs, such that any overall offsets in the
SFR estimates will not change our results.

\subsection{Spatial Selection Function}\label{sec:windowdata}
In order to perform accurate clustering measurements, we require that all of
the PRIMUS and DEEP2 sources used here are located within the area of each
survey that has a well-understood spatial selection function. This ensures that
any spatially-dependent density differences in the surveys that are due to
target selection or missing data, such as in CCD chip gaps or around bright
stars, as well accounted for In PRIMUS we require that sources fall within the
observed window function area targeted with at least two slitmasks.
\citet{Coil11} provides details on the spatial selection function of PRIMUS,
and \citet{Coil04a} and \citet{Newman12} provide details for the DEEP2 survey.
The \Xray, \Radio, and \IR samples are identified within the areas with
observed X-ray, radio, or mid-IR coverage. While there is generally overlap
between the multi-wavelength imaging coverage, there are some areas that lack
full multi-wavelength coverage.

%% file: table_01_numbers.tex
%!TEX root = ms.tex
\begin{deluxetable*}{lrrrrrrrrrr}
\tablewidth{0pc}
\tablecolumns{11}
\tablecaption{Field information, including multi-wavelength coverage area and number of sources.\label{table:numbers}}
\tablehead{
 \multicolumn{1}{c}{} &
 \multicolumn{4}{c}{Area [deg$^2$]} &
 \multicolumn{6}{c}{Number of Detected Sources} \\
  \colhead{\makecell[l]{Field}} &
  \colhead{X-ray} &
  \colhead{Radio} &
  \colhead{IRAC} &
  \colhead{WISE} &
  \colhead{$N_\textrm{Galaxy}$\tablenotemark{a}} &
  \colhead{$N_\textrm{Mass}$\tablenotemark{b}} &
  \colhead{$N_\textrm{X-ray}$\tablenotemark{c}} &
  \colhead{$N_\textrm{Radio}$\tablenotemark{d}} &
  \colhead{$N_\textrm{Donley}$} &
  \colhead{$N_\textrm{Assef}$} 
}
\startdata
CDFS-SWIRE & - & 1.77 & 1.77 & 1.77 & 20,423 & 20,380 & - & 37 (41) & 131 & 44\\
COSMOS & 0.93 & 0.93 & 0.93 & 0.93 & 12,284 & 12,265 & 203 & 94 (361) & 45 & 27\\
Elais - South 1 & 0.51 & 0.90 & 0.90 & 0.90 & 9,922 & 9,903 & 67 & 64 (133) & 59 & 18\\
Extended Groth Strip & 0.69 & 0.71 & 0.61 & 0.71 & 13,957 & 13,178 & 343 & 43 (181) & 64 & 15\\
DEEP2 02hr & 0.58 & 0.61 & 0.60 & 0.58 & 13,222 & 12,961 & 61 & 11 (11) & 33 & 23\\
DEEP2 16hr\tablenotemark{e} & 0.73 & 0.73 & - & 0.73 & 5,645 & 5,426 & 31 & 6 (6) & - & 9\\
DEEP2 23hr & 0.89 & 0.92 & - & 0.67 & 13,486 & 13,239 & 75 & 9 (10) & - & 19\\
XMM LSS/SXDS & 2.88 & 2.88 & 2.84 & 2.88 & 35,460 & 35,388 & 178 & 78 (141) & 157 & 79\\
\hline \\[-2ex]
Totals: & 7.21 & 9.45 & 7.64 & 9.17 & 124,399 & 122,740 & 958 & 342 (894) & 489 & 234\\
\enddata
\tablenotetext{a}{Redshifts limited to $0.2\leq z\leq1.2$ and $z_\textrm{quality} \geq 3$.}
\tablenotetext{b}{Broad-line AGN are excluded, as the optical light for these AGN is contaminated and prevents an accurate estimate of the stellar mass.}
\tablenotetext{c}{X-ray detected sources with \Lx{>}{41}.}
\tablenotetext{d}{Radio AGN with \PR{\geq}{24}. Number within parentheses represents all detected radio sources with a robust redshift.}
\tablenotetext{e}{PRIMUS did not survey the DEEP2 16hr field, and we include it only for samples at $0.7<z<1.2$.}
\end{deluxetable*}

%% file: samples.tex
%!TEX root = ms.tex

\input{table_02_samples}

\begin{figure}
  \epsscale{1.1}
  \plotone{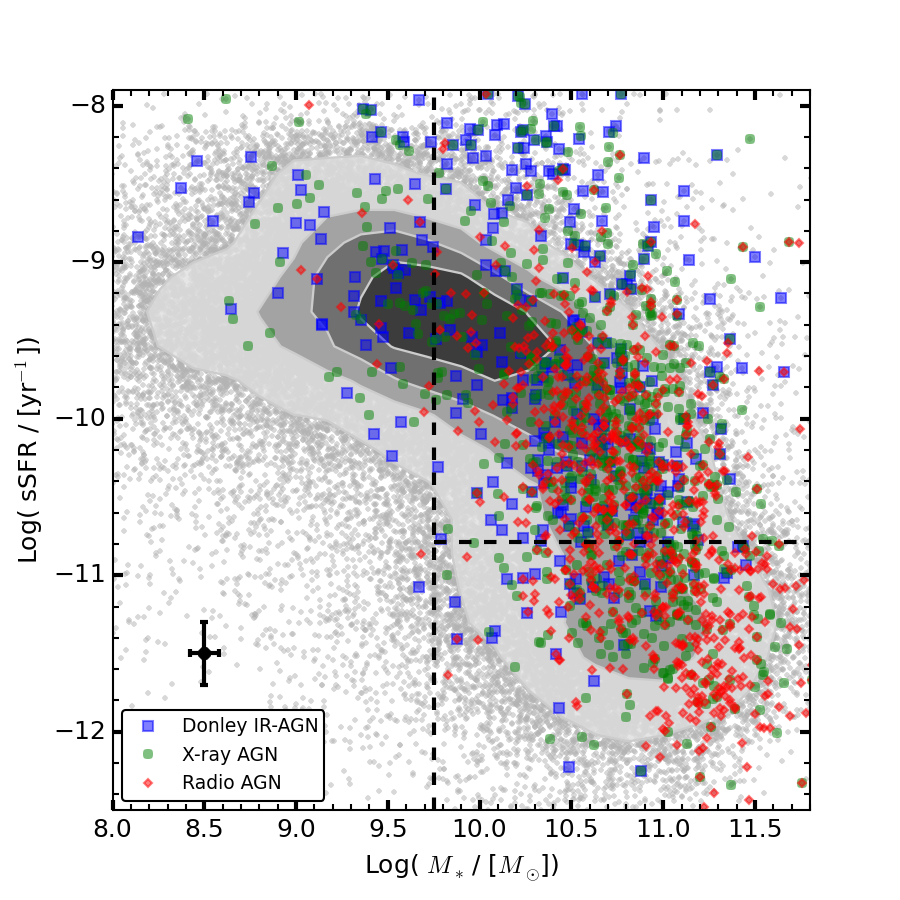}
  \caption{
    Specific star-formation rate (\ssfr) versus stellar mass for the \Xray,
    \Radio, and \Donley sources. Colors and contours are similar to
    Figure~\ref{fig:redshift}. Dashed black lines show the stellar mass and
    \ssfr cuts used to define the \ssfr selected samples. AGN
    identified at different wavelengths have different distributions in both
    \ssfr and stellar mass, though AGN identified at any 
    wavelengths are typically identified in massive host galaxies. 
    Median 1$\sigma$ errors on both parameters are shown in the lower left corner.
}
  \label{fig:properties}
\end{figure}

\section{AGN and Galaxy Samples}\label{sec:sample}
The goal of this paper is to quantify and compare the clustering properties of
X-ray-, radio-, and mid-IR-selected AGN at $z\sim0.7$ with each other, as well
as with inactive galaxies with stellar mass, SFR and redshift
distributions that match the AGN samples. To this end we select AGN and galaxy
samples using the PRIMUS and DEEP2 surveys in regions with either X-ray, radio
or mid-IR imaging coverage. Below we present our selection criteria for our AGN
and matched galaxy samples. Details of each AGN sample are given in
Table~\ref{table:samples}.

\subsection{X-ray AGN Samples}\label{sec:xraysample}
For the fiducial X-ray AGN sample we require that the detected X-ray sources have a
hard-band X-ray luminosity \Lx{>}{41} and a redshift in the range $0.2<z<1.2$. 
We choose to use an X-ray luminosity limit of \Lx{>}{41} rather than
a more conservative \Lx{>}{42} limit, as this leads to larger samples 
with smaller uncertainties and no significant differences in our results. 
We have applied X-ray K-corrections ( $(1+z)^{(\Gamma-2)}$; $\Gamma\sim1.7$ )
to estimate the hard-band X-ray luminosity. We create a `non-broadline'
subsample where we remove the sourced identified as BLAGN in their PRIMUS or
DEEP2 spectra. The fiducial \Xray sample is additionally divided into six
subsamples defined either by an AGN property (\LX, specific accretion rate, or
hardness ratio) or a host galaxy property (redshift, stellar mass, or \ssfr),
in order to investigate clustering trends with both AGN and host galaxy
properties.

For the X-ray AGN samples split by AGN luminosity, we divide the fiducial \Xray
sample into low luminosity (\medianLx{42.4}) and high luminosity (\medianLx{43.2})
samples using a luminosity cut, shown in the upper left panel of
Figure~\ref{fig:redshift}.

We also split the fiducial \Xray sample by specific accretion rate, defined as
\begin{eqnarray}
    \lambda_\mathrm{Edd} &=& \frac{L_\mathrm{Bol}}{L_\mathrm{Edd}} \\
    &=& \frac{L_\mathrm{Bol}}{1.3\,\times\,10^{38}\, \ergs\, 
                                   \times\,0.002\,\frac{\mstar}{\msun} }
\end{eqnarray}
\noindent where $L_\mathrm{Edd}$ is the Eddington limit, and $L_\mathrm{Bol}$
is the bolometric luminosity derived using the X-ray luminosity to bolometric
luminosity relationship of \citet{Hopkins08} in units of \ergs. The specific
accretion rate is a rough estimate of the Eddington ratio, assuming a constant
scaling relationship between black hole mass and host stellar mass \citep[e.g.,
$\mbh\sim0.002\,\mstar$; ][]{Marconi03}. While there is substantial scatter in
both the $M-\sigma$ relationship and in the scaling between bulge mass and
stellar mass of the galaxy, such that the specific accretion rate is not an
exact estimate of Eddington ratio, it is a robust tracer of the rate at which
the SMBH is growing relative to the stellar mass of the host galaxy
\citep{Aird12}.

We create high and low specific accretion rate ($\lambda$) samples only for AGN
with a host galaxy stellar mass above the stellar mass limit (\mass{=}{9.75})
for which we are complete for quiescent galaxies at the highest redshifts used
here and divide the fiducial X-ray sample at roughly the median specific accretion
rate of \lambdavalue{=}{-2} (see the lower right panel of
Figure~\ref{fig:redshift}). We also create X-ray AGN samples based on hardness
ratio, defining hard and soft samples by dividing the fiducial sample at $HR=0$ and
requiring that the AGN included are identified in both the soft and hard X-ray
bands. 
We adopt a simple cut of $HR=0$ as this approximately corresponds to $N_H = 3
\times 10^{22} cm^{-2}$, assuming a simple absorbed power-law X-ray spectrum
with $\Gamma=1.9$ at $z=0.6$ (the approximate median redshift of our sample),
and thus roughly divides the sample into obscured and unobscured populations
\citep{Szokoly04, Hasinger08}.

\begin{figure*}
  \epsscale{1.1}
  \plotone{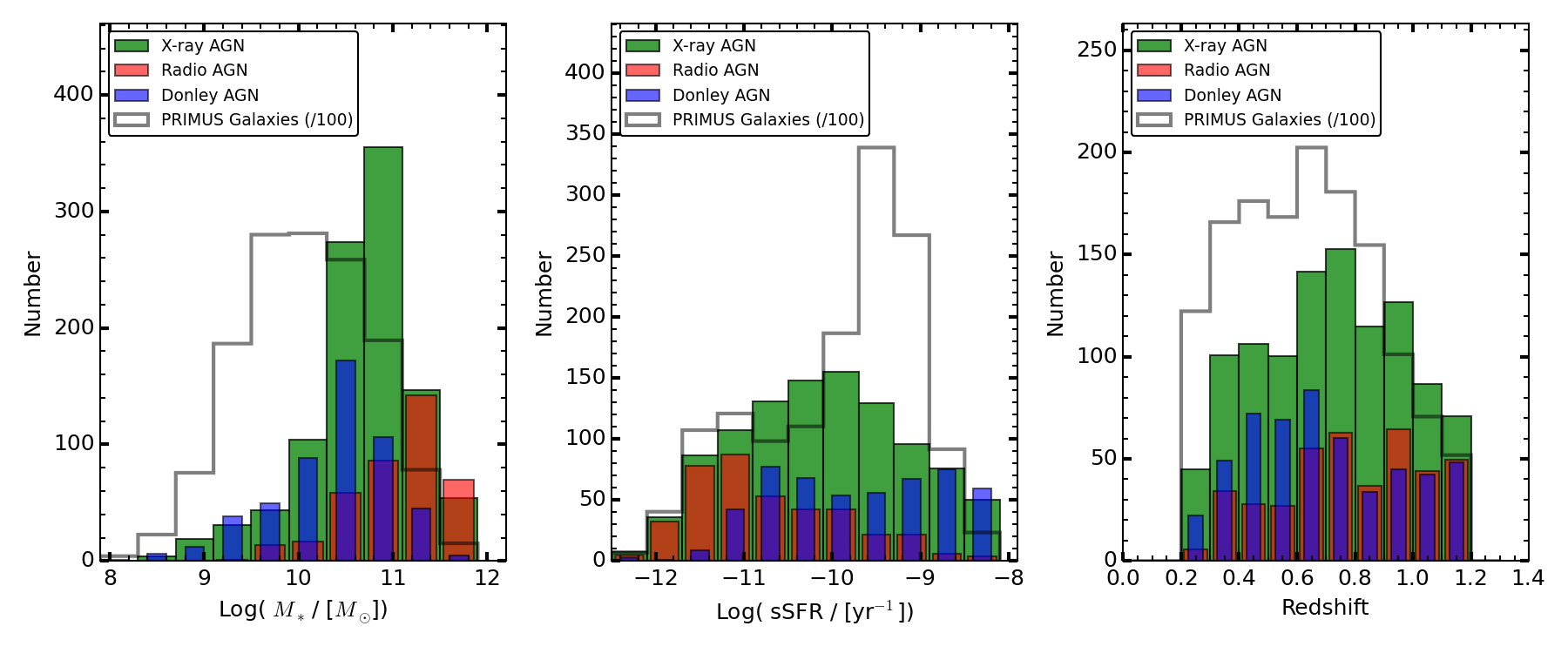}
  \caption{ 
    Stellar mass, \ssfr, and redshift distributions for the \Xray, \Radio,
    \Donley, and galaxy tracer samples. The green, red, and blue filled
    histograms show the distributions of the \Xray, \Radio, and \Donley,
    respectively. The grey line showing the distributions for the galaxy tracer
    sample is scaled down by a factor of 100 for ease of comparison. 
    Differences between the underlying galaxy sample
    and AGN identified at different wavelengths are dominated by the difficulty 
    in selecting AGN of a given specific accretion rate in low mass galaxies \citep[e.g.,][]{Aird12}.
    Additionally, there are
    substantial differences in the stellar mass, \ssfr, and redshift
    distributions between AGN identified at different wavelengths.}
  \label{fig:matched}
\end{figure*}

We further divide the fiducial \Xray sample by various host galaxy properties, to
quantify how the clustering of X-ray AGN depends on the host galaxy. We create
a high and low redshift sample by dividing the fiducial \Xray sample at $z\sim0.7$.
For both the stellar mass samples and \ssfr samples, we require that the host
galaxy has a stellar mass above \mass{=}{9.75} (see
Figure~\ref{fig:properties}). We define the high and low stellar mass samples
using the median stellar mass of the fiducial \Xray sample (\mass{\sim}{10.65}),
and we define high and low \ssfr samples by dividing the \ssfr at
$\ssfr\,=\,10.65\,\ssfrunit$ (see Figure~\ref{fig:properties}). This cut
roughly matches the evolving SFR-mass cut of \citet{Moustakas13} at $z\sim0.7$
that divides the galaxy sample into quiescent and star-forming galaxies.

\subsection{Radio AGN Samples}\label{sec:radiosample}
We define four \Radio samples based on the observed optical broad lines and
measured radio luminosity of each source. For our fiducial \Radio sample we require
$0.2<z<1.2$. We create a `non-broadline' subsample where we remove the BLAGN
identified by their PRIMUS or DEEP2 optical spectra. We have applied a radio
K-correction ($(1+z)^{(\alpha-1)}$; $\alpha\sim0.5$) when estimating the radio
luminosity. Radio continuum emission may contain contributions from thermal
bremsstrahlung (free-free) emission in star forming galaxies as well as from
non-thermal synchrotron emission associated with radio jets emanating from an
AGN. To separate these two populations, we follow \citet{Condon92} and
\citet{Murphy11} and define a high luminosity radio sample with \PR{>}{24}, to
remove any potential contamination from luminous starburst galaxies. Above this
luminosity the radio emission cannot be explained by even extreme star
formation (SFR $>10^3$\sfrunit) \citep{Goulding12,Hickox09}. This \Radio sample
reliably contains radio-loud \citep[Class FR-II; ][]{Fanaroff74} sources. The
small sample size in the high luminosity sample limits our analysis of the
\Radio sample to the fiducial redshift range ($0.2<z<1.2$), as we do not have
enough sources to create subsamples at different redshifts.

This sample of radio-loud sources necessarily does not contain radio-quiet AGN
\citep{Mullaney13}. A variety of optical, mid-IR, or far-IR to radio flux ratio
excess techniques have been suggested to identify more complete samples of
radio-quiet AGN while limiting contamination from star forming galaxies
\citep[e.g.][]{Smolcic08, Park08, Donley05, Appleton04}. To investigate the
clustering properties of radio-quiet AGN, we define a low luminosity radio
sample (\PR{<}{24}). This sample includes all radio-detected sources below the
luminosity limit, identifying all possible optical- or IR-excess selected
sources.

\begin{figure}
  \epsscale{1.2}
  \plotone{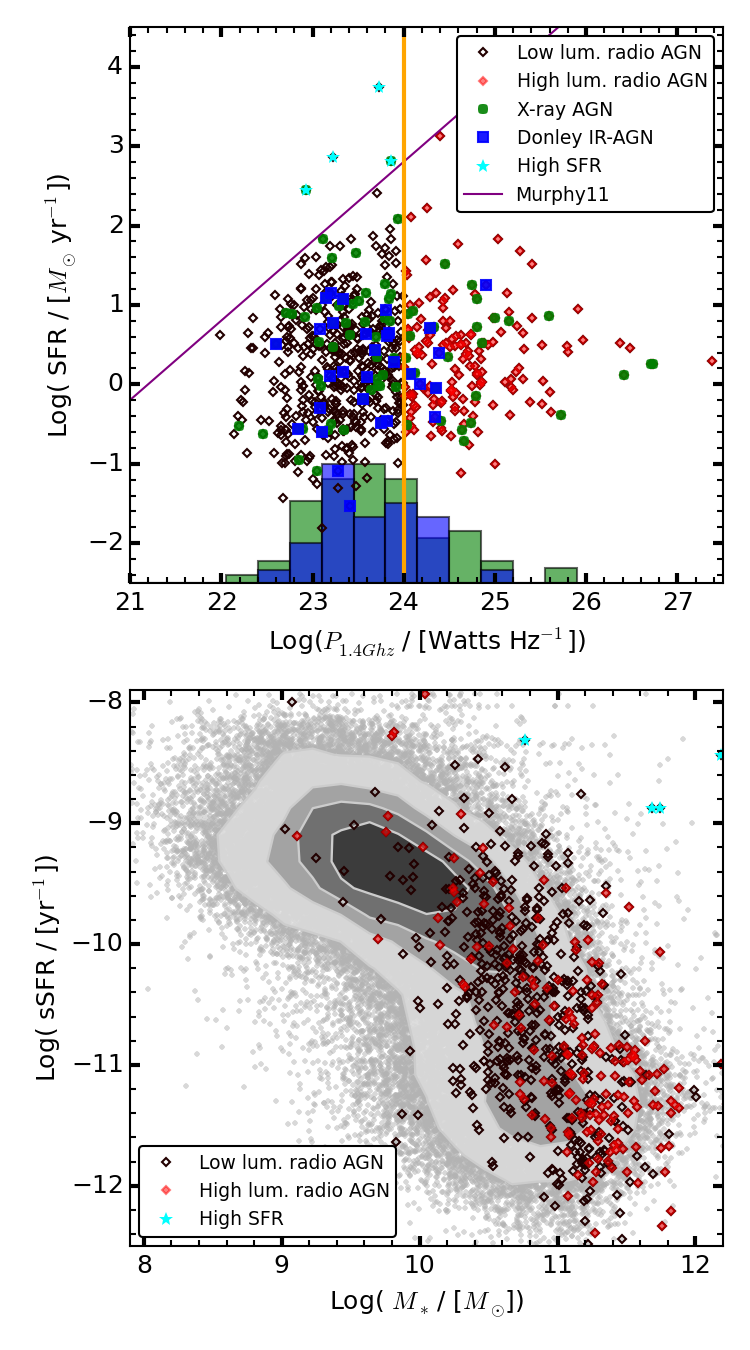}
  \caption{ Host galaxy properties for \Radio. \textbf{Top:} SFR versus radio
  luminosity for the fiducial radio-detected sample. High luminosity (\PR{>}{24})
  \Radio are highlighted as red points. \Xray and \Donley that are radio
  detected are shown with green and blue points, respectively, with relative
  distributions shown at the bottom. Radio AGN above the \citet{Murphy11}
  SFR-to-radio luminosity relationship (purple line) are highlighted in cyan.
  In only these four sources can the radio light be dominated by star formation
  and not AGN activity.
  \textbf{Bottom:} Comparison of low luminosity (black) and high luminosity
  (red) radio AGN in the sSFR and stellar mass plane. The PRIMUS galaxy sample
  is shown in grey contours, with outliers beyond the 90\% contour shown with
  grey points. The four sources above the SFR-to-radio luminosity relationship
  of \citet{Murphy11} are shown as cyan points. }
  \label{fig:radio_ssfr}
\end{figure}

In order to investigate possible contamination of this low luminosity radio
sample by star forming galaxies, in the upper panel of
Figure~\ref{fig:radio_ssfr} we show SFR versus radio luminosity for the radio
detected sample. We highlight high luminosity (\PR{>}{24}) sources in red,
\Xray that are radio detected in green, and \Donley that are radio detected in
blue. The radio luminosity distribution of the \Xray and \Donley samples are
shown as normalized histograms at the bottom of the panel. Cyan points
highlight the few highly star forming sources where their radio luminosity can
be explained solely as due to star formation using the \citet{Murphy11}
SFR-to-radio luminosity relationship (purple line). The small number of sources
($\lesssim1\%$) above this lin shows that neither the high nor low luminosity
radio samples are contaminated by star forming galaxies.
The median uncertainty on the SFR estimates is less than
0.2 dex, suggesting that only a few sources could be scattered above the 
\citet{Murphy11} line. 
We also find that most of the X-ray and mid-IR AGN that are
radio-detected are in the low luminosity radio sample \PR{<}{24}, supporting
the conclusion of minimal contamination from star forming galaxies.

In the bottom panel of Figure~\ref{fig:radio_ssfr} we show the sSFR versus
stellar mass diagram for all PRIMUS galaxies (grey contours), low luminosity
(black points) and high luminosity \Radio (\PR{>}{24}, red points). Radio
detected sources are preferentially identified in massive galaxies, with the
high luminosity sources found in the most massive galaxies. As we do not find
large differences in the sSFRs of the low luminosity radio 
sources compared to the high luminosity radio sources, 

\subsection{IR-AGN Samples}\label{sec:irsample}
We use the \citet{Donley12} IRAC color-color selection to identify mid-IR red
power law AGN. As shown in \citet{Mendez13}, in the PRIMUS survey 
this selection provides reliable
identification of luminous AGNs with minimal contamination by star
forming galaxies. We require
that objects are detected in all four IRAC bands and have colors such that
they lie within the following region in IRAC color-color space:
\begin{eqnarray}
  x={\rm log_{10}}\left( \frac{f_{\rm 5.8 \um}}{f_{\rm 3.6 \um}} \right), \quad
  y={\rm log_{10}}\left( \frac{f_{\rm 8.0 \um}}{f_{\rm 4.5 \um}} \right)
\end{eqnarray}
\begin{eqnarray}
  x &\ge& 0.08 \textrm{~ and ~} y \ge 0.15 \textrm{~ and ~} \\
  y &\ge& (1.21\times{x})-0.27 \textrm{~ and ~} \\
  y &\le& (1.21\times{x})+0.27 \textrm{~ and ~} \\
  f_{\rm 4.5 \um} &>& f_{\rm 3.6 \um} \textrm{~ and ~} % no new line here.
  f_{\rm 5.8 \um} > f_{\rm 4.5 \um} \textrm{~ and ~} \\
  f_{\rm 8.0 \um} &>& f_{\rm 5.8 \um}.
\end{eqnarray}
\noindent The small sample size of the \Donley sample limits our analysis to
the fiducial redshift range ($0.2<z<1.2$).

Additionally, we identify WISE-selected IR-AGN using the \citet{Assef13}
magnitude-dependent selection. We require sources to have measured W1 and W2
fluxes such that,
\begin{eqnarray}
    W1-W2 &>& 0.662~\textrm{exp}\left[0.232~(W2-13.97)^2  \right] \\
       W2 &<& 17.11
\end{eqnarray}
\noindent where W1 and W2 are in Vega magnitudes. \citet{Assef13} show that
this selection is $90\%$ reliable in its identification of IRAC selected AGN.
This selection extends the \citet{Stern12} WISE IR-AGN color selection to
fainter limiting magnitudes, while controlling for contamination (see
\citet{Assef13} for details).

\begin{figure}
  \epsscale{1.2}
  \plotone{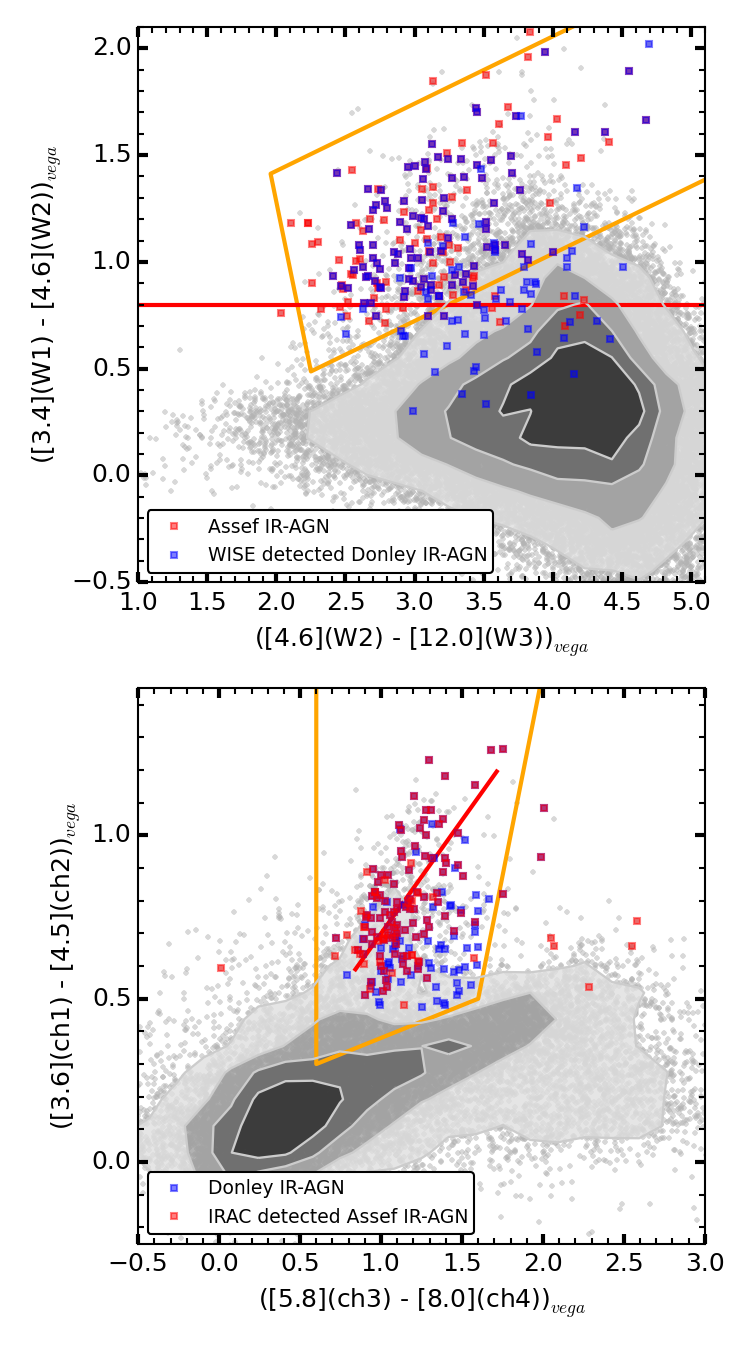}
  \caption{ Comparison of the \Donley (blue) and \Assef (red) samples.
  \textbf{Top:} WISE color-color diagram with the \citet{Mateos13} selection
  wedge shown in orange and the \citet{Stern12} color selection shown in red.
  WISE sources identified by the \Assef selection technique are shown with red
  points and \Donley that are also detected in WISE are shown with blue points.
  Grey contours show all PRIMUS galaxies that are detected in WISE.
  \textbf{Bottom:} IRAC color-color diagram with the \citet{Stern05} selection
  wedge shown in orange and the power law locus indicated in red. \Assef are
  shown with red points and \Donley are shown with blue points. A small number
  of the \Assef that are IRAC detected fall outside of the power law region and
  Stern wedge, due to the selection not utilizing the longer WISE wavelengths. }
  \label{fig:ir_selection}
\end{figure}

In Figure~\ref{fig:ir_selection} we compare the MIR colors of \Donley and
\Assef. In the top panel we show the WISE selection plane with the
\citet{Mateos12} color-color selection wedge in orange and the \citet{Stern12}unobscured
color selection limit in red. We show WISE-detected \Donley with blue points,
\Assef with red points, and PRIMUS galaxies with grey contours and grey outlier
points. In this color-color space the \Donley and \Assef generally have similar
MIR colors, with the \Donley extending to slightly lower [W1-W2] colors. In
the bottom panel we show the IRAC color-color selection wedge with the
\citet{Stern05} selection in orange and the power law locus in red. We show
\Donley with blue points, \Assef that are IRAC detected with red points, and
the PRIMUS galaxy sample in grey contours. We find that some \Assef have high
IRAC [Ch3-Ch4] colors, beyond the \citet{Stern05} wedge and power law locus.
This results from the \Assef selection using only the two shorter wavelength
WISE bands and not using the longer wavelength information. These few sources
are likely not AGN as they reside in the region of this diagram dominated by
low-redshift ($z\lesssim0.3$), star-forming galaxies \citep{Mendez13}. The
\Donley selection uses all four bands to ensure a red monotonically increasing
flux in the MIR, at the cost of requiring detections in the generally shallow
longer wavebands.

\begin{figure}
  \epsscale{1.2}
  \plotone{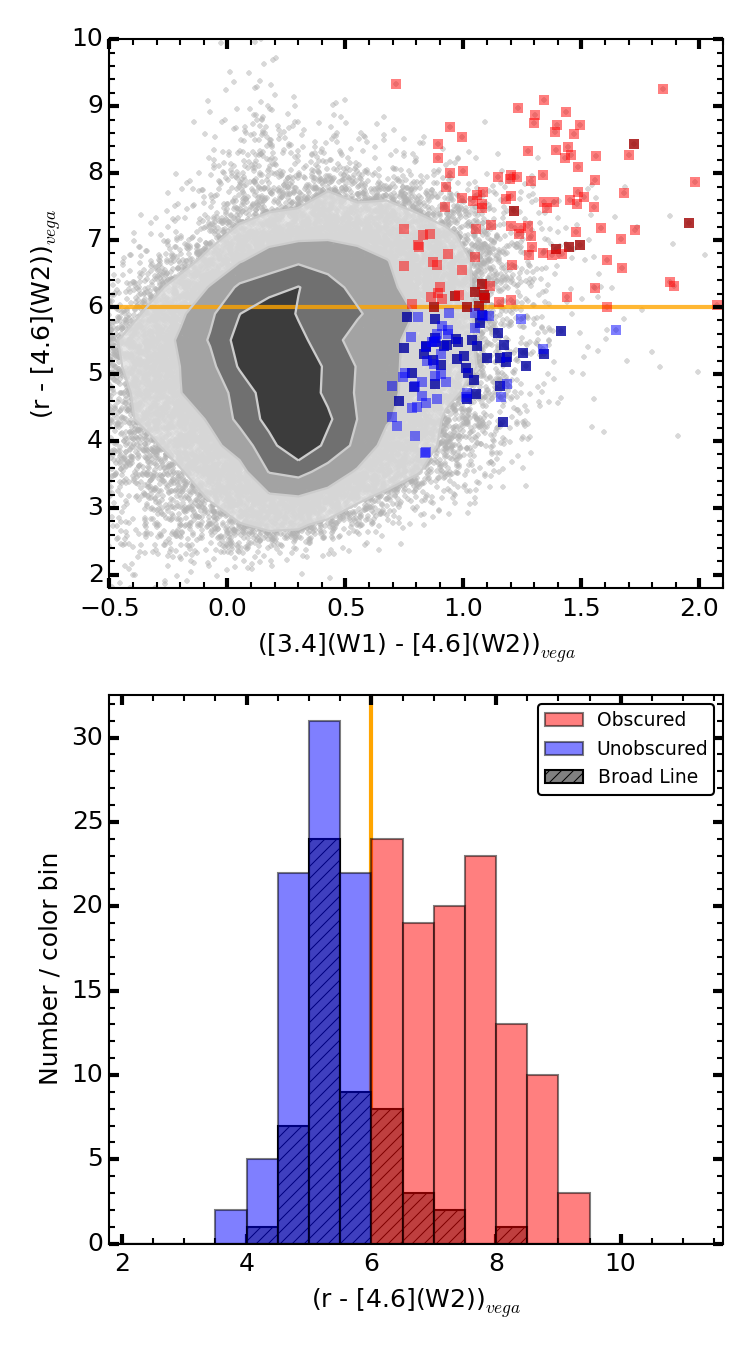}
  \caption{ Comparison of obscured and unobscured \Assef. \textbf{Top:}
  \citet{Yan13} optical and MIR color selection plot showing obscured (red) and
  unobscured (blue) \Assef. WISE-detected PRIMUS galaxies are shown with grey
  contours. \textbf{Bottom:} Optical to MIR color distribution for \Assef.
  Obscured \Assef are shown with a red histogram, unobscured \Assef with a blue
  histogram, and the broad line \Assef with a black histogram. Most broad line
  sources have optical-to-MIR colors of unobscured AGN. }
  \label{fig:obscured_selection}
\end{figure}

We divide the WISE IR-AGN sample into obscured and unobscured subsamples. We
use the criteria of \citet{Yan13}, who use obscured and unobscured templates at
$z<1.5$ to define a MIR-to-optical color cut of $(r-W2)\sim6$ to separate these
sources. Due to differences in the photometric filters in our measured r-band
magnitudes, we use a synthesized SDSS r-band magnitude from \texttt{K-Correct}
to ensure a uniform selection in each field. In the top panel of
Figure~\ref{fig:obscured_selection} we show the selection of our samples in the
optical and MIR color-color diagram, where obscured \Assef are shown in red and
unobscured \Assef in blue. In the bottom panel we show the optical to MIR color
distributions of the obscured (red) and unobscured (blue) samples, as well as
the optically identified broad line AGN (black) in these samples. We find that
most broad line AGN have colors that identify them as unobscured AGN.

We additionally test the \citet{Mateos12} WISE IR selection technique. Similar
to the \Donley selection, it identifies sources with a red power law in the
mid-IR. This technique is more robust than that of \citet{Assef13} as it uses
longer wavelength information (W3: 12\um) to ensure a monotonic mid-IR SED, but
it is less complete due to the relatively shallow W3 coverage in the WISE
survey. As we find no significant differences in the clustering properties of
the AGN samples defined using \citet{Mateos12} and \citet{Assef13}, we use the
slightly larger \Assef sample throughout this paper.

\subsection{Galaxy Tracer Samples}\label{sec:tracersample}
We use the dense galaxy samples provided by the PRIMUS and DEEP2 redshift
surveys to measure the clustering of AGN using a cross-correlation measurement
with galaxies. To do this we define galaxy ``tracer'' samples to trace the
cosmic web in the fields and at the redshifts of interest. For the fiducial galaxy
tracer sample, we use all galaxies with robust redshifts within the fiducial
redshift range used here, $0.2 < z < 1.2$. We do not require that the galaxy
tracer sample be volume limited, as we are using it only to trace the
large-scale structure in these fields; it therefore needs to span the same
volume as our AGN samples, but it does not need to have the same median
luminosity at all redshifts. We additionally split the fiducial galaxy tracer
sample into low and high redshift subsamples for the \Xray sample, split at the
redshift of $z=0.7$ (Figure~\ref{fig:redshift}, dashed orange line in each
panel). This redshift cut divides the number of \Xray into approximately equal
sized samples and results in 30\% more tracer galaxies at lower redshifts than
at higher redshifts.

\subsection{Galaxy Matched Control Samples}\label{sec:matchedsample}
We construct galaxy samples for each of the fiducial \Xray, \Radio, and \Donley
samples with matched stellar mass, sSFR, and redshift distributions. We use
these as control samples to compare the clustering of galaxies that host AGN to
similar galaxies without AGN.  This limits the effect of AGN 
selection biases such as AGN being preferentially identified in galaxies 
with high stellar mass \citep[e.g.,][]{Aird12}.
Additionally, comparing the clustering of AGN to matched galaxy samples 
cancels out any effective flux limits that
arise from either the multi-wavelength AGN selection or 
the spectroscopic redshift requirement \citep[e.g.,][]{Leauthaud15}.
Effectively, we use individual matched galaxy control samples for each of the
\Xray, \Radio, and \Donley samples to control for differences in the host
galaxy properties that each AGN selection identifies. \citet{Coil09} and
\citet{Hickox09} created similar matched galaxy samples, however they matched
rest-frame optical magnitude and color. While these are easily observed
properties, they are not as physically-motivated as stellar mass and sSFR.
While we find no significant differences in our results matching joint stellar
mass, sSFR, and redshift distributions rather than magnitude, color, and
redshift distributions, we use the former parameters as they reflect intrinsic
host galaxy physical properties.

We measure the joint stellar mass, sSFR, and redshift distribution for
each of the \Xray, \Radio, and \Donley samples in stellar mass bins of
$\Delta\mstar\,=\,0.2$ dex, sSFR bins of $\Delta{sSFR}\,=\,0.2$ dex, and
redshift bins of $\Delta{z}\,=\,0.1$. The projected distributions of each
sample are shown in Figure~\ref{fig:matched}.
Normalizing each distribution by the total number of galaxies in each bin results in 
the observed fraction of AGN in that bin, effectively estimating the
probability density in this three-dimensional parameter space. We use this as
an estimate of the probability density to weight each inactive galaxy when 
creating matched galaxy samples to compare to various AGN samples.
To ensure that we can robustly estimate the stellar
masses and SFRs of the AGN, we limit the comparison of matched
galaxies to AGN that do not have any broad-line emission in their
optical spectra, such that the optical light is dominated by the host
galaxy.

%% file: table_02_samples.tex
%!TEX root = ms.tex
\begin{deluxetable*}{llrrrrrr}
\tablewidth{0pc}
\tablecolumns{8}
\tablecaption{Information on the X-ray, radio, and \IR samples. \label{table:samples}}
\tablehead{
  \colhead{\makecell[l]{AGN Selection}} &
  \colhead{\makecell[l]{Sample Name}} &
  \colhead{\makecell[c]{Redshift\\Range}} &
  \colhead{Number\tablenotemark{a}} &
  \colhead{Density\tablenotemark{b}} &
  \colhead{\median{z}{}} &
  \colhead{\median{L}{}\tablenotemark{c}} &
  \colhead{\median{M}{*}\tablenotemark{d}} 
}
\startdata
X-ray AGN & Full & $0.2 - 1.2$ & 958 & 0.79 & 0.73 & 42.8 & 10.7\\
% Radio AGN & Full & $0.2 - 1.2$ & 327 & 0.21 & 0.80 & 24.3 & 11.17\\ % high lum
Radio AGN & Full & $0.2 - 1.2$ & 894 & 0.55 & 0.62 & 23.5 & 10.9\\
Donley IR-AGN & Full & $0.2 - 1.2$ & 583 & 0.44 & 0.68 & 44.2 & 10.5\\
\hline \\[-2ex]
X-ray AGN & Non-broadline & $0.2 - 1.2$ & 633 & 0.53 & 0.72 & 42.7 & 10.8\\
% Radio AGN & Non-broadline & $0.2 - 1.2$ & 272 & 0.17 & 0.80 & 24.3 & 11.1\\
Radio AGN & Non-Broadline  & $0.2 - 1.2$ & 768 & 0.48 & 0.61 & 23.5 & 10.91\\
Donley IR-AGN & Non-broadline & $0.2 - 1.2$ & 328 & 0.25 & 0.61 & 44.0 & 10.4\\
\hline \\[-2ex]
X-ray AGN & Low \LX & $0.2 - 1.2$ & 288 & 0.26 & 0.57 & 42.2 & 10.7\\
X-ray AGN & High \LX & $0.2 - 1.2$ & 328 & 0.33 & 0.87 & 43.3 & 10.8\\
X-ray AGN & Low \specific & $0.2 - 1.2$ & 329 & 0.27 & 0.65 & 42.4 & 10.9\\
X-ray AGN & High \specific & $0.2 - 1.2$ & 305 & 0.26 & 0.90 & 43.4 & 10.7\\
X-ray AGN & Low HR & $0.2 - 1.2$ & 671 & 0.54 & 0.73 & 42.9 & 10.7\\
X-ray AGN & High HR & $0.2 - 1.2$ & 287 & 0.25 & 0.73 & 42.7 & 10.8\\
\hline \\[-2ex]
Radio AGN & High \Pr & $0.2 - 1.2$ & 327 & 0.21 & 0.80 & 24.3 & 11.2\\
Radio AGN & Low \Pr & $0.2 - 1.2$ & 569 & 0.34 & 0.50 & 23.3 & 10.8\\
\hline \\[-2ex]
Assef IR-AGN & Full & $0.2 - 1.2$ & 234 & 0.14 & 0.74 & 44.5 & 10.6\\
Assef IR-AGN & Obscured WISE color & $0.2 - 1.2$ & 129 & 0.08 & 0.77 & 44.5 & 10.6\\
Assef IR-AGN & Unobscured WISE color & $0.2 - 1.2$ & 106 & 0.06 & 0.70 & 44.5 & 10.6\\
\enddata
\tablenotetext{a}{Number of sources in window function with applied selection cuts.}
\tablenotetext{b}{Density is in units of $[10^{-4} h^3\,\textrm{Mpc}^{-3}]$.}
\tablenotetext{c}{\median{L}{} is \median{L}{X} $[\textrm{log}(\textrm{erg}\,\textrm{s}^{-1})]$ for \Xray samples, 
\median{P}{1.4Ghz} $[\textrm{log}(\textrm{Watts}\, \textrm{Hz}^{-1})]$ for \Radio, and \median{L}{3.6\,\um} $[\textrm{log}(\textrm{erg}\,\textrm{s}^{-1})]$ for IR-AGN samples.}
\tablenotetext{d}{Units of $[\textrm{log}(h^{-1}\, \msun)]$}
\end{deluxetable*}

%% file: methods.tex
%!TEX root = ms.tex

\section{Clustering Analysis}\label{sec:clustering}
We measure the spatial distribution of AGN using the two-point correlation
function, which quantifies the excess probability above Poisson of finding two
sources with a given physical separation. While most studies measure the
auto-correlation function (ACF) of the AGN sample of interest, here we measure
the cross-correlation function (CCF) of AGNs with galaxies, from which we then
infer the ACF of the AGN alone. As discussed in \citet{Coil09} there are two
main advantages to this method. First, the CCF of AGN and galaxies has a much
greater statistical power due to the larger number density of galaxies, which
better probe the underlying dark matter distribution where AGN are located.
Second, it does not require a complete understanding of the AGN selection
function, which may not be well understood. Instead, all that is required is an
understanding of the selection function of the galaxy tracer sample.

\subsection{Measuring the Cross-Correlation Function}\label{sec:calccluster}
The two-point correlation function $\xi(r)$ is defined as the excess
probability density, $d\,P / d\,V$, above that of a Poisson random field, of a
second source being physically separated by a distance $r$ from a given source,
\begin{eqnarray}
  \frac{d\,P}{d\,V} &=& n[1 + \xi(r)]
\end{eqnarray}
\noindent where $n$ is the mean number density of the sample of interest
\citep{Peebles80}. The ACF measures the clustering of a single sample, where
the two sources are from the same sample, while the CCF measures the clustering
of one type of source, taken from one sample, around that of another type of
source, taken from a second sample. Here we measure the CCF of AGN (A) around
galaxies (G), which are used as a tracer sample, and find the excess
probability above random (R) of finding AGN and galaxies with a given
separation ($r$).  We use the \citet{Davis83} estimator:
\begin{equation} \label{eqn:davis}
  \xi(r) = \frac{AG(r)}{AR(r)} - 1 
\end{equation}
\noindent where $AG(r)$ is the sum of the weighted AGN-galaxy pairs and $AR(r)$
is the sum of the weighted AGN-random pairs, both as a function of separation.
Weights are used to account for target selection in the PRIMUS sample (see
Section~\ref{sec:primusdata}); by applying these weights we are able to create
a statistically-complete sample that is not subject to spatial biases. In the
DEEP2 fields the weights are included in the spatial selection function which
we use to generate the random catalogs, such that galaxies have unity weight.
We calculate the weighted number of pairs:
\begin{eqnarray}
  AG \, &=& \, \sum\limits_{i\in{A},\atop j\in{G}} 
                            \frac{ w_{\mathrm{AGN}; i} \times 
                                          w_{\mathrm{galaxy}; j} }
                                          {W_\mathrm{AGN} \times 
                                          W_\mathrm{galaxy}} \\
  AR \, &=& \,  \sum\limits_{i\in{A}} 
                              \frac{ w_{\mathrm{AGN}; i}  }
                                   {W_\mathrm{AGN} \times N_\mathrm{random}} 
\end{eqnarray}
\noindent where $w_\mathrm{AGN}$ is the weight of a given AGN,
$w_\mathrm{galaxy}$ is the weight of a given galaxy, $W_\mathrm{AGN}$
is the total AGN weight, $W_\mathrm{galaxy}$ is the total galaxy weight, and
$N_\mathrm{random}$ is the number of random objects. The AGN weight is the
multiplicative combination of the targeting weight and any additional
completeness weight such as the \Xray weight (see Section~\ref{sec:xraydata}
for details). For the DEEP2 fields the targeting weight is unity for each
source.

Peculiar velocities distort \xir\ measurements in the redshift direction,
along the line of sight. We therefore measure \xir\ in two dimensions,
\xisp, where \rp\ is the separation perpendicular to the line of sight, which
is unaffected by peculiar velocities, and $\pi$ is the separation along the
line of sight. Integrating \xisp\ along the $\pi$ dimension leads to a statistic
that is independent of redshift space distortions, the projected correlation
function:
\begin{eqnarray}
  \wprp &=& \; 2 \int_0^\infty d\pi\; \xisp \\
          &\approx& \; 2 \int_0^{\pimax} d\pi\; \xisp
\end{eqnarray}
\noindent where \pimax\ is the maximum $\pi$ separation to which we integrate.
As the signal to noise of \xisp\ declines quickly for large values of $\pi$, we
measure the projected correlation function by integrating to a given \pimax\ to
limit shot noise. We use a larger \pimax\ value in the PRIMUS fields compared
to the DEEP2 fields to account for the larger redshift uncertainty in the
PRIMUS survey. In the PRIMUS fields we use $\pimax\,=\,80\hMpc$, while in DEEP2
we use $\pimax\,=\,20\hMpc$.  \citet{Skibba14} and \citet{Coil08}
use similar values for these surveys, respectively.

\subsection{Jackknife Uncertainty Estimation}\label{sec:jackknife}
We estimate the uncertainty in our measurements using jackknife resampling of
the data \citep[e.g., ][]{Lupton93, Scranton02}. For reasonably large surveys
(including both PRIMUS and DEEP2) jackknife errors are generally similar to the
cosmic variance errors in \wp derived from simulated mock catalogs \citep[e.g.,
][]{Zehavi05, Norberg08, Coil08, Skibba14}. For each of our samples, we use
between 10 and 12 jackknife samples across our 8 fields, where we have
spatially subdivided the larger fields into two or more subfields. The
different number of jackknife samples is due to the multi-wavelength coverage
in each field (i.e. CDFS does not contain X-ray data; see
Table~\ref{table:numbers} for field details.) We subdivide the large fields
(CDFS and XMM) along lines of constant RA and declination such that the
resulting subsamples probe roughly similar volumes and cover an area on the sky
approximately equal to $\sim$1 \degsq.

The uncertainty in \wp is estimated by calculating the projected correlation
function using each jackknife sample. From this collection of \wp estimates we
calculate the variance in the projected correlation function,
\begin{eqnarray}
    \sigma_{\wp}^{2}(\rp) = \frac{N-1}{N}\sum_{j}^{N}(\wprp - 
                                                     \hat{w}_j(\rp))^{2},
\end{eqnarray}
\noindent where the $N$ is the number of jackknife samples, $j$ indexes each
jackknife sample, and $\hat{w}_j(\rp)$ is the projected correlation function
computed for a given jackknife sample. By measuring the projected correlation
function using multiple fields across the sky, the jackknife resampling of
fields estimates the uncertainty on our measurements due to cosmic variance.

\subsection{Inferring the AGN Auto-correlation Function}\label{sec:auto}
Following \citet{Coil09}, we infer the AGN ACF from measurements of the
AGN-galaxy CCF and the galaxy ACF. We calculate the galaxy ACF using the
\citet{Landy93} estimator:
\begin{equation}
  \xi(\rp,\, \pi) = \frac{GG(\rp,\, \pi) - 2\,GR(\rp,\, \pi) + RR(\rp,\, \pi)}{RR(\rp,\, \pi)}
\end{equation}
\noindent where $GG$, $GR$, $RR$ are the galaxy-galaxy, galaxy-random, and
random-random weighted pair counts, respectively, where we include the galaxy
targeting weights. We calculate the AGN-galaxy CCF and galaxy ACF in the same
volume. We integrate the galaxy ACF projected correlation function to the same
\pimax \ limits used for the AGN-galaxy CCF.

We then infer the autocorrelation function of the AGN sample using,
\begin{equation}
  \wAA(\rp) = \frac{\wAG^2(\rp)}{\wGG(\rp)}
\end{equation}
\noindent where \wAA\ is the projected AGN ACF, \wAG\ is the projected AGN CCF,
and \wGG\ is the projected galaxy ACF. Implicit is the assumption that the
spatial distributions of AGN and galaxies are linearly related to the
underlying dark matter spatial distribution (i.e., that the bias is linear, see
Section~\ref{sec:bias} below), and that galaxies and AGN are well mixed within
dark matter halos.

\subsection{Power Law Fits}\label{sec:powerlaw}
The correlation function can roughly be fit by a power law, with 
$\xir=(r/r_0)^\gamma$, where the scale factor $r_0$ is the scale at which 
there is unity excess probability and $\xi=1$. 
An analytic form can then be fit to \wprp:
\begin{eqnarray}
  \wprp = \, \rp \left(\frac{r_0}{\rp}\right)^\gamma 
                 \frac{\Gamma(\tfrac{1}{2})\Gamma(\tfrac{\gamma-1}{2})}
                                          {\Gamma(\tfrac{\gamma}{2})}
\end{eqnarray}
\noindent where $\Gamma$ is the Gamma function. We fit this analytic function
to our clustering measurements in the approximately linear regime of
\rpvalue{=}{1-10}.  On larger scales the size of our fields limits 
the number of pair counts, which artificially lowers the measured 
correlation function and leads to large statistical fluctuations.

\begin{figure*}
\epsscale{1.2}
\epstrim{0.1in 0.2in 0.1in 0.1in}
\plotone{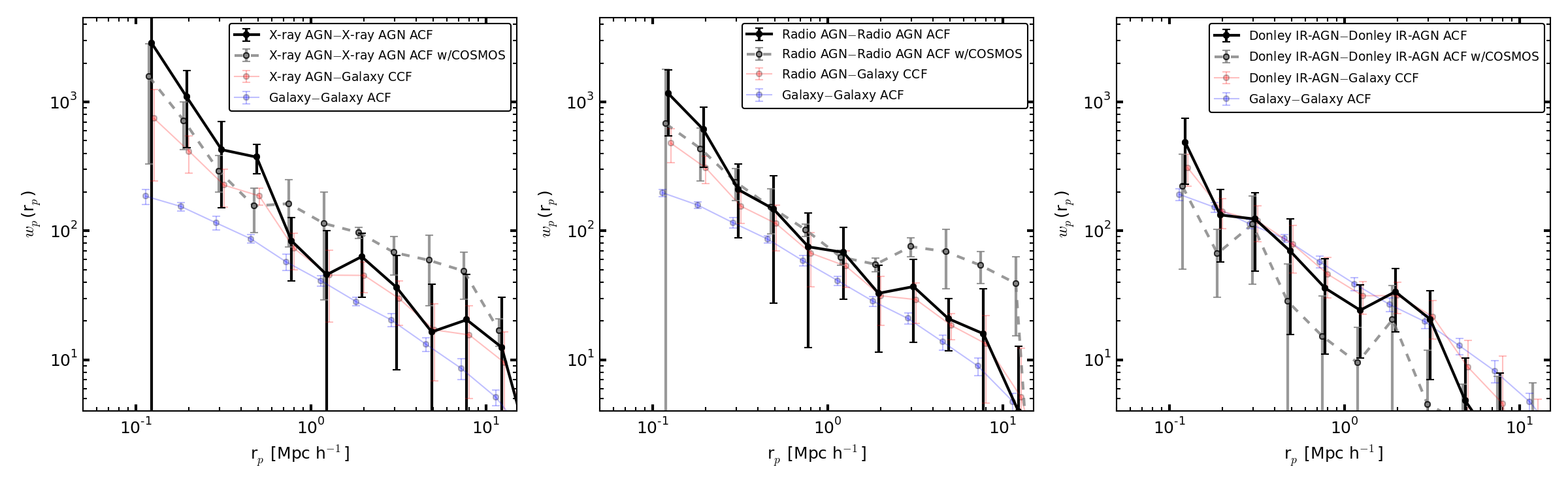}
\caption{%
Projected correlation functions for the \Xray (left), \Radio (center), and \Donley (right) samples. 
In each panel we show the AGN auto-correlation function (ACF; black), the galaxy-AGN cross-correlation function (CCF; red) and the galaxy tracer ACF (blue), as well as the uncertainties from jackknife resampling of the fields. 
The black line shows our fiducial sample which excludes the COSMOS field, 
while the light grey dashed line shows the ACF including the COSMOS field.
As discussed in Section~\ref{sec:data} the large over-densities in the COSMOS field at $z<1$ can systematically affect the measured clustering amplitudes, as seen here.}
\label{fig:auto}
\end{figure*}

\subsection{Bias Estimation}\label{sec:bias}

We use the projected correlation function to estimate
the dark matter bias of the AGN ACF. The bias $b$ measures the relative
clustering strength of the AGN sample to that of dark matter particles. We
estimate the bias at the median redshift of each AGN sample using the publicly
available code of \citet{Smith03}. We integrate the dark matter correlation
function to a $\pimax\,=\,80\,\hMpc$ and then calculate the bias as 
\begin{equation}
  b = \sqrt{\frac{w_\textrm{AGN}}{w_\textrm{DM}}}
\end{equation}
\noindent where $w_\textrm{AGN}$ is the AGN ACF and $w_\textrm{DM}$ is the dark
matter ACF on scales of \rpvalue{=}{1-10}. When comparing the clustering of
different samples it is useful to compare the bias values instead of the 
clustering scale lengths, as the bias accounts
for differences in the median redshift of each sample and further 
does not assume that $\xi$ is a power law.

Additionally, the relative bias between two AGN or galaxy samples is defined as
the square root of the ratio of their respective projected correlation
functions. This allows for a simple comparison of the clustering strength of
two samples and is akin to comparing their absolute bias (relative to dark
matter) values. We estimate the relative bias on scales of \rpvalue{=}{1-10}.
We use the ratio of CCFs which does not increase the fractional uncertainty of
the resulting bias due to the common galaxy-tracer ACF term in the AGN ACF.
Below we present the mean and $1\sigma$ uncertainty of the relative bias across
the jackknife samples when comparing two samples.

\subsection{Halo Mass Estimation}\label{sec:halomass}

We estimate the median dark matter halo mass
($M_\mathrm{DM}$) that hosts AGN selected at different wavelengths from the
absolute bias measured on scales of \rpvalue{=}{1-10}. 
We convert the bias to the quantity $\nu = \delta_c /
\sigma(M)$, where $\delta_c \sim 1.69$ is the critical density for collapse 
and $\sigma(M)$ is the mass density fluctuation in a sphere of radius
$r^3=(3M\Delta/4\pi\bar{\rho})$ from linear theory. We use Equation (8) of
\citet{Sheth01} to convert the absolute bias to $\nu$ and Equations (A8-A10) of
\citet{Vandenbosch02} to infer the median $M_\mathrm{DM}$ of the sample.

%% file: results.tex
%!TEX root = ms.tex

\section{Results}\label{sec:results}
In this section we present the results of our correlation function analysis. As
discussed above, we measure the CCF of our AGN samples relative to the galaxy
tracer sample and the ACF of the galaxy tracer sample, in order to infer the
ACF of the various AGN samples. We first present the ACF of the \Xray, \Radio,
and \Donley in Section~\ref{sec:absoluteclustering}. We then compare the
clustering properties of various subsamples within each of the \Xray, \Radio,
and \Donley samples in Section~\ref{sec:sampleclustering}, to investigate
whether the clustering amplitude depends on AGN luminosity or hardness ratio.
In Section~\ref{sec:obscuredclustering} we present the clustering of obscured
versus unobscured IR AGN selected in WISE. We compare the relative clustering
between \Xray, \Radio, and \Donley in Section~\ref{sec:relativeclustering}, to
determine whether AGN selected at different wavelengths have different
clustering amplitudes. Finally, we compare the clustering strength of each AGN
sample relative to matched galaxy control samples in
Section~\ref{sec:matchedclustering}.

\subsection{Clustering of X-ray, Radio, and IR AGN}\label{sec:absoluteclustering}

% xray
In the left panel of Figure~\ref{fig:auto} we show the \Xray ACF (black), which
is derived from the AGN-galaxy CCF (red) and the galaxy tracer ACF (blue). In
this figure we present results including (grey) and excluding (black) the
COSMOS field. The \Xray ACF is more clustered than the galaxy tracer sample.
Table~\ref{table:clustering} lists the measured $r_0$ and $\gamma$ values
excluding the COSMOS field, as well as the absolute bias and inferred median
dark matter halo mass both with and without the COSMOS field included. As can
be seen in Figure~\ref{fig:auto}, the clustering amplitude of \Xray increases
when the COSMOS field is included; the bias on large scales increases by 26\%.
As discussed above, the COSMOS field contains several large structures at
$z<1$, which both systematically increases the clustering amplitude when
including this field and increases the jackknife error. We therefore prefer to
focus on results that exclude COSMOS when discussing the absolute bias of our
AGN samples. The bias of \Xray that we measure (\bias{1.5}{0.2}) corresponds to
a median dark matter halo mass of \medianMhalo{12.9}, which is generally
associated with galaxy groups. The \Xray ACF rises sharply at small projected
separations (\rpvalue{<}{0.7}), indicating that on small scales there is a
increase in the number of pairs of objects within the same dark matter halo.

\input{table_03_clustering}

\begin{figure*}
\epsscale{1.1}
\epstrim{0.1in 0.2in 0.1in 0.1in}
\plotone{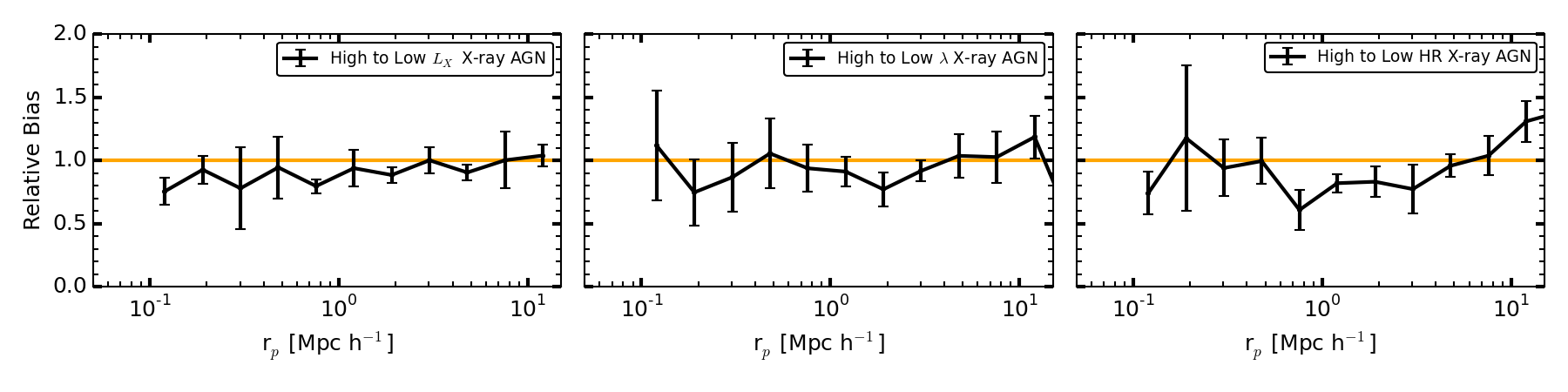}
\caption{%
The scale dependent relative bias between various \Xray samples defined by
X-ray luminosity (\LX, left), specific accretion rate (\specific, center), and
hardness ratio (HR, right). 
% For all relative bias measurements the COSMOS field
% is included, as the relative bias measured in the same volume is not as
% impacted by large over-densities.
Error bars are derived using jackknife
resampling of the fields.
}\label{fig:xraysamples}
\end{figure*}

% radio
In the center panel of Figure~\ref{fig:auto} we show the \Radio ACF, again with
(grey) and without (black) the COSMOS field included. Including COSMOS causes
the large-scale bias to increase substantially by 50\%. We find that similar to
\Xray, \Radio are more clustered than the galaxy tracer sample and have a large
bias value (\bias{1.8}{0.1}, excluding COSMOS), which suggest that they reside
in massive dark matter halos (\medianMhalo{13.3}) typically associated with
massive groups or small clusters.

% IR
In the right panel of Figure~\ref{fig:auto} we show the \Donley ACF. Unlike the
\Xray and \Radio, we find that \Donley are {\it less} clustered than the galaxy
tracer sample. The \Donley sample has a lower bias value (\bias{1.3}{0.5},
excluding COSMOS) than either of the \Xray or \Radio, from which we estimate a
median dark matter halo mass of (\medianMhalo{12.8}). Unlike for the
\Xray and \Radio samples, including COSMOS for \Donley results in a lower
clustering amplitude; the bias decreases by 30\%.

While not shown here, we additionally measure the \Assef ACF and list the
derived clustering parameters in Table~\ref{table:clustering}. Similar to the
\Donley sample, the \Assef is less clustered than the galaxy tracer sample and
has a lower bias (\bias{0.8}{0.1}) than either the \Xray or \Radio. The large
scale bias of the \Assef sample does not change when including the COSMOS
field, though there is an increase in the clustering strength on small scales
which substantially increases the slope ($\gamma=2.2$). We also measured the
clustering properties of \Mateos and find similar results to both \Donley and
\Assef.

\subsection{Relative Bias Within AGN Samples at a Given Wavelength}\label{sec:sampleclustering}

As discussed above, the COSMOS field contains large over-densities at
\zsim{0.3} and \zsim{0.7} \citep[e.g., ][]{Lilly07,McCracken07, Meneux09,
Kovac10, delaTorre10, Skibba14} In our sample, which combines the PRIMUS and
DEEP2 fields and therefore covers a large volume and probes a range of cosmic
densities, we find including the COSMOS field systematically impacts our
clustering results (somewhat akin to the Sloan Great Wall, \citep[e.g.,
][]{Zehavi11, McBride11}), both in terms of the measured amplitude and the
jackknife errors, which increase when COSMOS is included.

However, including the COSMOS field is more robust when comparing the {\it
relative} bias between two samples, as the same volume is used for both
measurements and overall changes in the density (i.e., cosmic variance) cancel
to first order. This is reflected in the fact that the fractional jackknife
errors on the relative bias values measured here {\it decrease} when the COSMOS
field is included (which results in larger samples and volumes probed).
Therefore, when presenting relative bias measurements throughout the paper, we
include the COSMOS field. We note that if the COSMOS field is excluded from our
relative bias analysis, the significance of our results lowers (due to the
larger errors) but the actual relative bias values do not change substantially.

\input{table_04_relative}

\begin{figure*}
\epsscale{1.1}
\epstrim{0.1in 0.2in 0.1in 0.1in}
\plotone{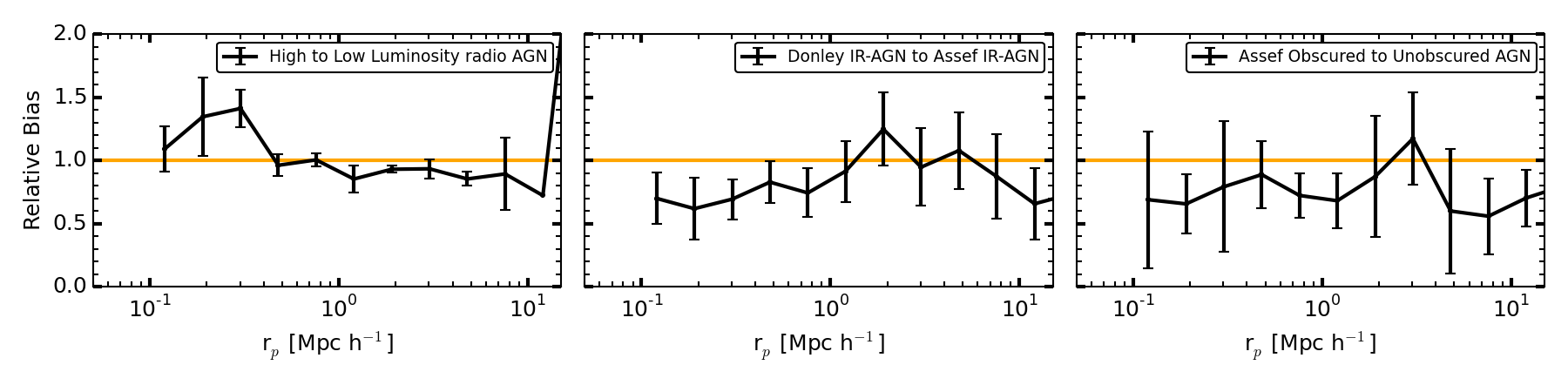}
\caption{%
The relative bias between the high and low luminosity \Radio samples (left), 
the \Donley and \Assef samples (center), and the obscured \Assef and 
unobscured \Assef (right).
There is no significant difference between the clustering of these samples;
all relative biases are consistent with unity.
} \label{fig:subsamples}
\end{figure*}

% Xray
As discussed in Section~\ref{sec:sample}, we divide the full \Xray sample into
subsamples depending on various AGN properties: X-ray luminosity \LX, specific
accretion rate \specific, and hardness ratio. The measured clustering
parameters for each subsample are given in Table~\ref{table:clustering}. The
relative bias between these samples is shown in Figure~\ref{fig:xraysamples}
and listed in Table~\ref{table:relative} for scales associated with the one
halo term (\smallrp) and the two halo term (\largerp). As above, uncertainties
are derived from jackknife resampling of the fields, and COSMOS is included
here. We find a 2.6$\sigma$ difference on small scales with \LX, such that the
lower \LX sources are more clustered in the one-halo regime. For these X-ray
samples defined by \LX, \specific, and hardness ratio we find no statistically
significant differences ($>3\sigma$) in the clustering amplitudes on either
small or larger scales. This implies that the mass of dark matter host halo
does not correlate with any of these properties, within the ranges that we
probe.

% Radio
In the left panel of Figure~\ref{fig:subsamples} we show the relative bias
between high and low luminosity \Radio. We do not find significant differences
in the clustering of high and low luminosity \Radio on either small or large
scales (see Table~\ref{table:relative}). The lack of a dependence of the
clustering amplitude on radio luminosity may be surprising, given that the
highest luminosity \Radio are found in the most massive quiescent galaxies
compared to lower luminosity \Radio (bottom panel of
Figure~\ref{fig:radio_ssfr}). While the lower luminosity radio AGN sample
contains more galaxies with slightly lower mass (\mass{\sim}{10.8}, compared to
\mass{\sim}{11.2} for the high luminosity sample), the clustering signal is
dominated by the most massive objects in the sample. Additionally, we find no
significant difference in the sSFR distribution of the host galaxies as a
function of the radio luminosity, which suggests that high and low luminosity
\Radio have fairly similar host galaxies.

For each of the AGN properties tested above (i.e., luminosity, specific
accretion rate) we test the significance of the clustering differences when
comparing sources in the upper and lower quartiles, as opposed to the upper and
lower halves. This allows us to look for clustering differences between the
most extreme sources in each parameter of interest. This does not change any of
our results, however, the larger uncertainties that result from using smaller
samples may prevent us from detecting any underlying differences.

% IR
In the center panel of Figure~\ref{fig:subsamples} we show the relative bias
between the \Donley and \Assef samples. We list the measured clustering
parameters in Table~\ref{table:clustering} and relative bias between these
samples in Table~\ref{table:relative}. We find that these samples have
consistent clustering properties, given the error bars, however the bias of
\Donley is 37\% higher than that of the \Assef. Both samples have lower
estimates for the median dark matter halo mass than the \Xray or \Radio.

\begin{figure}
\epsscale{1.1}
\epstrim{0.1in 0.2in 0.1in 0.1in}
\plotone{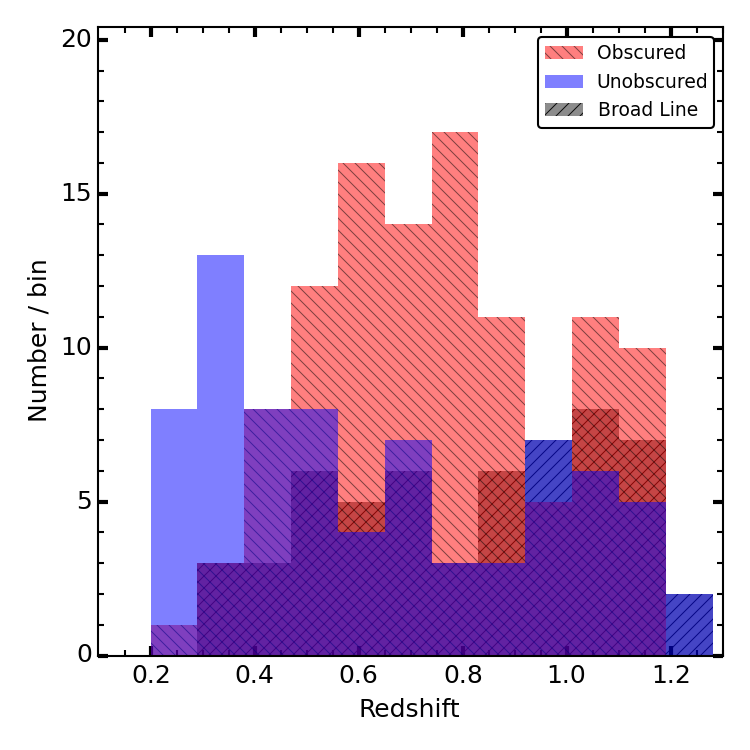}
\caption{%
Redshift distributions for the obscured (red hatched), unobscured (blue), and
broad line (dark grey hatched) \Assef. The obscured and unobscured IR-AGN
populations have substantially different redshift distributions; the obscured
sample peaks at \zsim{0.7}, while the unobscured sample has a much flatter
distribution, including more sources are low redshift. The higher redshift
sources in both samples are more likely to have broad optical lines.
}\label{fig:obscured_redshift}
\end{figure}

\subsection{Relative Bias of Obscured versus Unobscured WISE IR-AGN}%
\label{sec:obscuredclustering}

Following \citet{Donoso13} and \citet{Dipompeo14}, we compare the clustering of
obscured and unobscured WISE-selected AGN. In the right panel of
Figure~\ref{fig:subsamples} we show the scale dependent relative bias between
obscured and unobscured \Assef. The obscured \Assef are less clustered than the
unobscured \Assef at the $\sim2\sigma$ level on small scales and at the $\sim1\sigma$
level on larger scales. Within the uncertainties, therefore, we do not find a
significant difference in the clustering amplitudes of these samples. This
suggests that the differences found by \citet{Donoso13}, where the COSMOS field
alone was used to determine the redshift distributions of these two
populations, was impacted by using a single field. Here, using eight fields
that cover $\sim9.1\,\degsq$ of the sky, and using spectroscopic redshifts for
each source, we do not find a significant difference in their clustering.

To understand this further, we show in Figure~\ref{fig:obscured_redshift} the
redshift distributions of the obscured (red), unobscured (blue), and broad line
\Assef (black) in our sample. While the median redshift of the obscured \Assef
(\medianzsim{0.77}) and unobscured \Assef (\medianzsim{0.70}) are similar, the
samples have very different redshift distributions. The obscured AGN peak at
\medianzsim{0.7}, whereas the unobscured AGN have a flatter distribution and
peak at both lower and higher redshift, at \medianzsim{0.3} and
\medianzsim{1.0}. Within the unobscured sample, the broad line sources
typically have higher redshifts than the non-broad line sources.

The differences in the redshift distributions of obscured and unobscured \Assef
strongly limits the interpretation of angular clustering measurements of these
two populations. While there are broad similarities between our redshift
distribution and that of \citet{Donoso13}, our obscured sample shows a much
broader redshift distribution extending out to \medianzsim{1.2}. Our unobscured
sample also has a larger fraction of sources at low redshift
(\medianzsim{0.3}). Our redshift distribution is more similar to the redshift
distributions found in the \bootes Survey \citep{Dipompeo14}.
%The spectroscopic redshift sample used by \citet{Donoso13} is dominated by the COSMOS field, which is strongly impacted by cosmic variance at $z\lesssim1$ and systematically affects the clustering amplitude. 
%Here we use seven independent fields, covering $\sim9.2\degsq$ of the sky.

Similarly, we test both the \Donley sample and \Mateos samples for any
dependence of the clustering amplitude with obscuration and find no significant
differences. For \Donley we test both an optical-to-WISE color cut ($r-W2=6.0$)
and an optical-to-IRAC color cut ($r-[4.6]=6.1$) and find no significant
differences using either cut.

\begin{figure}
\epsscale{1.1}
\epstrim{0.1in 0.2in 0.1in 0.1in}
\plotone{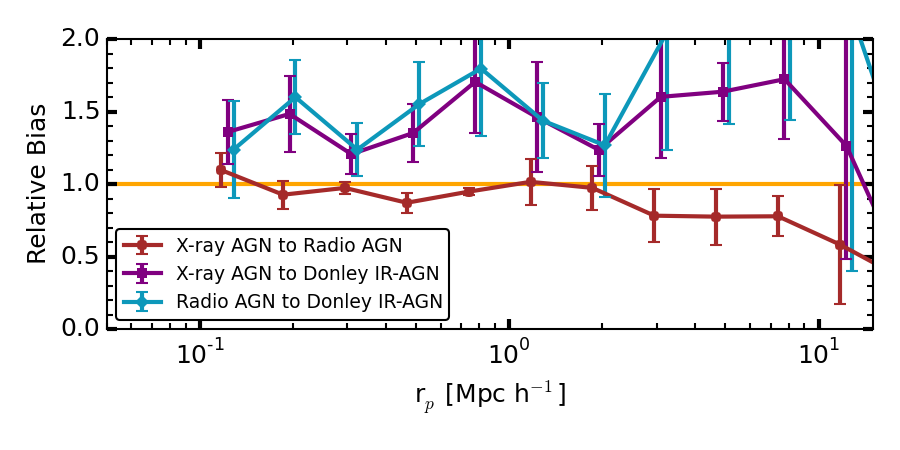}
\caption{%
The relative bias between the X-ray, radio, and \Donley samples with error
bars determined from jackknife resampling of the fields. We show the relative
bias between X-ray to \Radio (dark red), X-ray to \Donley (purple), and radio 
to \Donley (cyan). We find that the X-ray and \Radio have similar clustering
amplitudes, both of which are larger than that of \Donley.
}\label{fig:agncompare}
\end{figure}

\subsection{Comparison of the Clustering of X-ray, Radio, and IR AGN}\label{sec:relativeclustering}

In Figure~\ref{fig:agncompare} we show the relative bias between AGN identified
at different wavelengths; the results are given in Table~\ref{table:relative}.
We find that \Radio are more clustered than \Xray (red line) on large
scales (15\% higher bias), but the difference is not significant (1.3$\sigma$).
Comparing the \Xray and \Donley samples (purple line), there is a significant
difference (4.4 $\sigma$) on small scales, where \Xray have a 38\% higher bias,
while on large scales \Xray having a 44\% higher bias (2.4$\sigma$). Comparing
the \Radio and \Donley samples (cyan line), \Radio have a 40\% higher bias on
small scales (3.0$\sigma$) and an 58\% higher bias on large scales
(1.7$\sigma$). The relative bias averaged over all scales results in a
difference at the 2.5$\sigma$ level.

\begin{figure*}[t]
  \epsscale{0.9}
  \epstrim{0.1in 0.1in 0.1in 0.1in}
  \plotone{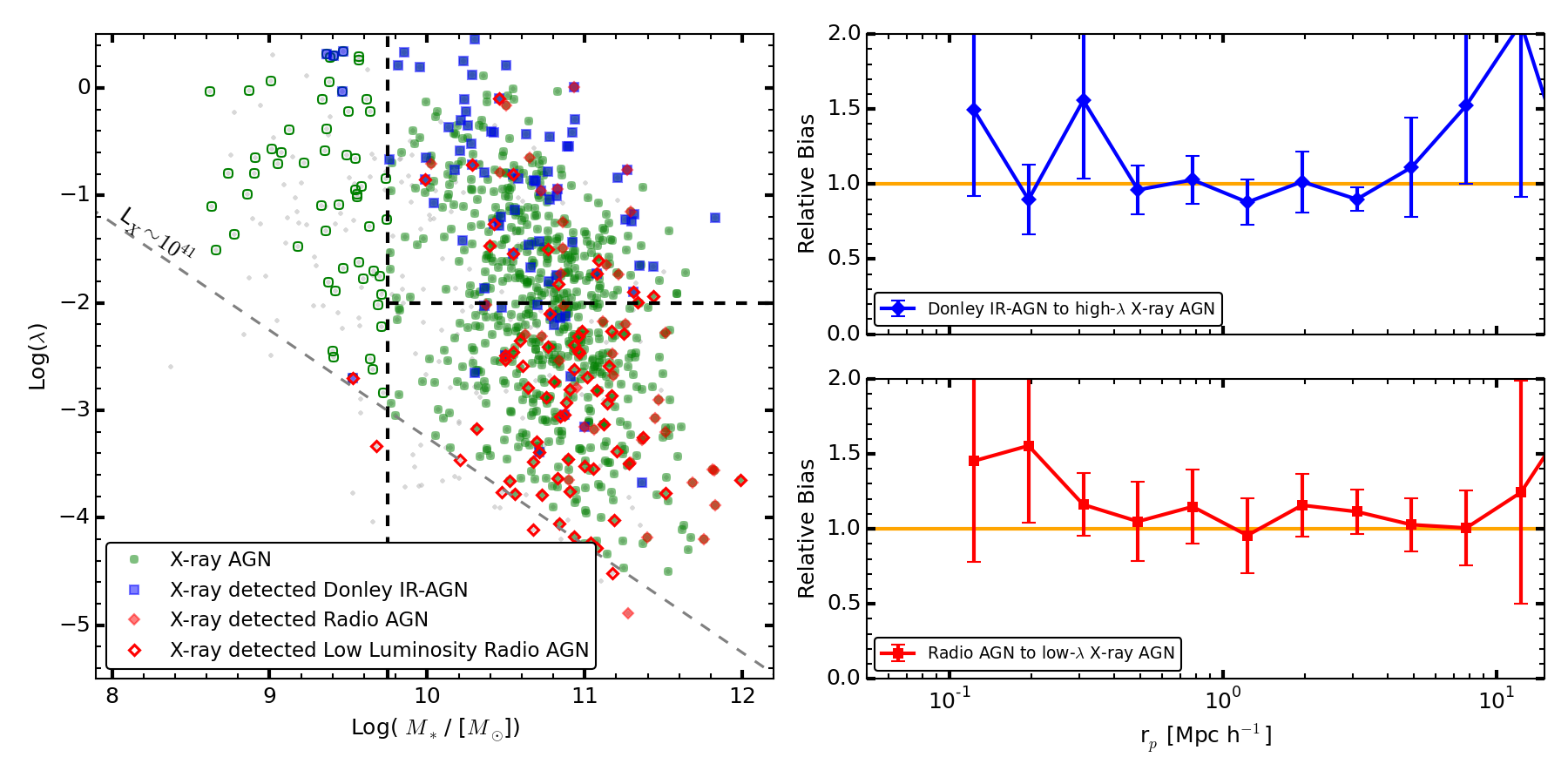}
  \caption{ Dependence of clustering amplitude on specific accretion rate
  (\specific) for X-ray, radio, and \Donley. \textbf{Left panel}: Specific
  accretion rate versus stellar mass for X-ray detected AGN. \Xray that are
  also radio AGN are shown with red diamonds, and those that are also \Donley
  are shown with blue squares. The X-ray luminosity limit for our sample is
  shown as a light-grey dashed diagonal line. \Xray below the \mass{=}{9.75}
  stellar mass limit (vertical dashed line) are shown with open symbols, while
  \Xray with broad lines in their optical spectra are shown as small grey
  points. The \SpecificP{=}{-2} line defines the high \specific\ and low
  \specific\ \Xray. \textbf{Top right panel:} Relative bias between the high
  \specific\ \Xray and \Donley for the non-broadline samples. \textbf{Bottom
  right panel:} Relative bias between the low \specific\ \Xray and \Radio for
  the non-broadline samples. We find that \Donley are similarly clustered as
  high \specific\ \Xray while \Radio are clustered similarly as low \specific\
  \Xray. }
  \label{fig:lambdacompare}
\end{figure*}

On small scales (\smallrp) the correlation functions of both the \Xray and
\Radio are significantly higher than that of the \Donley. This may be due in
part to the difficulty in identifying \Donley in quiescent galaxies, due to the
$1.4\,\um$ stellar bump entering into the mid-IR photometry \citep{Mendez13}.
This selection effect limits the number of \Donley that can be identified in
quiescent host galaxies, which would decrease the clustering amplitude on all
scales, though particularly on small scales \citep[reflecting differences in
color-dependent clustering; e.g., ][]{Zehavi05, Coil09, Skibba14}.

\subsection{Dependence on Specific Accretion Rate}%
\label{sec:lambdadiscussion}
As shown in \citet{Hickox09} and \citet{Mendez13}, there is a substantial
difference in the specific accretion rate (\specific) distributions of AGN
selected at different wavelengths. To account for these differences in
\specific, here we compare \Radio and \Donley to \Xray with similar \specific\
values. In the large left panel of Figure~\ref{fig:lambdacompare} we show
\specific\ versus stellar mass for \Xray (green circles). \Xray that are also
\Radio are shown with red diamonds, and those that are also \Donley are shown
with blue squares. We divide \Xray into a high \specific\ sample and a low
\specific\ sample at \lambdavalue{=}{-2}. Most \Donley lie above this line,
while most \Radio lie below this line. Here we do not include sources below
\mass{\sim}{9.75}, to ensure that we are roughly complete in stellar mass at
all redshifts. We also remove all broad-line sources from this comparison, as
we require a stellar mass estimate for the high-\specific\ \Xray and
low-\specific\ \Xray samples.

In the upper-right panel of Figure~\ref{fig:lambdacompare} we show the scale
dependent relative bias between the high-\specific\ \Xray and \Donley. In the
lower-right panel of Figure~\ref{fig:lambdacompare} we show the relative bias
between the low-\specific\ \Xray (green) and \Radio (red). We find no
significant differences between either sample. The limited significance of
these results are dominated by the low number of non-broadline AGN in each
sample.

\subsection{Comparison with Matched Galaxy Control Samples}%
\label{sec:matchedclustering}

As discussed above, AGN samples identified at different wavelengths are 
biased in terms of identifying specific types of AGN in specific types 
of host galaxies.  In general, AGN are more easily identified in more 
massive galaxies \citep[e.g.,][]{Silverman11, Aird12}. Additionally, 
there can be substantial differences in the sSFR and redshift 
distributions of the host galaxies of AGN selected at X-ray, radio, 
and IR wavelengths \citep[e.g., ][]{Hickox09, Coil09, Mendez13, Goulding14}.

Differences in the host galaxy populations can influence
the observed AGN clustering amplitude, which must be understood before AGN
clustering can be used to constrain the AGN triggering mechanism. To account
for this, we compare the clustering of each of our AGN samples to that of
matched galaxy control samples that have the same stellar mass, sSFR, and
redshift distributions as the AGN samples.

As this requires robust estimates of the stellar masses and SFRs, we remove
optical broad-line AGN for these comparisons, in order to limit AGN
contamination of the optical broadband photometry used in the SED fits. While
broad line AGN are a substantial fraction (34\%) of the \Xray population,
excluding them does not significantly change the measured clustering
properties. As seen in Table~\ref{table:clustering}, excluding broad line AGN
from the fiducial \Xray sample leads to a 1\% change in the large scale bias.
This implies that at least for the \LX range and redshift range considered
here, the clustering of narrow line and broad line \Xray are not significantly
different. The \Radio sample contains the smallest fraction of broad line
sources (14\%). Excluding the broad line radio sources results in only a 3\%
change in the bias on large scales. The broad line fraction of the \Donley
sample is also substantial (31\%), similar to that of the \Xray sample, and
excluding these sources leads to a 14\% change in the bias. The relative biases
{\it between} the X-ray, radio, and IR AGN samples after the broad line sources
have been removed do not change too substantially, but the fractional errors do
increase, due to the smaller sample sizes.

\begin{figure}
\epsscale{1.1}
  \epstrim{0.1in 0.2in 0.1in 0.1in}
\plotone{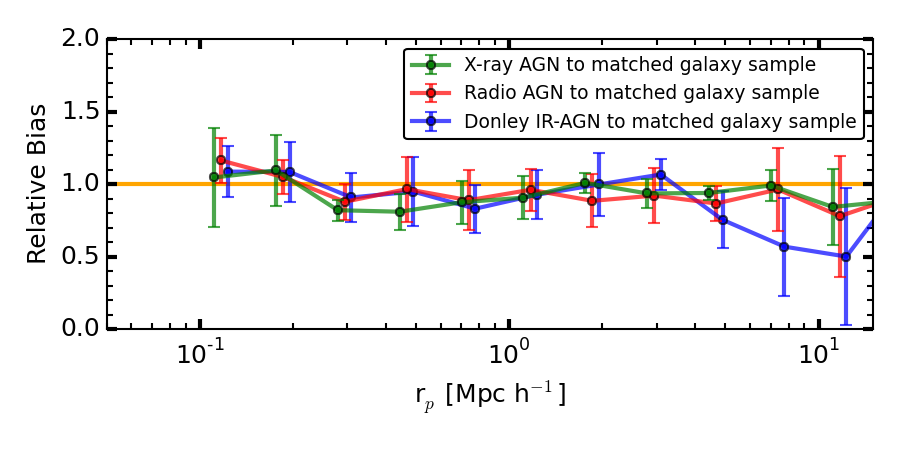}
\caption{%
The relative bias between the \Xray (green), \Radio (red), and \Donley (blue)
and their respective stellar mass, sSFR, and redshift matched galaxy control
samples. We do not find any significant differences in the clustering
properties of any of the AGN and matched galaxy samples; all relative biases
are consistent with unity.
}\label{fig:matchedagn}
\end{figure}

In Figure~\ref{fig:matchedagn} we show the relative biases between
(non-broadline) AGN samples identified at different wavelengths and their
matched control galaxy samples; the results are listed in
Table~\ref{table:relative}. We find no significant differences in the
clustering amplitude of either the \Xray, \Radio, or \Donley and their matched
galaxy control samples on small or large scales; all differences are
significant at 1.5$\sigma$ or less.

This suggests that the physical effect(s) of the host galaxy large-scale
environment is either sub-dominant in AGN triggering or is not separable from
the host galaxy properties. The strong agreement between the clustering of AGN
host galaxies and similarly-selected inactive galaxies, on both small and large
scales corresponding to the one- and two-halo terms, indicates that the AGN
triggering mechanism either does not act on these scales or correlates with the
properties of the galaxies in which they are identified. It could be possible,
for example, that an environmental effect that triggers AGN also causes changes
in the host galaxy's sSFR, such that active and inactive galaxies with the same
sSFR distribution have the same clustering properties. However, as shown
elsewhere, there are not substantial correlations between host galaxy sSFR and
AGN incidence, once selection effects are taken into account \citep[e.g.,
][]{Aird12, Azadi14}.

%% file: table_03_clustering.tex
%!TEX root = ms.tex
\begin{deluxetable*}{llcccccc}
\tablewidth{0pc}
\tablecolumns{8}
\tablecaption{Clustering results for X-ray, radio, and \IR.\label{table:clustering}}
\tablehead{
  \colhead{\makecell[l]{AGN Selection}} &
  \colhead{\makecell[l]{Sample Name}} &
  \colhead{$r_0$\tablenotemark{a}} &
  \colhead{$\gamma$\tablenotemark{b}} &
  \colhead{Bias\tablenotemark{c}} &
  \colhead{\median{M}{DM}\tablenotemark{d}} &
  \colhead{Bias\tablenotemark{c}} &
  \colhead{\median{M}{DM}\tablenotemark{d}} \\
  \multicolumn{2}{c}{} &
  \multicolumn{4}{|c}{without the COSMOS field} &
  \multicolumn{2}{|c}{with the COSMOS field} 
}
\startdata 
\iffalse                                      r_0                     gamma                   bias             m_halo           bias w/ cosmos        m_halo \fi
X-ray AGN      & Fiducial              & 4.0 $\pm$ 1.9 &          1.5 $\pm$ 0.2 &        1.5 $\pm$ 0.2 &       12.9 &         1.9 $\pm$ 0.1 &       13.3 \\
Radio AGN      & Fiducial              & 5.5 $\pm$ 1.9 &          1.9 $\pm$ 0.7 &        1.8 $\pm$ 0.1 &       13.3 &         2.7 $\pm$ 0.7 &       13.8 \\
Donley IR-AGN  & Fiducial              & 4.8 $\pm$ 1.3 &          1.9 $\pm$ 0.3 &        1.3 $\pm$ 0.5 &       12.8 &         0.9 $\pm$ 0.2 &       11.6 \\
\hline \\[-2ex]                                                                         
X-ray AGN      & Non-broadline         & 7.0 $\pm$ 6.0 &          1.4 $\pm$ 0.3 &        2.5 $\pm$ 0.4 &       13.6 &         1.9 $\pm$ 0.2 &       13.4 \\
Radio AGN      & Non-broadline         & 5.7 $\pm$ 2.8 &          1.6 $\pm$ 0.6 &        1.9 $\pm$ 0.3 &       13.3 &         2.6 $\pm$ 0.7 &       13.8 \\
Donley IR-AGN  & Non-broadline         & 4.9 $\pm$ 1.5 &          1.7 $\pm$ 0.3 &        1.4 $\pm$ 0.5 &       12.9 &         1.1 $\pm$ 0.3 &       12.5 \\
\hline \\[-2ex]                                                                         
X-ray AGN      & Low \LX               & 4.5 $\pm$ 4.3 &          1.3 $\pm$ 0.3 &        1.9 $\pm$ 0.3 &       13.3 &         2.0 $\pm$ 0.2 &       13.4 \\
X-ray AGN      & High \LX              & 4.1 $\pm$ 1.8 &          1.6 $\pm$ 0.4 &        1.5 $\pm$ 0.1 &       12.9 &         1.8 $\pm$ 0.2 &       13.2 \\
X-ray AGN      & Low \specific         & 6.0 $\pm$ 3.6 &          1.6 $\pm$ 0.4 &        2.0 $\pm$ 0.2 &       13.3 &         2.1 $\pm$ 0.2 &       13.5 \\
X-ray AGN      & High \specific        & 3.8 $\pm$ 1.0 &          1.5 $\pm$ 0.3 &        1.4 $\pm$ 0.2 &       12.9 &         1.8 $\pm$ 0.3 &       13.3 \\
X-ray AGN      & Low HR                & 3.2 $\pm$ 1.6 &          1.4 $\pm$ 0.3 &        1.4 $\pm$ 0.2 &       12.8 &         2.0 $\pm$ 0.0 &       13.4 \\
X-ray AGN      & High HR               & 5.0 $\pm$ 2.1 &          1.7 $\pm$ 0.7 &        1.6 $\pm$ 0.2 &       13.1 &         1.6 $\pm$ 0.4 &       13.1 \\
X-ray AGN      & Low Redshift          & 2.7 $\pm$ 1.8 &          1.6 $\pm$ 0.4 &        0.9 $\pm$ 0.2 &       12.0 &         1.6 $\pm$ 0.2 &       13.4 \\
X-ray AGN      & High Redshift         & 6.1 $\pm$ 2.1 &          2.3 $\pm$ 0.7 &        2.0 $\pm$ 0.5 &       13.3 &         2.7 $\pm$ 0.6 &       13.6 \\
\hline \\[-2ex]                                                                         
Radio AGN      & High \Pr              & 5.7 $\pm$ 2.0 &          2.3 $\pm$ 1.0 &        1.9 $\pm$ 0.3 &       13.3 &         2.2 $\pm$ 0.6 &       13.6 \\
Radio AGN      & Low \Pr               & 4.3 $\pm$ 5.0 &          1.4 $\pm$ 1.2 &        1.7 $\pm$ 0.3 &       13.2 &         2.9 $\pm$ 0.7 &       13.9 \\
\hline \\[-2ex]                                                                         
Assef IR-AGN   & Fiducial              & 2.3 $\pm$ 1.3 &          1.8 $\pm$ 1.0 &        0.8 $\pm$ 0.1 &       11.2 &         0.8 $\pm$ 0.1 &       11.4 \\
Assef IR-AGN   & Obscured WISE color   & 0.8 $\pm$ 1.6 &          1.3 $\pm$ 0.5 &        0.7 $\pm$ 0.2 &       11.1 &         0.6 $\pm$ 0.2 &       11.1 \\
Assef IR-AGN   & Unobscured WISE color & 2.7 $\pm$ 0.9 &          3.1 $\pm$ 3.8 &        0.9 $\pm$ 0.2 &       11.7 &         1.1 $\pm$ 0.3 &       12.4 \\
\enddata
\tablenotetext{a}{Correlation scale length, $r_0$, in units of $[h^{-1}\, \textrm{Mpc}]$.}
\tablenotetext{b}{Correlation power-law index $\gamma$.}
\tablenotetext{c}{Absolute bias estimated on scales of $1<r_p/$ $[h^{-1}\, \textrm{Mpc}]$ $\leq10$.}
\tablenotetext{d}{Dark matter halo median mass, $M_{DM}$, in units of $[\textrm{log}(h^{-1}\, \msun)]$.}
\end{deluxetable*}

%% file: table_04_relative.tex
%!TEX root = ms.tex
\begin{deluxetable*}{lcrcr}
\tablewidth{0pc}
\tablecolumns{5}
\tablecaption{Relative clustering bias for X-ray, radio, and \IR  samples, including the COSMOS field.\label{table:relative}}
\tablehead{
  \colhead{\makecell[l]{AGN Sample Comparison}} &
  \colhead{\makecell[c]{Relative Bias \\ $0.1 < r_p < 1$}} &
  \colhead{$N_\sigma$} &
  \colhead{\makecell[c]{Relative Bias \\ $1 < r_p < 10$}} &
  \colhead{$N_\sigma$} 
}
\startdata
\iffalse  Name                                                                   bias smallrp        N_sigma         bias largerp      N_sigma    \fi
Full X-ray AGN to Full Radio AGN ratio                                 &       0.96 $\pm$ 0.04 &      -1.0 &       0.87 $\pm$ 0.10 &      -1.3     \\
Full X-ray AGN to Full Donley IR-AGN ratio                             &       1.42 $\pm$ 0.10 &       4.4 &       1.53 $\pm$ 0.22 &       2.4     \\
Full Radio AGN to Full Donley IR-AGN ratio                             &       1.49 $\pm$ 0.16 &       3.0 &       1.81 $\pm$ 0.49 &       1.7     \\
Full Assef IR-AGN to Full Donley IR-AGN ratio                          &       1.41 $\pm$ 0.24 &       1.7 &       1.00 $\pm$ 0.19 &       0.0     \\
[1ex]\hline \\[-1ex]                                                                                               
No Broadline X-ray AGN to Full X-ray AGN ratio                         &       0.95 $\pm$ 0.05 &      -1.0 &       1.01 $\pm$ 0.02 &       0.6     \\
No Broadline Radio AGN to Full Radio AGN ratio                         &       0.97 $\pm$ 0.04 &      -0.8 &       0.97 $\pm$ 0.03 &      -1.0     \\
No Broadline Donley IR-AGN to Full Donley IR-AGN ratio                 &       1.15 $\pm$ 0.09 &       1.6 &       1.16 $\pm$ 0.07 &       2.2     \\
[1ex]\hline \\[-1ex]                                                                                               
No Broadline X-ray AGN to No Broadline Radio AGN ratio                 &       0.94 $\pm$ 0.09 &      -0.7 &       0.91 $\pm$ 0.08 &      -1.2     \\
No Broadline X-ray AGN to No Broadline Donley IR-AGN ratio             &       1.17 $\pm$ 0.11 &       1.6 &       1.36 $\pm$ 0.18 &       2.0     \\
No Broadline Radio AGN to No Broadline Donley IR-AGN ratio             &       1.25 $\pm$ 0.20 &       1.3 &       1.56 $\pm$ 0.51 &       1.1     \\
[1ex]\hline \\[-1ex]                                                                                               
High \LX to Low \LX X-ray AGN ratio                                    &       0.84 $\pm$ 0.06 &      -2.6 &       0.95 $\pm$ 0.05 &      -1.0     \\
High \specific to Low \specific X-ray AGN ratio                        &       0.95 $\pm$ 0.18 &      -0.3 &       0.93 $\pm$ 0.09 &      -0.7     \\
High HR to Low HR X-ray AGN ratio                                      &       0.89 $\pm$ 0.15 &      -0.7 &       0.88 $\pm$ 0.09 &      -1.3     \\
[1ex]\hline \\[-1ex]                                                                                               
High \Pr to Low \Pr Radio AGN ratio                                    &       1.20 $\pm$ 0.15 &       1.4 &       0.87 $\pm$ 0.11 &      -1.2     \\
[1ex]\hline \\[-1ex]                                                                                               
Obscured to Unobscured Assef IR-AGN ratio                              &       0.75 $\pm$ 0.12 &      -2.1 &       0.78 $\pm$ 0.27 &      -0.8     \\
[1ex]\hline \\[-1ex]                                                                                               
Low \specific X-ray AGN to No Broadline Radio AGN ratio                &       1.00 $\pm$ 0.16 &       0.0 &       0.95 $\pm$ 0.08 &      -0.7     \\
High \specific X-ray AGN to No Broadline Donley IR-AGN ratio           &       1.18 $\pm$ 0.13 &       1.4 &       1.33 $\pm$ 0.26 &       1.3     \\
[1ex]\hline \\[-1ex]                                                                                               
No Broadline X-ray AGN to Matched Galaxy ratio                         &       0.93 $\pm$ 0.12 &      -0.6 &       0.96 $\pm$ 0.05 &      -0.8     \\
No Broadline Radio AGN to Matched Galaxy ratio                         &       0.91 $\pm$ 0.17 &      -0.5 &       1.09 $\pm$ 0.07 &       1.3     \\
No Broadline Donley IR-AGN to Matched Galaxy ratio                     &       0.89 $\pm$ 0.08 &      -1.5 &       0.80 $\pm$ 0.16 &      -1.3     \\

\enddata
\end{deluxetable*}

%% file: discussion.tex
%!TEX root = ms.tex

\begin{figure*}[t]
  \epsscale{1.1}
  \epstrim{0.1in 0.1in 0.1in 0.1in}
  \plotone{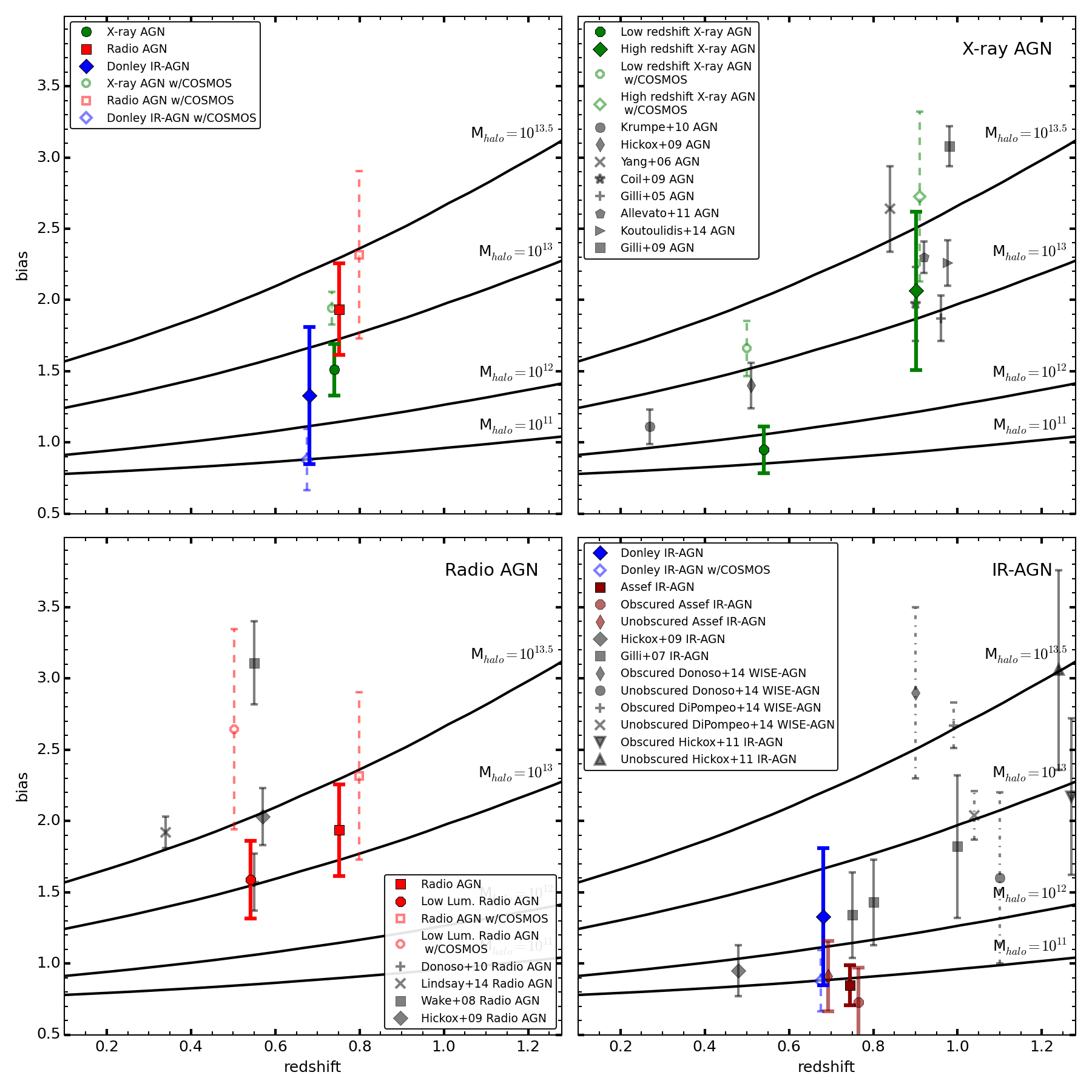}
  \caption{ Comparison of the absolute bias as a function of redshift for AGNs
  identified at different wavelengths in the literature. \textbf{Top left
  panel:} Bias of the \Xray (green circles), \Radio (red squares), and \Donley
  (blue diamonds) presented in this paper, where the uncertainties are derived
  using jackknife resampling of the fields. In all panels, we show the
  estimated bias for samples in this paper. 
  The filled symbols show the measured values excluding the COSMOS field.
  Black lines show constant dark matter halo mass (\Mhalo{=}{11} -
  \Mhalo{=}{13.5}). \textbf{Top right panel:} Comparison of bias values in the
  literature for \Xray. Our results are shown in green, while grey points show
  results from the literature to \zsim{1}. \textbf{Bottom left panel:} Bias of
  the high luminosity (red squares) and low luminosity \Radio samples (red
  circles). Grey points show results from the literature for \Radio.
  \textbf{Bottom right panel:} Bias of the \Donley (blue diamond) and \Assef
  samples (red square). We show both the \Assef WISE obscured subsample (red
  thin diamond) and unobscured subsample (red thin circle). Grey points show
  results from the literature for IR-AGN. }
  \label{fig:bias}
\end{figure*}

\section{Discussion}\label{sec:discussion}
We have combined spectroscopic redshifts with multi-wavelength imaging to
quantify the clustering properties of AGN selected at different wavelengths and
compared their clustering with matched galaxy samples. In this section, we
discuss the implications of these findings. In Section~\ref{sec:litdiscussion}
we compare our results to other multi-wavelength AGN clustering studies in the
literature. In Section~\ref{sec:lambdadiscussion} we investigate whether AGN
clustering amplitude depends on specific accretion rate, and in
Section~\ref{sec:environdiscussion} we discuss the halo mass dependence of AGN
activity.

\subsection{Comparison with the literature}\label{sec:litdiscussion}
In Figure~\ref{fig:bias} we compare the measured clustering amplitude of AGN
identified at different wavelengths using the bias parameter estimated on
scales $1\hMpc<\rp<10\hMpc$. We do not compare $r_0$ and $\gamma$ values in the
literature, due to the degeneracy between these parameters. In the top left
panel we compare the bias parameters for our three full samples selected at
different wavelengths, while each of the other panels compares our results at a
given wavelength with other results from the literature. In all panels the
filled, colored symbols show our results excluding the COSMOS field, while the
open, colored symbols include the COSMOS field.

% X-ray

In the top right panel we compare the bias values of various \Xray clustering
studies. For each of the results shown that include the COSMOS field (our open
symbol samples, \citet{Gilli09}, \citet{Allevato11}, and
\citet{Koutoulidis13}), the median dark matter halo mass of \Xray is above
\medianMhalo{13}, while those that do not include the COSMOS field
(\citet{Gilli05}, \citet{Yang06}, \citet{Krumpe10}, \citet{Hickox09}, and
\citet{Coil09}) find a median mass below \medianMhalo{13.0}. Both
\citet{Gilli05} and \citet{Yang06} use one or two small fields ($<0.5\,\degsq$)
to measure the clustering of \Xray, which leads to an underestimate of the
cosmic variance in their measurements. Including the COSMOS field
systematically raises the bias due to the large over-densities found in the
field at $z<1$. \citet{Gilli09} find an increase of $\sim24\%$ in the bias when
including the over-density found at \zsim{0.36} within the COSMOS field. We
find a similar $\sim20\%$ increase.
% (over our full redshift range) in the clustering
% of X-ray AGN when we include COSMOS, as our results rely on many fields,
% somewhat mitigating the effects of a single very overdense field.
When we exclude COSMOS, we find similar results to \citet{Coil09} and \cite{Hickox09}.

Interestingly, \citet{Allevato11} and \citet{Koutoulidis13} include the COSMOS
field in their results and find \Xray bias values similar to our fiducial 
results, which exclude COSMOS.  However, \citet{Allevato11} derive their results
using {\it only} the COSMOS field and bootstrap errors within that field,
such that their errors due to cosmic variance are underestimated. The analysis
in \citet{Koutoulidis13} spans a very wide redshift range, $0<z<3$, and the 
results shown here are for the median redshift of their sample.  This makes 
a direct comparison with results derived in smaller redshift bins 
somewhat difficult.

% \medianpower{\LX}{=}{42.5}{\ergs}

Taken together, all of these results show that \Xray are typically found in
somewhat more massive dark matter halos at \medianzsim{0.9} compared
to \medianzsim{0.4}. The lower redshift \Xray have a lower median
X-ray luminosity (\medianLx{42.3}) compared to the higher
redshift \Xray (\medianLx{43.2}). However, we do not find a
correlation between clustering amplitude and X-ray luminosity in our samples,
which suggests that the luminosity differences between the redshift samples is
not driving the difference in clustering strength. We also do not find a
significant difference in the median stellar mass of the lower redshift AGN
hosts in our sample (\medianMass{10.7}) compared to the
higher redshift hosts (\medianMass{10.8}). This
difference is similar to that of the low X-ray luminosity AGN
(\medianMass{10.7}) and high X-ray luminosity AGN
(\medianMass{10.8}) samples, suggesting that differences
in stellar mass are not driving the redshift-dependent results seen here.

% Radio

In the bottom left panel we show the bias of \Radio compared to results in the
literature. Generally, the bias we measure agrees well with other published
studies and indicates a relatively high dark matter halo mass of
\medianMhalo{13.3}. Including the COSMOS field again
increases the bias by $\sim50$\% but also substantially increases the error
bars, due to that one field having a systematically different clustering
amplitude compared to the other fields. The \citet{Wake08} results are higher
than other measurements and are derived from the cross-correlation of
radio-loud luminous red galaxies (LRGs) (\PR{>}{24}) with the main 2SLAQ LRG
survey. Their radio-detected LRGs have luminosities of 3-5$L_*$, far higher
than the average \Radio, which may account for their high bias value. Our low
luminosity \Radio have a consistent bias value as the \citet{Donoso10} and
\citet{Hickox09} samples.

% IR

In the bottom right panel we show the \IR selected using the \Donley selection
technique (blue diamond) and the \Assef selection technique (red square). We
additionally show the \Assef obscured (red diamond) and unobscured (red circle)
subsamples. We compare our results to those of \citet[grey diamond,
\zsim{0.5}]{Hickox09} and \citet[grey square, $z>0.7$]{Gilli07}, as well as
angular clustering estimates from \citet{Donoso13} and \citet{Dipompeo14} for
obscured and unobscured sources. While the individual selection techniques
compared (e.g. \Donley, \Stern, $f_{24\um}$-selected) differ, we generally find
that the \IR samples have lower bias values than the \Xray or \Radio and
therefore typically reside in lower mass dark matter halos
(\medianMhalo{11.5}). We do not find a significant
difference between the clustering amplitudes of samples identified using the
\Donley or \Assef techniques. This is in agreement with \citet{Hickox11}, who
consider somewhat higher redshift (\medianzsim{1.2}) and higher
luminosity (\medianLx{44}) IR-AGN than we do here. However,
our results coupled with theirs suggest that there is not a significant
difference in the clustering of obscured and unobscured IR-AGN for a range of
different redshifts and luminosities.

We also find no significant difference in the bias of the obscured and
unobscured \Assef. While there is a $\sim29\%$ higher bias for the unobscured
sources compared to the obscured sources, this is not significant
($\sim1\sigma$). Our results do not agree with the angular clustering
measurements from \citet{Dipompeo14} or \citet{Donoso13}. Since we use
spectroscopic redshifts, our results are more robust to differences in the
redshift distributions of the two samples, which could be driving the
difference in the clustering amplitude of the obscured and unobscured samples
in these other papers, as discussed above. Additionally, we find that the
unobscured \Assef have a brighter median W1 flux ($\sim0.5$dex) than obscured
\Assef, which suggest that these samples have different effective luminosities,
which will result in different redshift distributions. Since we require
spectroscopic redshifts our samples are smaller and therefore our statistical
error bars are larger, however the systematic errors associated with our
spectroscopic samples should be much lower.

We note that the redshift success rate (the fraction of PRIMUS targets for
which we derive a robust redshift) is very similar for the \WISE sample
($\sim72\%$) as for the full PRIMUS sample ($\sim75\%$ \citet{Cool13}), when we
account for the number of \WISE sources outside the redshift range of PRIMUS,
using the \citet{Dipompeo14} \WISE redshift distribution from the AGES survey
\citep[the AGN and Galaxy Evolution Survey][]{Kochanek11}. We find a small
trend (10\% difference from the median) between the redshift success fraction
and the observed W2 magnitude, where brighter \WISE have a higher redshift
success fraction. We find a similar trend for both the obscured and unobscured
sources, which would bias both samples to somewhat more luminous sources.

% Previous section 6.2 is now in section 5

 \begin{figure}
   \epsscale{1.1}
   \epstrim{0.1in 0.1in 0.1in 0.1in}
   \plotone{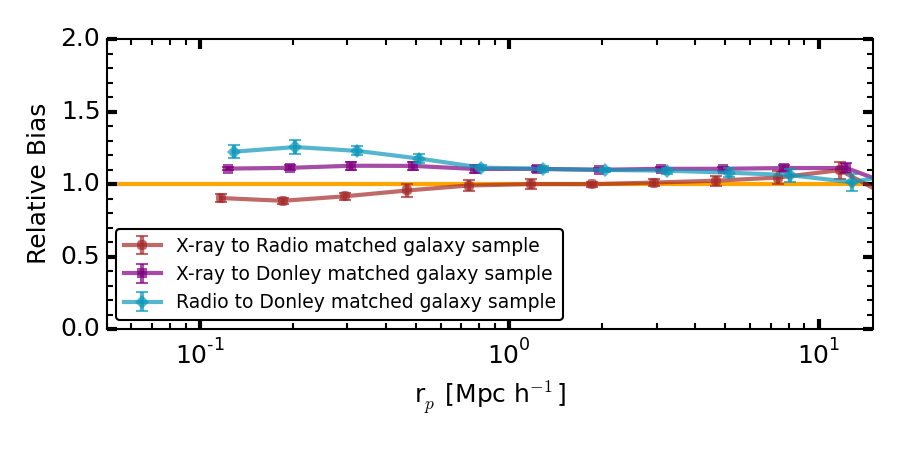}
   \caption{ The relative bias for galaxy samples matched in stellar mass,
   sSFR, and redshift to AGN samples selected at different wavelengths. Error
   bars are from jackknife resampling of all fields. We show the
   scale-dependent relative bias for the \Xray to \Radio matched galaxies (dark
   red), \Xray to \Donley matched galaxies (purple), and \Radio to \Donley
   matched galaxies (cyan).}
   \label{fig:matchedratio}
 \end{figure}

%  (Better subtitle?)
\subsection{Does AGN activity depend on halo mass?}%
\label{sec:environdiscussion}
We show that differences in the host galaxy populations of AGN identified at
different wavelengths likely contribute to (if not fully account for) the
differences in the inferred host dark matter halo masses of these AGN. To
account for the known host galaxy selection biases in AGN identification, we
compare the clustering of each AGN sample to galaxy control samples with the
same distribution of stellar mass, sSFR, and redshift in
Section~\ref{sec:matchedclustering} and find no significant differences.
Likewise, \citet{Leauthaud15} find that \Xray do not reside in different halos
than inactive galaxies, when they control for the stellar mass distribution of
their detected \Xray. The lack of significant differences in the clustering
amplitudes of AGN and their galaxy control samples at even the $<2\,\sigma$
level strongly suggests that the physical mechanisms that are fueling and
triggering AGN either correlate with galaxy environment on scales much smaller
than those that we probe here (\rpvalue{\sim}{0.1}) or that AGN triggering is
not correlated with the mass of the dark matter halo. For example, our results
are consistent with \citet{Ellison11}, who use pairs of optically-selected AGN
to identify a sharp increase in the activity of AGN at close separations ($<80$
\hKpc); here we are sensitive to $>100\,\hKpc$ only.

The validity of these statements relies on our ability to estimate the relevant
galaxy properties and the fractional uncertainty in the clustering
measurements. Both the stellar mass and sSFR estimates that we use improve upon
previous techniques found in the literature, as we fit full stellar population
synthesis models to the broad-band SEDs of the AGN and galaxy samples in a
consistent manner, to limit biases that result from using individual bands or
simple color-to-SFR correlations \citep[e.g.,][]{Mostek13}. Additionally, we
use relatively large samples of AGN and galaxy control samples with
spectroscopic redshifts and maximize their measurement power by
cross-correlating each with the PRIMUS and DEEP2 galaxy samples. This lowers
the fractional uncertainty in our measurements and increases the significance
of our results.

While we do not find a significant difference between the clustering of \Radio
and matched galaxy control samples, \citet{Wake08} and \citet{Donoso10} find
that \Radio are more clustered than their stellar mass-matched samples. Both
compare mass-matched luminous red galaxies (LRGs) to \Radio at \zsim{0.5} and
find that \Radio are significantly more clustered than the matched sample.
While both authors examine \Radio with stellar masses and halo masses higher
than those probed by our sample, neither of these papers explicitly control for
differences in the sSFR distributions, which we find to be important in
comparing different AGN samples to their host galaxies.

Finally, the matched galaxy control samples account for the individual
selection biases from the AGN samples identified at different wavelengths.
These biases depend on the depth of the sample and the different wavelengths
that are used to identify the AGN, both of which lead to differences in host
galaxy properties (e.g., stellar mass, sSFR, and redshift). In
Figure~\ref{fig:matchedagn} we show the scale dependent relative bias between
matched galaxy control samples for our AGN identified at different wavelengths.
Similar to the relative biases found between the AGN samples, we find that the
clustering strength of both the \Xray and \Radio matched galaxy control samples
are higher than the \Donley matched galaxy control sample on all scales. The
clustering strength of the \Radio matched galaxy control sample is higher than
the \Xray matched galaxy control sample on small scales but is statistically
similar on large scales.

The consistency between the matched galaxy control samples and the AGN
identified at different wavelengths can be used to better measure the
clustering of AGN. Differences in the host galaxy properties of AGN selected at
different wavelengths can be understood entirely as being due to selection
effects \citep[i.e.,][]{Mendez13}. We have shown here that these selection
biases can entirely account for differences in the observed clustering
properties of AGN selected at different wavelengths. This confirms and extends
the \Xray results of \citet{Leauthaud15} to AGN detected in the radio and IR as
well. The clustering of AGN can therefore be understood in terms of the
clustering of their host galaxy populations.

%% file: conclusions.tex
%!TEX root = ms.tex

\section{Conclusions}\label{sec:conclusions}
In this paper we measure the clustering properties of X-ray, radio, and
IR-selected AGN in the PRIMUS and DEEP2 spectroscopic surveys. Within the \Xray
sample we measure the dependence of clustering on X-ray luminosity, specific
accretion rate, and hardness ratio. Within the \Radio sample we measure the
dependence of clustering on radio luminosity, and within the \IR sample we
measure the dependence of clustering on obscuration. As the AGN in these
samples span a wide range of specific accretion rates (which roughly traces
Eddington ratio), we also investigate the dependence of clustering on specific
accretion rate. We quantify the relative clustering strength (or relative bias)
between each AGN sample, as well as between the AGN samples and galaxy control
samples that are matched in stellar mass, sSFR, and redshift. The main results
from our work are as follows:

\begin{enumerate}

\item The clustering amplitude of observed \Xray, \Radio and \Donley at $0.2 <
z < 1.2$ differ, indicating that they reside in different mass dark matter
halos. \Xray and \Radio cluster similarly, and both are more clustered (at
$\gtrsim$2$\sigma$) than \Donley, especially on scales \rpvalue{<}{1.0}. We
estimate that our \Xray, \Radio, and \Donley samples have median dark matter
halo masses of \Mhalo{\sim}{12.9}, \Mhalo{\sim}{13.3}, and \Mhalo{\sim}{12.8},
respectively.

\item We find no significant dependencies ($<2\sigma$) on the clustering
amplitude with X-ray luminosity, specific accretion rate, or hardness ratio. We
also find no significant difference in the clustering amplitude of radio-loud
AGN (\PR{>}{24}) compared to low luminosity radio-detected AGN (\PR{<}{24}).

\item There is no significant difference in the clustering of \IR samples
selected using either the \citet{Donley12} or \citet{Assef13} selection
techniques. Using either selection we find no significant difference in the
clustering amplitude of obscured versus unobscured IR-AGN, using WISE-optical
colors to define obscuration. This suggests that previously published
differences determined using angular clustering are dominated by differences
and uncertainties in the redshift distributions of these sources.

\item The clustering amplitudes of the \Xray, \Radio and \Donley samples agree
well with those of the matched galaxy control samples, which have the same
distribution in stellar mass, sSFR, and redshift of the AGN host galaxy samples.

\end{enumerate}

It is now understood that all AGN selection techniques have inherent biases no
matter which waveband or technique is used. For example, AGN identified using
X-ray, radio, or MIR emission, as used here, are all more easily detected in
massive host galaxies. This means that clustering results should always be
interpreted as the clustering of \textit{observed} AGN samples, above a given
flux limit and therefore corresponding to a given stellar mass limit. It is
clear that in addition to selection biases with stellar mass, there are
additional biases with respect to the SFR of the host galaxy, where \Radio tend
to be identified in quiescent galaxies and \IR have a bias towards being
detected in star-forming host galaxies. This affects the observed clustering of
the AGN, which should only be interpreted relative to matched galaxy samples.
The full population of AGN amongst galaxies of all stellar masses is likely to
exhibit different clustering properties that the observed AGN, which are more
easily identified in massive galaxies. When we match samples with respect to
stellar mass, SFR, and redshift we find excellent agreement between the
clustering of AGN host galaxies and inactive galaxies. Therefore AGN clustering
can be understood entirely in terms of galaxy clustering (and the dependence of
clustering on galaxy properties) and AGN selection effects. Looking forward,
theoretical models that include AGN evolution and predict the clustering of AGN
must include AGN selection biases in order to accurately constrain the physical
mechanisms triggering AGN. Future observational results from surveys such as 
the Dark Energy Survey (DES) and the Hyper Suprime-Cam (HSC) Survey will 
also need to account for these AGN selection effects as they push to 
lower AGN luminosities in order to tighten constraints on theoretical models.

%% Support and Acknoledgements
\vspace{4em}
We thank Andy Goulding for reducing the DEEP2 02hr IRAC data and providing a
source catalog.

Funding for PRIMUS has been provided by NSF grants AST-0607701, 0908246,
0908442, 0908354, and NASA grant 08-ADP08-0019. ALC acknowledges support from
NSF CAREER award AST-1055081. AJM and JA acknowledge support from NASA grant
NNX12AE23G through the Astrophysics Data Analysis Program. AMD acknowledges
support from The Grainger Foundation.

We thank the CFHTLS, COSMOS, DLS, and SWIRE teams for their public data
releases and/or access to early releases. This paper includes data gathered
with the 6.5 m Magellan Telescopes located at Las Campanas Observatory, Chile.
We thank the support staff at LCO for their help during our observations, and
we acknowledge the use of community access through NOAO observing time. We use
data from the DEEP2 survey, which was supported by NSF AST grants AST00-71048,
AST00-71198, AST05-07428, AST05-07483, AST08-07630, AST08-08133. This study
makes use of data from AEGIS Survey and in particular uses data from $\galex$,
Keck, and CFHT. The AEGIS Survey was supported in part by the NSF, NASA, and
the STFC. Some of the data used for this project are from the CFHTLS public
data release, which includes observations obtained with MegaPrime/MegaCam, a
joint project of CFHT and CEA/DAPNIA, at the Canada-France-Hawaii Telescope
(CFHT) which is operated by the National Research Council (NRC) of Canada, the
Institut National des Science de l'Univers of the Centre National de la
Recherche Scientifique (CNRS) of France, and the University of Hawaii. This
work is based in part on data products produced at TERAPIX and the Canadian
Astronomy Data Centre as part of the Canada-France-Hawaii Telescope Legacy
Survey, a collaborative project of NRC and CNRS. We also thank those who have
built and operate the Chandra and XMM-Newton X-Ray Observatories. This research
has made use of the NASA/IPAC Infrared Science Archive, which is operated by
the Jet Propulsion Laboratory, California Institute of Technology, under
contract with the National Aeronautics and Space Administration.

%% file: ms.bbl
\begin{thebibliography}{141}
\expandafter\ifx\csname natexlab\endcsname\relax\def\natexlab#1{#1}\fi

\bibitem[{{Aird} {et~al.}(2012){Aird}, {Coil}, {Moustakas}, {Blanton},
  {Burles}, {Cool}, {Eisenstein}, {Smith}, {Wong}, \& {Zhu}}]{Aird12}
{Aird}, J., {et~al.} 2012, \href {http://dx.doi.org/10.1088/0004-637X/746/1/90}
  {\apj}, 746, 90

\bibitem[{{Aird} {et~al.}(2015){Aird}, {Alexander}, {Ballantyne}, {Civano},
  {Del-Moro}, {Hickox}, {Lansbury}, {Mullaney}, {Bauer}, {Brandt}, {Comastri},
  {Fabian}, {Gandhi}, {Harrison}, {Luo}, {Stern}, {Treister}, {Zappacosta},
  {Ajello}, {Assef}, {Balokovi{\'c}}, {Boggs}, {Brightman}, {Christensen},
  {Craig}, {Elvis}, {Forster}, {Grefenstette}, {Hailey}, {Koss}, {LaMassa},
  {Madsen}, {Puccetti}, {Saez}, {Urry}, {Wik}, \& {Zhang}}]{Aird15}
{Aird}, J., {et~al.} 2015, \href {http://dx.doi.org/10.1088/0004-637X/815/1/66}
  {\apj}, 815, 66

\bibitem[{{Allevato} {et~al.}(2011){Allevato}, {Finoguenov}, {Cappelluti},
  {Miyaji}, {Hasinger}, {Salvato}, {Brusa}, {Gilli}, {Zamorani}, {Shankar},
  {James}, {McCracken}, {Bongiorno}, {Merloni}, {Peacock}, {Silverman}, \&
  {Comastri}}]{Allevato11}
{Allevato}, V., {et~al.} 2011, \href
  {http://dx.doi.org/10.1088/0004-637X/736/2/99} {\apj}, 736, 99

\bibitem[{{Allevato} {et~al.}(2012){Allevato}, {Finoguenov}, {Hasinger},
  {Miyaji}, {Cappelluti}, {Salvato}, {Zamorani}, {Gilli}, {George}, {Tanaka},
  {Brusa}, {Silverman}, {Civano}, {Elvis}, \& {Shankar}}]{Allevato12}
{Allevato}, V., {et~al.} 2012, \href
  {http://dx.doi.org/10.1088/0004-637X/758/1/47} {\apj}, 758, 47

\bibitem[{{Antonucci} \& {Ulvestad}(1985)}]{Antonucci85}
{Antonucci}, R.~R.~J., \& {Ulvestad}, J.~S. 1985, \href
  {http://dx.doi.org/10.1086/163284} {\apj}, 294, 158

\bibitem[{{Appleton} {et~al.}(2004){Appleton}, {Fadda}, {Marleau}, {Frayer},
  {Helou}, {Condon}, {Choi}, {Yan}, {Lacy}, {Wilson}, {Armus}, {Chapman},
  {Fang}, {Heinrichson}, {Im}, {Jannuzi}, {Storrie-Lombardi}, {Shupe},
  {Soifer}, {Squires}, \& {Teplitz}}]{Appleton04}
{Appleton}, P.~N., {et~al.} 2004, \href {http://dx.doi.org/10.1086/422425}
  {\apjs}, 154, 147

\bibitem[{{Assef} {et~al.}(2013){Assef}, {Stern}, {Kochanek}, {Blain},
  {Brodwin}, {Brown}, {Donoso}, {Eisenhardt}, {Jannuzi}, {Jarrett}, {Stanford},
  {Tsai}, {Wu}, \& {Yan}}]{Assef13}
{Assef}, R.~J., {et~al.} 2013, \href
  {http://dx.doi.org/10.1088/0004-637X/772/1/26} {\apj}, 772, 26

\bibitem[{{Azadi} {et~al.}(2015){Azadi}, {Aird}, {Coil}, {Moustakas}, {Mendez},
  {Blanton}, {Cool}, {Eisenstein}, {Wong}, \& {Zhu}}]{Azadi14}
{Azadi}, M., {et~al.} 2015, \href {http://arxiv.org/abs/1407.1975} {\apj}, 806,
  187

\bibitem[{{Barmby} {et~al.}(2008){Barmby}, {Huang}, {Ashby}, {Eisenhardt},
  {Fazio}, {Willner}, \& {Wright}}]{Barmby08}
{Barmby}, P., {et~al.} 2008, \href {http://dx.doi.org/10.1086/588583} {\apjs},
  177, 431

\bibitem[{{Barro} {et~al.}(2011){Barro}, {P{\'e}rez-Gonz{\'a}lez}, {Gallego},
  {Ashby}, {Kajisawa}, {Miyazaki}, {Villar}, {Yamada}, \& {Zamorano}}]{Barro11}
{Barro}, G., {et~al.} 2011, \href
  {http://dx.doi.org/10.1088/0067-0049/193/1/13} {\apjs}, 193, 13

\bibitem[{{Becker} {et~al.}(1994){Becker}, {White}, \& {Helfand}}]{Becker94}
{Becker}, R.~H., {White}, R.~L., \& {Helfand}, D.~J. 1994, in Astronomical
  Society of the Pacific Conference Series, Vol.~61, Astronomical Data Analysis
  Software and Systems III, ed. D.~R. {Crabtree}, R.~J. {Hanisch}, \&
  J.~{Barnes}, 165

\bibitem[{{Becker} {et~al.}(1995){Becker}, {White}, \& {Helfand}}]{Becker95}
{Becker}, R.~H., {White}, R.~L., \& {Helfand}, D.~J. 1995, \href
  {http://dx.doi.org/10.1086/176166} {\apj}, 450, 559

\bibitem[{{Best} {et~al.}(2005){Best}, {Kauffmann}, {Heckman}, {Brinchmann},
  {Charlot}, {Ivezi{\'c}}, \& {White}}]{Best05}
{Best}, P.~N., {et~al.} 2005, \href
  {http://dx.doi.org/10.1111/j.1365-2966.2005.09192.x} {\mnras}, 362, 25

\bibitem[{{Bigelow} \& {Dressler}(2003)}]{Bigelow03}
{Bigelow}, B.~C., \& {Dressler}, A.~M. 2003, in Society of Photo-Optical
  Instrumentation Engineers (SPIE) Conference Series, Vol. 4841, Society of
  Photo-Optical Instrumentation Engineers (SPIE) Conference Series, ed.
  M.~{Iye} \& A.~F.~M. {Moorwood}, 1727--1738.
\newblock \href {http://dx.doi.org/10.1117/12.461870} {[link]}

\bibitem[{{Blanton} \& {Roweis}(2007)}]{Blanton07}
{Blanton}, M.~R., \& {Roweis}, S. 2007, \href
  {http://dx.doi.org/10.1086/510127} {\aj}, 133, 734

\bibitem[{{Booth} \& {Schaye}(2010)}]{Booth10}
{Booth}, C.~M., \& {Schaye}, J. 2010, \href
  {http://dx.doi.org/10.1111/j.1745-3933.2010.00832.x} {\mnras}, 405, L1

\bibitem[{{Brusa} {et~al.}(2007){Brusa}, {Zamorani}, {Comastri}, {Hasinger},
  {Cappelluti}, {Civano}, {Finoguenov}, {Mainieri}, {Salvato}, {Vignali},
  {Elvis}, {Fiore}, {Gilli}, {Impey}, {Lilly}, {Mignoli}, {Silverman}, {Trump},
  {Urry}, {Bender}, {Capak}, {Huchra}, {Kneib}, {Koekemoer}, {Leauthaud},
  {Lehmann}, {Massey}, {Matute}, {McCarthy}, {McCracken}, {Rhodes}, {Scoville},
  {Taniguchi}, \& {Thompson}}]{Brusa07}
{Brusa}, M., {et~al.} 2007, \href {http://dx.doi.org/10.1086/516575} {\apjs},
  172, 353

\bibitem[{{Brusa} {et~al.}(2010){Brusa}, {Civano}, {Comastri}, {Miyaji},
  {Salvato}, {Zamorani}, {Cappelluti}, {Fiore}, {Hasinger}, {Mainieri},
  {Merloni}, {Bongiorno}, {Capak}, {Elvis}, {Gilli}, {Hao}, {Jahnke},
  {Koekemoer}, {Ilbert}, {Le Floc'h}, {Lusso}, {Mignoli}, {Schinnerer},
  {Silverman}, {Treister}, {Trump}, {Vignali}, {Zamojski}, {Aldcroft},
  {Aussel}, {Bardelli}, {Bolzonella}, {Cappi}, {Caputi}, {Contini},
  {Finoguenov}, {Fruscione}, {Garilli}, {Impey}, {Iovino}, {Iwasawa},
  {Kampczyk}, {Kartaltepe}, {Kneib}, {Knobel}, {Kovac}, {Lamareille},
  {Leborgne}, {Le Brun}, {Le Fevre}, {Lilly}, {Maier}, {McCracken}, {Pello},
  {Peng}, {Perez-Montero}, {de Ravel}, {Sanders}, {Scodeggio}, {Scoville},
  {Tanaka}, {Taniguchi}, {Tasca}, {de la Torre}, {Tresse}, {Vergani}, \&
  {Zucca}}]{Brusa10}
{Brusa}, M., {et~al.} 2010, \href
  {http://dx.doi.org/10.1088/0004-637X/716/1/348} {\apj}, 716, 348

\bibitem[{{Bruzual} \& {Charlot}(2003)}]{Bruzual03}
{Bruzual}, G., \& {Charlot}, S. 2003, \href
  {http://dx.doi.org/10.1046/j.1365-8711.2003.06897.x} {\mnras}, 344, 1000

\bibitem[{{Cappelluti} {et~al.}(2009){Cappelluti}, {Brusa}, {Hasinger},
  {Comastri}, {Zamorani}, {Finoguenov}, {Gilli}, {Puccetti}, {Miyaji},
  {Salvato}, {Vignali}, {Aldcroft}, {B{\"o}hringer}, {Brunner}, {Civano},
  {Elvis}, {Fiore}, {Fruscione}, {Griffiths}, {Guzzo}, {Iovino}, {Koekemoer},
  {Mainieri}, {Scoville}, {Shopbell}, {Silverman}, \& {Urry}}]{Cappelluti09}
{Cappelluti}, N., {et~al.} 2009, \href
  {http://dx.doi.org/10.1051/0004-6361/200810794} {\aap}, 497, 635

\bibitem[{{Cappelluti} {et~al.}(2010){Cappelluti}, {Ajello}, {Burlon},
  {Krumpe}, {Miyaji}, {Bonoli}, \& {Greiner}}]{Cappelluti10}
{Cappelluti}, N., {et~al.} 2010, \href
  {http://dx.doi.org/10.1088/2041-8205/716/2/L209} {\apjl}, 716, L209

\bibitem[{{Chabrier}(2003)}]{Chabrier03}
{Chabrier}, G. 2003, \href {http://dx.doi.org/10.1086/376392} {\pasp}, 115, 763

\bibitem[{{Charlot} \& {Fall}(2000)}]{Charlot00}
{Charlot}, S., \& {Fall}, S.~M. 2000, \href {http://dx.doi.org/10.1086/309250}
  {\apj}, 539, 718

\bibitem[{{Chiappetti} {et~al.}(2012){Chiappetti}, {Clerc}, {Pacaud}, {Pierre},
  {Gueguen}, {Paioro}, {Polletta}, {Melnyk}, {Elyiv}, {Surdej}, \&
  {Faccioli}}]{Chiappetti12}
{Chiappetti}, L., {et~al.} 2012, \href {http://arxiv.org/abs/1211.4492} {ArXiv
  e-prints}

\bibitem[{{Ciliegi} {et~al.}(2003){Ciliegi}, {Zamorani}, {Hasinger}, {Lehmann},
  {Szokoly}, \& {Wilson}}]{Ciliegi03}
{Ciliegi}, P., {et~al.} 2003, \href
  {http://dx.doi.org/10.1051/0004-6361:20021721} {\aap}, 398, 901

\bibitem[{{Civano} {et~al.}(2012){Civano}, {Elvis}, {Brusa}, {Comastri},
  {Salvato}, {Zamorani}, {Aldcroft}, {Bongiorno}, {Capak}, {Cappelluti},
  {Cisternas}, {Fiore}, {Fruscione}, {Hao}, {Kartaltepe}, {Koekemoer}, {Gilli},
  {Impey}, {Lanzuisi}, {Lusso}, {Mainieri}, {Miyaji}, {Lilly}, {Masters},
  {Puccetti}, {Schawinski}, {Scoville}, {Silverman}, {Trump}, {Urry},
  {Vignali}, \& {Wright}}]{Civano12}
{Civano}, F., {et~al.} 2012, VizieR Online Data Catalog, 220, 10030

\bibitem[{{Coil} {et~al.}(2004){Coil}, {Newman}, {Kaiser}, {Davis}, {Ma},
  {Kocevski}, \& {Koo}}]{Coil04a}
{Coil}, A.~L., {et~al.} 2004, \href {http://dx.doi.org/10.1086/425676} {\apj},
  617, 765

\bibitem[{{Coil} {et~al.}(2008){Coil}, {Newman}, {Croton}, {Cooper}, {Davis},
  {Faber}, {Gerke}, {Koo}, {Padmanabhan}, {Wechsler}, \& {Weiner}}]{Coil08}
{Coil}, A.~L., {et~al.} 2008, \href {http://dx.doi.org/10.1086/523639} {\apj},
  672, 153

\bibitem[{{Coil} {et~al.}(2009){Coil}, {Georgakakis}, {Newman}, {Cooper},
  {Croton}, {Davis}, {Koo}, {Laird}, {Nandra}, {Weiner}, {Willmer}, \&
  {Yan}}]{Coil09}
{Coil}, A.~L., {et~al.} 2009, \href
  {http://dx.doi.org/10.1088/0004-637X/701/2/1484} {\apj}, 701, 1484

\bibitem[{{Coil} {et~al.}(2011){Coil}, {Blanton}, {Burles}, {Cool},
  {Eisenstein}, {Moustakas}, {Wong}, {Zhu}, {Aird}, {Bernstein}, {Bolton}, \&
  {Hogg}}]{Coil11}
{Coil}, A.~L., {et~al.} 2011, \href
  {http://dx.doi.org/10.1088/0004-637X/741/1/8} {\apj}, 741, 8

\bibitem[{{Condon}(1992)}]{Condon92}
{Condon}, J.~J. 1992, \href
  {http://dx.doi.org/10.1146/annurev.aa.30.090192.003043} {\araa}, 30, 575

\bibitem[{{Condon} {et~al.}(1998){Condon}, {Cotton}, {Greisen}, {Yin},
  {Perley}, {Taylor}, \& {Broderick}}]{Condon98}
{Condon}, J.~J., {et~al.} 1998, \href {http://dx.doi.org/10.1086/300337} {\aj},
  115, 1693

\bibitem[{{Cool} {et~al.}(2013){Cool}, {Moustakas}, {Blanton}, {Burles},
  {Coil}, {Eisenstein}, {Wong}, {Zhu}, {Aird}, {Bernstein}, {Bolton}, {Hogg},
  \& {Mendez}}]{Cool13}
{Cool}, R.~J., {et~al.} 2013, \href
  {http://dx.doi.org/10.1088/0004-637X/767/2/118} {\apj}, 767, 118

\bibitem[{{Cress} {et~al.}(1996){Cress}, {Helfand}, {Becker}, {Gregg}, \&
  {White}}]{Cress96}
{Cress}, C.~M., {et~al.} 1996, \href {http://dx.doi.org/10.1086/178122} {\apj},
  473, 7

\bibitem[{{Croton}(2009)}]{Croton09}
{Croton}, D.~J. 2009, \href
  {http://dx.doi.org/10.1111/j.1365-2966.2009.14429.x} {\mnras}, 394, 1109

\bibitem[{{Cutri} {et~al.}(2011){Cutri}, {Wright}, {Conrow}, {Bauer},
  {Benford}, {Brandenburg}, {Dailey}, {Eisenhardt}, {Evans}, {Fajardo-Acosta},
  {Fowler}, {Gelino}, {Grillmair}, {Harbut}, {Hoffman}, {Jarrett},
  {Kirkpatrick}, {Liu}, {Mainzer}, {Marsh}, {Masci}, {McCallon}, {Padgett},
  {Ressler}, {Royer}, {Skrutskie}, {Stanford}, {Wyatt}, {Tholen}, {Tsai},
  {Wachter}, {Wheelock}, {Yan}, {Alles}, {Beck}, {Grav}, {Masiero}, {McCollum},
  {McGehee}, \& {Wittman}}]{Cutri11}
{Cutri}, R.~M., {et~al.} 2011, {Explanatory Supplement to the WISE Preliminary
  Data Release Products}, Tech. rep., IPAC/Caltech

\bibitem[{{Davis} \& {Peebles}(1983)}]{Davis83}
{Davis}, M., \& {Peebles}, P.~J.~E. 1983, \href
  {http://dx.doi.org/10.1086/160884} {\apj}, 267, 465

\bibitem[{{Davis} {et~al.}(2003){Davis}, {Faber}, {Newman}, {Phillips},
  {Ellis}, {Steidel}, {Conselice}, {Coil}, {Finkbeiner}, {Koo}, {Guhathakurta},
  {Weiner}, {Schiavon}, {Willmer}, {Kaiser}, {Luppino}, {Wirth}, {Connolly},
  {Eisenhardt}, {Cooper}, \& {Gerke}}]{Davis03}
{Davis}, M., {et~al.} 2003, in Society of Photo-Optical Instrumentation
  Engineers (SPIE) Conference Series, Vol. 4834, Society of Photo-Optical
  Instrumentation Engineers (SPIE) Conference Series, ed. P.~{Guhathakurta},
  161--172.
\newblock \href {http://dx.doi.org/10.1117/12.457897} {[link]}

\bibitem[{{de la Torre} {et~al.}(2010){de la Torre}, {Guzzo}, {Kova{\v c}},
  {Porciani}, {Abbas}, {Meneux}, {Carollo}, {Contini}, {Kneib}, {Le F{\`e}vre},
  {Lilly}, {Mainieri}, {Renzini}, {Sanders}, {Scodeggio}, {Scoville},
  {Zamorani}, {Bardelli}, {Bolzonella}, {Bongiorno}, {Caputi}, {Coppa},
  {Cucciati}, {de Ravel}, {Franzetti}, {Garilli}, {Iovino}, {Kampczyk},
  {Knobel}, {Koekemoer}, {Lamareille}, {Le Borgne}, {Le Brun}, {Maier},
  {Mignoli}, {Pell{\'o}}, {Peng}, {Perez-Montero}, {Ricciardelli}, {Silverman},
  {Tanaka}, {Tasca}, {Tresse}, {Vergani}, {Welikala}, {Zucca}, {Bottini},
  {Cappi}, {Cassata}, {Cimatti}, {Fumana}, {Ilbert}, {Leauthaud}, {Maccagni},
  {Marinoni}, {McCracken}, {Memeo}, {Nair}, {Oesch}, {Pozzetti}, {Presotto}, \&
  {Scaramella}}]{delaTorre10}
{de la Torre}, S., {et~al.} 2010, \href
  {http://dx.doi.org/10.1111/j.1365-2966.2010.17352.x} {\mnras}, 409, 867

\bibitem[{{DiPompeo} {et~al.}(2014){DiPompeo}, {Myers}, {Hickox}, {Geach}, \&
  {Hainline}}]{Dipompeo14}
{DiPompeo}, M.~A., {et~al.} 2014, \href
  {http://dx.doi.org/10.1093/mnras/stu1115} {\mnras}, 442, 3443

\bibitem[{{Donley} {et~al.}(2005){Donley}, {Rieke}, {Rigby}, \&
  {P{\'e}rez-Gonz{\'a}lez}}]{Donley05}
{Donley}, J.~L., {et~al.} 2005, \href {http://dx.doi.org/10.1086/491668}
  {\apj}, 634, 169

\bibitem[{{Donley} {et~al.}(2012){Donley}, {Koekemoer}, {Brusa}, {Capak},
  {Cardamone}, {Civano}, {Ilbert}, {Impey}, {Kartaltepe}, {Miyaji}, {Salvato},
  {Sanders}, {Trump}, \& {Zamorani}}]{Donley12}
{Donley}, J.~L., {et~al.} 2012, \href
  {http://dx.doi.org/10.1088/0004-637X/748/2/142} {\apj}, 748, 142

\bibitem[{{Donoso} {et~al.}(2010){Donoso}, {Li}, {Kauffmann}, {Best}, \&
  {Heckman}}]{Donoso10}
{Donoso}, E., {et~al.} 2010, \href
  {http://dx.doi.org/10.1111/j.1365-2966.2010.16907.x} {\mnras}, 407, 1078

\bibitem[{{Donoso} {et~al.}(2014){Donoso}, {Yan}, {Stern}, \&
  {Assef}}]{Donoso13}
{Donoso}, E., {et~al.} 2014, \href
  {http://dx.doi.org/10.1088/0004-637X/789/1/44} {\apj}, 789, 44

\bibitem[{{Ellison} {et~al.}(2011){Ellison}, {Patton}, {Mendel}, \&
  {Scudder}}]{Ellison11}
{Ellison}, S.~L., {et~al.} 2011, \href
  {http://dx.doi.org/10.1111/j.1365-2966.2011.19624.x} {\mnras}, 418, 2043

\bibitem[{{Elvis} {et~al.}(2009){Elvis}, {Civano}, {Vignali}, {Puccetti},
  {Fiore}, {Cappelluti}, {Aldcroft}, {Fruscione}, {Zamorani}, {Comastri},
  {Brusa}, {Gilli}, {Miyaji}, {Damiani}, {Koekemoer}, {Finoguenov}, {Brunner},
  {Urry}, {Silverman}, {Mainieri}, {Hasinger}, {Griffiths}, {Carollo}, {Hao},
  {Guzzo}, {Blain}, {Calzetti}, {Carilli}, {Capak}, {Ettori}, {Fabbiano},
  {Impey}, {Lilly}, {Mobasher}, {Rich}, {Salvato}, {Sanders}, {Schinnerer},
  {Scoville}, {Shopbell}, {Taylor}, {Taniguchi}, \& {Volonteri}}]{Elvis09}
{Elvis}, M., {et~al.} 2009, \href
  {http://dx.doi.org/10.1088/0067-0049/184/1/158} {\apjs}, 184, 158

\bibitem[{{Faber} {et~al.}(2003){Faber}, {Phillips}, {Kibrick}, {Alcott},
  {Allen}, {Burrous}, {Cantrall}, {Clarke}, {Coil}, {Cowley}, {Davis}, {Deich},
  {Dietsch}, {Gilmore}, {Harper}, {Hilyard}, {Lewis}, {McVeigh}, {Newman},
  {Osborne}, {Schiavon}, {Stover}, {Tucker}, {Wallace}, {Wei}, {Wirth}, \&
  {Wright}}]{Faber03}
{Faber}, S.~M., {et~al.} 2003, in Society of Photo-Optical Instrumentation
  Engineers (SPIE) Conference Series, Vol. 4841, Society of Photo-Optical
  Instrumentation Engineers (SPIE) Conference Series, ed. M.~{Iye} \& A.~F.~M.
  {Moorwood}, 1657--1669.
\newblock \href {http://dx.doi.org/10.1117/12.460346} {[link]}

\bibitem[{{Fanaroff} \& {Riley}(1974)}]{Fanaroff74}
{Fanaroff}, B.~L., \& {Riley}, J.~M. 1974, \mnras, 167, 31P

\bibitem[{{Fanidakis} {et~al.}(2013){Fanidakis}, {Georgakakis}, {Mountrichas},
  {Krumpe}, {Baugh}, {Lacey}, {Frenk}, {Miyaji}, \& {Benson}}]{Fanidakis13}
{Fanidakis}, N., {et~al.} 2013, \href {http://dx.doi.org/10.1093/mnras/stt1327}
  {\mnras}, 435, 679

\bibitem[{{Ferrarese} \& {Ford}(2005)}]{Ferrarese05}
{Ferrarese}, L., \& {Ford}, H. 2005, \href
  {http://dx.doi.org/10.1007/s11214-005-3947-6} {\ssr}, 116, 523

\bibitem[{{Franceschini} {et~al.}(1999){Franceschini}, {Hasinger}, {Miyaji}, \&
  {Malquori}}]{Franceschini99}
{Franceschini}, A., {et~al.} 1999, \href
  {http://dx.doi.org/10.1046/j.1365-8711.1999.03078.x} {\mnras}, 310, L5

\bibitem[{{Gebhardt} {et~al.}(2000){Gebhardt}, {Bender}, {Bower}, {Dressler},
  {Faber}, {Filippenko}, {Green}, {Grillmair}, {Ho}, {Kormendy}, {Lauer},
  {Magorrian}, {Pinkney}, {Richstone}, \& {Tremaine}}]{Gebhardt00}
{Gebhardt}, K., {et~al.} 2000, \href {http://dx.doi.org/10.1086/312840}
  {\apjl}, 539, L13

\bibitem[{{Georgakakis} {et~al.}(2008){Georgakakis}, {Nandra}, {Laird}, {Aird},
  \& {Trichas}}]{Georgakakis08}
{Georgakakis}, A., {et~al.} 2008, \href
  {http://dx.doi.org/10.1111/j.1365-2966.2008.13423.x} {\mnras}, 388, 1205

\bibitem[{{Gilli} {et~al.}(2005){Gilli}, {Daddi}, {Zamorani}, {Tozzi},
  {Borgani}, {Bergeron}, {Giacconi}, {Hasinger}, {Mainieri}, {Norman},
  {Rosati}, {Szokoly}, \& {Zheng}}]{Gilli05}
{Gilli}, R., {et~al.} 2005, \href
  {http://dx.doi.org/10.1051/0004-6361:20041375} {\aap}, 430, 811

\bibitem[{{Gilli} {et~al.}(2007){Gilli}, {Daddi}, {Chary}, {Dickinson},
  {Elbaz}, {Giavalisco}, {Kitzbichler}, {Stern}, \& {Vanzella}}]{Gilli07}
{Gilli}, R., {et~al.} 2007, \href
  {http://dx.doi.org/10.1051/0004-6361:20077506} {\aap}, 475, 83

\bibitem[{{Gilli} {et~al.}(2009){Gilli}, {Zamorani}, {Miyaji}, {Silverman},
  {Brusa}, {Mainieri}, {Cappelluti}, {Daddi}, {Porciani}, {Pozzetti}, {Civano},
  {Comastri}, {Finoguenov}, {Fiore}, {Salvato}, {Vignali}, {Hasinger}, {Lilly},
  {Impey}, {Trump}, {Capak}, {McCracken}, {Scoville}, {Taniguchi}, {Carollo},
  {Contini}, {Kneib}, {Le Fevre}, {Renzini}, {Scodeggio}, {Bardelli},
  {Bolzonella}, {Bongiorno}, {Caputi}, {Cimatti}, {Coppa}, {Cucciati}, {de La
  Torre}, {de Ravel}, {Franzetti}, {Garilli}, {Iovino}, {Kampczyk}, {Knobel},
  {Kova{\v c}}, {Lamareille}, {Le Borgne}, {Le Brun}, {Maier}, {Mignoli},
  {Pell{\`o}}, {Peng}, {Perez Montero}, {Ricciardelli}, {Tanaka}, {Tasca},
  {Tresse}, {Vergani}, {Zucca}, {Abbas}, {Bottini}, {Cappi}, {Cassata},
  {Fumana}, {Guzzo}, {Leauthaud}, {Maccagni}, {Marinoni}, {Memeo}, {Meneux},
  {Oesch}, {Scaramella}, \& {Walcher}}]{Gilli09}
{Gilli}, R., {et~al.} 2009, \href
  {http://dx.doi.org/10.1051/0004-6361:200810821} {\aap}, 494, 33

\bibitem[{{Goulding} {et~al.}(2012){Goulding}, {Forman}, {Hickox}, {Jones},
  {Kraft}, {Murray}, {Vikhlinin}, {Coil}, {Cooper}, {Davis}, \&
  {Newman}}]{Goulding12}
{Goulding}, A.~D., {et~al.} 2012, \href
  {http://dx.doi.org/10.1088/0067-0049/202/1/6} {\apjs}, 202, 6

\bibitem[{{Goulding} {et~al.}(2014){Goulding}, {Forman}, {Hickox}, {Jones},
  {Murray}, {Paggi}, {Ashby}, {Coil}, {Cooper}, {Huang}, {Kraft}, {Newman},
  {Weiner}, \& {Willner}}]{Goulding14}
{Goulding}, A.~D., {et~al.} 2014, \href
  {http://dx.doi.org/10.1088/0004-637X/783/1/40} {\apj}, 783, 40

\bibitem[{{Hasinger}(2008)}]{Hasinger08}
{Hasinger}, G. 2008, \href {http://dx.doi.org/10.1051/0004-6361:200809839}
  {\aap}, 490, 905

\bibitem[{{Hickox} {et~al.}(2009){Hickox}, {Jones}, {Forman}, {Murray},
  {Kochanek}, {Eisenstein}, {Jannuzi}, {Dey}, {Brown}, {Stern}, {Eisenhardt},
  {Gorjian}, {Brodwin}, {Narayan}, {Cool}, {Kenter}, {Caldwell}, \&
  {Anderson}}]{Hickox09}
{Hickox}, R.~C., {et~al.} 2009, \href
  {http://dx.doi.org/10.1088/0004-637X/696/1/891} {\apj}, 696, 891

\bibitem[{{Hickox} {et~al.}(2011){Hickox}, {Myers}, {Brodwin}, {Alexander},
  {Forman}, {Jones}, {Murray}, {Brown}, {Cool}, {Kochanek}, {Dey}, {Jannuzi},
  {Eisenstein}, {Assef}, {Eisenhardt}, {Gorjian}, {Stern}, {Le Floc'h},
  {Caldwell}, {Goulding}, \& {Mullaney}}]{Hickox11}
{Hickox}, R.~C., {et~al.} 2011, \href
  {http://dx.doi.org/10.1088/0004-637X/731/2/117} {\apj}, 731, 117

\bibitem[{{Hopkins} \& {Hernquist}(2009)}]{Hopkins09}
{Hopkins}, P.~F., \& {Hernquist}, L. 2009, \href
  {http://dx.doi.org/10.1088/0004-637X/698/2/1550} {\apj}, 698, 1550

\bibitem[{{Hopkins} {et~al.}(2006){Hopkins}, {Hernquist}, {Cox}, {Di Matteo},
  {Robertson}, \& {Springel}}]{Hopkins06}
{Hopkins}, P.~F., {et~al.} 2006, \href {http://dx.doi.org/10.1086/499298}
  {\apjs}, 163, 1

\bibitem[{{Hopkins} {et~al.}(2008){Hopkins}, {Hernquist}, {Cox}, \& {Kere{\v
  s}}}]{Hopkins08}
{Hopkins}, P.~F., {et~al.} 2008, \href {http://dx.doi.org/10.1086/524362}
  {\apjs}, 175, 356

\bibitem[{{H{\"u}tsi} {et~al.}(2014){H{\"u}tsi}, {Gilfanov}, \&
  {Sunyaev}}]{Hutsi14}
{H{\"u}tsi}, G., {Gilfanov}, M., \& {Sunyaev}, R. 2014, \href
  {http://dx.doi.org/10.1051/0004-6361/201321689} {\aap}, 561, A58

\bibitem[{{Ivison} {et~al.}(2007){Ivison}, {Chapman}, {Faber}, {Smail},
  {Biggs}, {Conselice}, {Wilson}, {Salim}, {Huang}, \& {Willner}}]{Ivison07}
{Ivison}, R.~J., {et~al.} 2007, \href {http://dx.doi.org/10.1086/517917}
  {\apjl}, 660, L77

\bibitem[{{Kauffmann} {et~al.}(2003){Kauffmann}, {Heckman}, {White}, {Charlot},
  {Tremonti}, {Brinchmann}, {Bruzual}, {Peng}, {Seibert}, {Bernardi},
  {Blanton}, {Brinkmann}, {Castander}, {Cs{\'a}bai}, {Fukugita}, {Ivezic},
  {Munn}, {Nichol}, {Padmanabhan}, {Thakar}, {Weinberg}, \&
  {York}}]{Kauffmann03}
{Kauffmann}, G., {et~al.} 2003, \href
  {http://dx.doi.org/10.1046/j.1365-8711.2003.06291.x} {\mnras}, 341, 33

\bibitem[{{Kochanek} {et~al.}(2012){Kochanek}, {Eisenstein}, {Cool},
  {Caldwell}, {Assef}, {Jannuzi}, {Jones}, {Murray}, {Forman}, {Dey}, {Brown},
  {Eisenhardt}, {Gonzalez}, {Green}, \& {Stern}}]{Kochanek11}
{Kochanek}, C.~S., {et~al.} 2012, \href
  {http://dx.doi.org/10.1088/0067-0049/200/1/8} {\apjs}, 200, 8

\bibitem[{{Kormendy} \& {Ho}(2013)}]{Kormendy13}
{Kormendy}, J., \& {Ho}, L.~C. 2013, \href
  {http://dx.doi.org/10.1146/annurev-astro-082708-101811} {\araa}, 51, 511

\bibitem[{{Kormendy} \& {Richstone}(1995)}]{KormendyRichstone95}
{Kormendy}, J., \& {Richstone}, D. 1995, \href
  {http://dx.doi.org/10.1146/annurev.aa.33.090195.003053} {\araa}, 33, 581

\bibitem[{{Koutoulidis} {et~al.}(2013){Koutoulidis}, {Plionis},
  {Georgantopoulos}, \& {Fanidakis}}]{Koutoulidis13}
{Koutoulidis}, L., {et~al.} 2013, \href
  {http://dx.doi.org/10.1093/mnras/sts119} {\mnras}, 428, 1382

\bibitem[{{Kova{\v c}} {et~al.}(2010){Kova{\v c}}, {Lilly}, {Cucciati},
  {Porciani}, {Iovino}, {Zamorani}, {Oesch}, {Bolzonella}, {Knobel},
  {Finoguenov}, {Peng}, {Carollo}, {Pozzetti}, {Caputi}, {Silverman}, {Tasca},
  {Scodeggio}, {Vergani}, {Scoville}, {Capak}, {Contini}, {Kneib}, {Le
  F{\`e}vre}, {Mainieri}, {Renzini}, {Bardelli}, {Bongiorno}, {Coppa}, {de la
  Torre}, {de Ravel}, {Franzetti}, {Garilli}, {Guzzo}, {Kampczyk},
  {Lamareille}, {Le Borgne}, {Le Brun}, {Maier}, {Mignoli}, {Pello}, {Perez
  Montero}, {Ricciardelli}, {Tanaka}, {Tresse}, {Zucca}, {Abbas}, {Bottini},
  {Cappi}, {Cassata}, {Cimatti}, {Fumana}, {Koekemoer}, {Maccagni}, {Marinoni},
  {McCracken}, {Memeo}, {Meneux}, \& {Scaramella}}]{Kovac10}
{Kova{\v c}}, K., {et~al.} 2010, \href
  {http://dx.doi.org/10.1088/0004-637X/708/1/505} {\apj}, 708, 505

\bibitem[{{Krumpe} {et~al.}(2010){Krumpe}, {Miyaji}, \& {Coil}}]{Krumpe10}
{Krumpe}, M., {Miyaji}, T., \& {Coil}, A.~L. 2010, \href
  {http://dx.doi.org/10.1088/0004-637X/713/1/558} {\apj}, 713, 558

\bibitem[{{Laird} {et~al.}(2009){Laird}, {Nandra}, {Georgakakis}, {Aird},
  {Barmby}, {Conselice}, {Coil}, {Davis}, {Faber}, {Fazio}, {Guhathakurta},
  {Koo}, {Sarajedini}, \& {Willmer}}]{Laird09}
{Laird}, E.~S., {et~al.} 2009, \href
  {http://dx.doi.org/10.1088/0067-0049/180/1/102} {\apjs}, 180, 102

\bibitem[{{Landy} \& {Szalay}(1993)}]{Landy93}
{Landy}, S.~D., \& {Szalay}, A.~S. 1993, \href
  {http://dx.doi.org/10.1086/172900} {\apj}, 412, 64

\bibitem[{{Le F{\`e}vre} {et~al.}(2005){Le F{\`e}vre}, {Vettolani}, {Garilli},
  {Tresse}, {Bottini}, {Le Brun}, {Maccagni}, {Picat}, {Scaramella},
  {Scodeggio}, {Zanichelli}, {Adami}, {Arnaboldi}, {Arnouts}, {Bardelli},
  {Bolzonella}, {Cappi}, {Charlot}, {Ciliegi}, {Contini}, {Foucaud},
  {Franzetti}, {Gavignaud}, {Guzzo}, {Ilbert}, {Iovino}, {McCracken}, {Marano},
  {Marinoni}, {Mathez}, {Mazure}, {Meneux}, {Merighi}, {Paltani}, {Pell{\`o}},
  {Pollo}, {Pozzetti}, {Radovich}, {Zamorani}, {Zucca}, {Bondi}, {Bongiorno},
  {Busarello}, {Lamareille}, {Mellier}, {Merluzzi}, {Ripepi}, \&
  {Rizzo}}]{LeFevre05}
{Le F{\`e}vre}, O., {et~al.} 2005, \href
  {http://dx.doi.org/10.1051/0004-6361:20041960} {\aap}, 439, 845

\bibitem[{{Leauthaud} {et~al.}(2015){Leauthaud}, {J.~Benson}, {Civano},
  {L.~Coil}, {Bundy}, {Massey}, {Schramm}, {Schulze}, {Capak}, {Elvis},
  {Kulier}, \& {Rhodes}}]{Leauthaud15}
{Leauthaud}, A., {et~al.} 2015, \href {http://dx.doi.org/10.1093/mnras/stu2210}
  {\mnras}, 446, 1874

\bibitem[{{Lilly} {et~al.}(2007){Lilly}, {Le F{\`e}vre}, {Renzini}, {Zamorani},
  {Scodeggio}, {Contini}, {Carollo}, {Hasinger}, {Kneib}, {Iovino}, {Le Brun},
  {Maier}, {Mainieri}, {Mignoli}, {Silverman}, {Tasca}, {Bolzonella},
  {Bongiorno}, {Bottini}, {Capak}, {Caputi}, {Cimatti}, {Cucciati}, {Daddi},
  {Feldmann}, {Franzetti}, {Garilli}, {Guzzo}, {Ilbert}, {Kampczyk}, {Kovac},
  {Lamareille}, {Leauthaud}, {Borgne}, {McCracken}, {Marinoni}, {Pello},
  {Ricciardelli}, {Scarlata}, {Vergani}, {Sanders}, {Schinnerer}, {Scoville},
  {Taniguchi}, {Arnouts}, {Aussel}, {Bardelli}, {Brusa}, {Cappi}, {Ciliegi},
  {Finoguenov}, {Foucaud}, {Franceschini}, {Halliday}, {Impey}, {Knobel},
  {Koekemoer}, {Kurk}, {Maccagni}, {Maddox}, {Marano}, {Marconi}, {Meneux},
  {Mobasher}, {Moreau}, {Peacock}, {Porciani}, {Pozzetti}, {Scaramella},
  {Schiminovich}, {Shopbell}, {Smail}, {Thompson}, {Tresse}, {Vettolani},
  {Zanichelli}, \& {Zucca}}]{Lilly07}
{Lilly}, S.~J., {et~al.} 2007, \href {http://dx.doi.org/10.1086/516589}
  {\apjs}, 172, 70

\bibitem[{{Lonsdale} {et~al.}(2003){Lonsdale}, {Smith}, {Rowan-Robinson},
  {Surace}, {Shupe}, {Xu}, {Oliver}, {Padgett}, {Fang}, {Conrow},
  {Franceschini}, {Gautier}, {Griffin}, {Hacking}, {Masci}, {Morrison},
  {O'Linger}, {Owen}, {P{\'e}rez-Fournon}, {Pierre}, {Puetter}, {Stacey},
  {Castro}, {Polletta}, {Farrah}, {Jarrett}, {Frayer}, {Siana}, {Babbedge},
  {Dye}, {Fox}, {Gonzalez-Solares}, {Salaman}, {Berta}, {Condon}, {Dole}, \&
  {Serjeant}}]{Lonsdale03}
{Lonsdale}, C.~J., {et~al.} 2003, \href {http://dx.doi.org/10.1086/376850}
  {\pasp}, 115, 897

\bibitem[{{Lupton}(1993)}]{Lupton93}
{Lupton}, R. 1993, {Statistics in theory and practice} (Princeton University
  Press)

\bibitem[{{Madau} {et~al.}(1996){Madau}, {Ferguson}, {Dickinson}, {Giavalisco},
  {Steidel}, \& {Fruchter}}]{Madau96}
{Madau}, P., {et~al.} 1996, \href {http://arxiv.org/abs/arXiv:astro-ph/9607172}
  {\mnras}, 283, 1388

\bibitem[{{Magliocchetti} {et~al.}(2004){Magliocchetti}, {Maddox}, {Hawkins},
  {Peacock}, {Bland-Hawthorn}, {Bridges}, {Cannon}, {Cole}, {Colless},
  {Collins}, {Couch}, {Dalton}, {de Propris}, {Driver}, {Efstathiou}, {Ellis},
  {Frenk}, {Glazebrook}, {Jackson}, {Jones}, {Lahav}, {Lewis}, {Lumsden},
  {Norberg}, {Peterson}, {Sutherland}, {Taylor}, \& {2dFGRS
  Team}}]{Magliocchetti04}
{Magliocchetti}, M., {et~al.} 2004, \href
  {http://dx.doi.org/10.1111/j.1365-2966.2004.07751.x} {\mnras}, 350, 1485

\bibitem[{{Magorrian} {et~al.}(1998){Magorrian}, {Tremaine}, {Richstone},
  {Bender}, {Bower}, {Dressler}, {Faber}, {Gebhardt}, {Green}, {Grillmair},
  {Kormendy}, \& {Lauer}}]{Magorrian98}
{Magorrian}, J., {et~al.} 1998, \href {http://dx.doi.org/10.1086/300353} {\aj},
  115, 2285

\bibitem[{{Mandelbaum} {et~al.}(2009){Mandelbaum}, {Li}, {Kauffmann}, \&
  {White}}]{Mandelbaum09}
{Mandelbaum}, R., {et~al.} 2009, \href
  {http://dx.doi.org/10.1111/j.1365-2966.2008.14235.x} {\mnras}, 393, 377

\bibitem[{{Marconi} \& {Hunt}(2003)}]{Marconi03}
{Marconi}, A., \& {Hunt}, L.~K. 2003, \href {http://dx.doi.org/10.1086/375804}
  {\apjl}, 589, L21

\bibitem[{{Mateos} {et~al.}(2012){Mateos}, {Alonso-Herrero}, {Carrera},
  {Blain}, {Watson}, {Barcons}, {Braito}, {Severgnini}, {Donley}, \&
  {Stern}}]{Mateos12}
{Mateos}, S., {et~al.} 2012, \href
  {http://dx.doi.org/10.1111/j.1365-2966.2012.21843.x} {\mnras}, 426, 3271

\bibitem[{{Mateos} {et~al.}(2013){Mateos}, {Alonso-Herrero}, {Carrera},
  {Blain}, {Severgnini}, {Caccianiga}, \& {Ruiz}}]{Mateos13}
{Mateos}, S., {et~al.} 2013, \href {http://dx.doi.org/10.1093/mnras/stt953}
  {\mnras}, 434, 941

\bibitem[{{Matthews} {et~al.}(2013){Matthews}, {Newman}, {Coil}, {Cooper}, \&
  {Gwyn}}]{Matthews13}
{Matthews}, D.~J., {et~al.} 2013, \href
  {http://dx.doi.org/10.1088/0067-0049/204/2/21} {\apjs}, 204, 21

\bibitem[{{McBride} {et~al.}(2011){McBride}, {Connolly}, {Gardner}, {Scranton},
  {Newman}, {Scoccimarro}, {Zehavi}, \& {Schneider}}]{McBride11}
{McBride}, C.~K., {et~al.} 2011, \href
  {http://dx.doi.org/10.1088/0004-637X/726/1/13} {\apj}, 726, 13

\bibitem[{{McCracken} {et~al.}(2007){McCracken}, {Peacock}, {Guzzo}, {Capak},
  {Porciani}, {Scoville}, {Aussel}, {Finoguenov}, {James}, {Kitzbichler},
  {Koekemoer}, {Leauthaud}, {Le F{\`e}vre}, {Massey}, {Mellier}, {Mobasher},
  {Norberg}, {Rhodes}, {Sanders}, {Sasaki}, {Taniguchi}, {Thompson}, {White},
  \& {El-Zant}}]{McCracken07}
{McCracken}, H.~J., {et~al.} 2007, \href {http://dx.doi.org/10.1086/518693}
  {\apjs}, 172, 314

\bibitem[{{Mendez} {et~al.}(2013){Mendez}, {Coil}, {Aird}, {Diamond-Stanic},
  {Moustakas}, {Blanton}, {Cool}, {Eisenstein}, {Wong}, \& {Zhu}}]{Mendez13}
{Mendez}, A.~J., {et~al.} 2013, \href
  {http://dx.doi.org/10.1088/0004-637X/770/1/40} {\apj}, 770, 40

\bibitem[{{Meneux} {et~al.}(2009){Meneux}, {Guzzo}, {de la Torre}, {Porciani},
  {Zamorani}, {Abbas}, {Bolzonella}, {Garilli}, {Iovino}, {Pozzetti}, {Zucca},
  {Lilly}, {Le F{\`e}vre}, {Kneib}, {Carollo}, {Contini}, {Mainieri},
  {Renzini}, {Scodeggio}, {Bardelli}, {Bongiorno}, {Caputi}, {Coppa},
  {Cucciati}, {de Ravel}, {Franzetti}, {Kampczyk}, {Knobel}, {Kova{\v c}},
  {Lamareille}, {Le Borgne}, {Le Brun}, {Maier}, {Pell{\`o}}, {Peng}, {Perez
  Montero}, {Ricciardelli}, {Silverman}, {Tanaka}, {Tasca}, {Tresse},
  {Vergani}, {Bottini}, {Cappi}, {Cimatti}, {Cassata}, {Fumana}, {Koekemoer},
  {Leauthaud}, {Maccagni}, {Marinoni}, {McCracken}, {Memeo}, {Oesch}, \&
  {Scaramella}}]{Meneux09}
{Meneux}, B., {et~al.} 2009, \href
  {http://dx.doi.org/10.1051/0004-6361/200912314} {\aap}, 505, 463

\bibitem[{{Middelberg} {et~al.}(2007){Middelberg}, {Norris}, {Cornwell},
  {Voronkov}, {Siana}, {Boyle}, {Ciliegi}, {Jackson}, {Huynh}, {Berta},
  {Rubele}, {Lonsdale}, {Ivison}, {Smail}, \& {Oliver}}]{Middelberg07}
{Middelberg}, E., {et~al.} 2007, \href {http://arxiv.org/abs/0712.1409} {ArXiv
  e-prints}

\bibitem[{{Mo} \& {White}(1996)}]{Mo96}
{Mo}, H.~J., \& {White}, S.~D.~M. 1996, \href
  {http://arxiv.org/abs/astro-ph/9512127} {\mnras}, 282, 347

\bibitem[{{Mostek} {et~al.}(2013){Mostek}, {Coil}, {Cooper}, {Davis}, {Newman},
  \& {Weiner}}]{Mostek13}
{Mostek}, N., {et~al.} 2013, \href
  {http://dx.doi.org/10.1088/0004-637X/767/1/89} {\apj}, 767, 89

\bibitem[{{Moustakas} {et~al.}(2013){Moustakas}, {Coil}, {Aird}, {Blanton},
  {Cool}, {Eisenstein}, {Mendez}, {Wong}, {Zhu}, \& {Arnouts}}]{Moustakas13}
{Moustakas}, J., {et~al.} 2013, \href
  {http://dx.doi.org/10.1088/0004-637X/767/1/50} {\apj}, 767, 50

\bibitem[{{Mullaney} {et~al.}(2013){Mullaney}, {Alexander}, {Fine}, {Goulding},
  {Harrison}, \& {Hickox}}]{Mullaney13}
{Mullaney}, J.~R., {et~al.} 2013, \href
  {http://dx.doi.org/10.1093/mnras/stt751} {\mnras}, 433, 622

\bibitem[{{Murphy} {et~al.}(2011){Murphy}, {Condon}, {Schinnerer}, {Kennicutt},
  {Calzetti}, {Armus}, {Helou}, {Turner}, {Aniano}, {Beir{\~a}o}, {Bolatto},
  {Brandl}, {Croxall}, {Dale}, {Donovan Meyer}, {Draine}, {Engelbracht},
  {Hunt}, {Hao}, {Koda}, {Roussel}, {Skibba}, \& {Smith}}]{Murphy11}
{Murphy}, E.~J., {et~al.} 2011, \href
  {http://dx.doi.org/10.1088/0004-637X/737/2/67} {\apj}, 737, 67

\bibitem[{{Newman} {et~al.}(2013){Newman}, {Cooper}, {Davis}, {Faber}, {Coil},
  {Guhathakurta}, {Koo}, {Phillips}, {Conroy}, {Dutton}, {Finkbeiner}, {Gerke},
  {Rosario}, {Weiner}, {Willmer}, {Yan}, {Harker}, {Kassin}, {Konidaris},
  {Lai}, {Madgwick}, {Noeske}, {Wirth}, {Connolly}, {Kaiser}, {Kirby},
  {Lemaux}, {Lin}, {Lotz}, {Luppino}, {Marinoni}, {Matthews}, {Metevier}, \&
  {Schiavon}}]{Newman12}
{Newman}, J.~A., {et~al.} 2013, \href
  {http://dx.doi.org/10.1088/0067-0049/208/1/5} {\apjs}, 208, 5

\bibitem[{{Norberg} {et~al.}(2008){Norberg}, {Frenk}, \& {Cole}}]{Norberg08}
{Norberg}, P., {Frenk}, C.~S., \& {Cole}, S. 2008, \href
  {http://dx.doi.org/10.1111/j.1365-2966.2007.12583.x} {\mnras}, 383, 646

\bibitem[{{Norris} {et~al.}(2006){Norris}, {Afonso}, {Appleton}, {Boyle},
  {Ciliegi}, {Croom}, {Huynh}, {Jackson}, {Koekemoer}, {Lonsdale},
  {Middelberg}, {Mobasher}, {Oliver}, {Polletta}, {Siana}, {Smail}, \&
  {Voronkov}}]{Norris06}
{Norris}, R.~P., {et~al.} 2006, \href {http://dx.doi.org/10.1086/508275} {\aj},
  132, 2409

\bibitem[{{Oliver} {et~al.}(2000){Oliver}, {Rowan-Robinson}, {Alexander},
  {Almaini}, {Balcells}, {Baker}, {Barcons}, {Barden}, {Bellas-Velidis},
  {Cabrera-Guerra}, {Carballo}, {Cesarsky}, {Ciliegi}, {Clements}, {Crockett},
  {Danese}, {Dapergolas}, {Drolias}, {Eaton}, {Efstathiou}, {Egami}, {Elbaz},
  {Fadda}, {Fox}, {Franceschini}, {Genzel}, {Goldschmidt}, {Graham},
  {Gonzalez-Serrano}, {Gonzalez-Solares}, {Granato}, {Gruppioni},
  {Herbstmeier}, {H{\'e}raudeau}, {Joshi}, {Kontizas}, {Kontizas},
  {Kotilainen}, {Kunze}, {La Franca}, {Lari}, {Lawrence}, {Lemke},
  {Linden-V{\o}rnle}, {Mann}, {M{\'a}rquez}, {Masegosa}, {Mattila}, {McMahon},
  {Miley}, {Missoulis}, {Mobasher}, {Morel}, {N{\o}rgaard-Nielsen}, {Omont},
  {Papadopoulos}, {Perez-Fournon}, {Puget}, {Rigopoulou}, {Rocca-Volmerange},
  {Serjeant}, {Silva}, {Sumner}, {Surace}, {Vaisanen}, {van der Werf}, {Verma},
  {Vigroux}, {Villar-Martin}, \& {Willott}}]{Oliver00}
{Oliver}, S., {et~al.} 2000, \href
  {http://dx.doi.org/10.1046/j.1365-8711.2000.03550.x} {\mnras}, 316, 749

\bibitem[{{Park} {et~al.}(2008){Park}, {Barmby}, {Fazio}, {Nandra}, {Laird},
  {Georgakakis}, {Rosario}, {Willner}, {Rieke}, {Ashby}, {Ivison}, {Coil}, \&
  {Miyazaki}}]{Park08}
{Park}, S.~Q., {et~al.} 2008, \href {http://dx.doi.org/10.1086/587136} {\apj},
  678, 744

\bibitem[{{Peebles}(1980)}]{Peebles80}
{Peebles}, P.~J.~E. 1980, {The large-scale structure of the universe}
  (Princeton University Press)

\bibitem[{{Pierre} {et~al.}(2004){Pierre}, {Valtchanov}, {Altieri}, {Andreon},
  {Bolzonella}, {Bremer}, {Disseau}, {Dos Santos}, {Gandhi}, {Jean}, {Pacaud},
  {Read}, {Refregier}, {Willis}, {Adami}, {Alloin}, {Birkinshaw}, {Chiappetti},
  {Cohen}, {Detal}, {Duc}, {Gosset}, {Hjorth}, {Jones}, {Le F{\`e}vre},
  {Lonsdale}, {Maccagni}, {Mazure}, {McBreen}, {McCracken}, {Mellier},
  {Ponman}, {Quintana}, {Rottgering}, {Smette}, {Surdej}, {Starck}, {Vigroux},
  \& {White}}]{Pierre04}
{Pierre}, M., {et~al.} 2004, \href
  {http://dx.doi.org/10.1088/1475-7516/2004/09/011} {\jcap}, 9, 11

\bibitem[{{Puccetti} {et~al.}(2006){Puccetti}, {Fiore}, {D'Elia}, {Pillitteri},
  {Feruglio}, {Grazian}, {Brusa}, {Ciliegi}, {Comastri}, {Gruppioni},
  {Mignoli}, {Vignali}, {Zamorani}, {La Franca}, {Sacchi}, {Franceschini},
  {Berta}, {Buttery}, \& {Dias}}]{Puccetti06}
{Puccetti}, S., {et~al.} 2006, \href
  {http://dx.doi.org/10.1051/0004-6361:20064904} {\aap}, 457, 501

\bibitem[{{Richstone} {et~al.}(1998){Richstone}, {Ajhar}, {Bender}, {Bower},
  {Dressler}, {Faber}, {Filippenko}, {Gebhardt}, {Green}, {Ho}, {Kormendy},
  {Lauer}, {Magorrian}, \& {Tremaine}}]{Richstone98}
{Richstone}, D., {et~al.} 1998, \href {http://arxiv.org/abs/astro-ph/9810378}
  {\nat}, 395, A14

\bibitem[{{Salim} {et~al.}(2007){Salim}, {Rich}, {Charlot}, {Brinchmann},
  {Johnson}, {Schiminovich}, {Seibert}, {Mallery}, {Heckman}, {Forster},
  {Friedman}, {Martin}, {Morrissey}, {Neff}, {Small}, {Wyder}, {Bianchi},
  {Donas}, {Lee}, {Madore}, {Milliard}, {Szalay}, {Welsh}, \& {Yi}}]{Salim07}
{Salim}, S., {et~al.} 2007, \href {http://dx.doi.org/10.1086/519218} {\apjs},
  173, 267

\bibitem[{{Schinnerer} {et~al.}(2007){Schinnerer}, {Smol{\v c}i{\'c}},
  {Carilli}, {Bondi}, {Ciliegi}, {Jahnke}, {Scoville}, {Aussel}, {Bertoldi},
  {Blain}, {Impey}, {Koekemoer}, {Le Fevre}, \& {Urry}}]{Schinnerer07}
{Schinnerer}, E., {et~al.} 2007, \href {http://dx.doi.org/10.1086/516587}
  {\apjs}, 172, 46

\bibitem[{{Schinnerer} {et~al.}(2010){Schinnerer}, {Sargent}, {Bondi}, {Smol{\v
  c}i{\'c}}, {Datta}, {Carilli}, {Bertoldi}, {Blain}, {Ciliegi}, {Koekemoer},
  \& {Scoville}}]{Schinnerer10}
{Schinnerer}, E., {et~al.} 2010, \href
  {http://dx.doi.org/10.1088/0067-0049/188/2/384} {\apjs}, 188, 384

\bibitem[{{Scoville} {et~al.}(2007){Scoville}, {Aussel}, {Brusa}, {Capak},
  {Carollo}, {Elvis}, {Giavalisco}, {Guzzo}, {Hasinger}, {Impey}, {Kneib},
  {LeFevre}, {Lilly}, {Mobasher}, {Renzini}, {Rich}, {Sanders}, {Schinnerer},
  {Schminovich}, {Shopbell}, {Taniguchi}, \& {Tyson}}]{Scoville07}
{Scoville}, N., {et~al.} 2007, \href {http://dx.doi.org/10.1086/516585}
  {\apjs}, 172, 1

\bibitem[{{Scranton} {et~al.}(2002){Scranton}, {Johnston}, {Dodelson},
  {Frieman}, {Connolly}, {Eisenstein}, {Gunn}, {Hui}, {Jain}, {Kent},
  {Loveday}, {Narayanan}, {Nichol}, {O'Connell}, {Scoccimarro}, {Sheth},
  {Stebbins}, {Strauss}, {Szalay}, {Szapudi}, {Tegmark}, {Vogeley}, {Zehavi},
  {Annis}, {Bahcall}, {Brinkman}, {Csabai}, {Hindsley}, {Ivezic}, {Kim},
  {Knapp}, {Lamb}, {Lee}, {Lupton}, {McKay}, {Munn}, {Peoples}, {Pier},
  {Richards}, {Rockosi}, {Schlegel}, {Schneider}, {Stoughton}, {Tucker},
  {Yanny}, \& {York}}]{Scranton02}
{Scranton}, R., {et~al.} 2002, \href {http://dx.doi.org/10.1086/342786} {\apj},
  579, 48

\bibitem[{{Serjeant} {et~al.}(2010){Serjeant}, {Negrello}, {Pearson},
  {Mortier}, {Austermann}, {Aretxaga}, {Clements}, {Chapman}, {Dye}, {Dunlop},
  {Dunne}, {Farrah}, {Hughes}, {Lee}, {Matsuhara}, {Ibar}, {Im}, {Jeong},
  {Kim}, {Oyabu}, {Takagi}, {Wada}, {Wilson}, {Vaccari}, \& {Yun}}]{Serjeant10}
{Serjeant}, S., {et~al.} 2010, \href
  {http://dx.doi.org/10.1051/0004-6361/200913483} {\aap}, 514, A10

\bibitem[{{Sheth} {et~al.}(2001){Sheth}, {Mo}, \& {Tormen}}]{Sheth01}
{Sheth}, R.~K., {Mo}, H.~J., \& {Tormen}, G. 2001, \href
  {http://dx.doi.org/10.1046/j.1365-8711.2001.04006.x} {\mnras}, 323, 1

\bibitem[{{Sheth} \& {Tormen}(1999)}]{Sheth99}
{Sheth}, R.~K., \& {Tormen}, G. 1999, \href
  {http://dx.doi.org/10.1046/j.1365-8711.1999.02692.x} {\mnras}, 308, 119

\bibitem[{{Sijacki} {et~al.}(2007){Sijacki}, {Springel}, {Di Matteo}, \&
  {Hernquist}}]{Sijacki07}
{Sijacki}, D., {et~al.} 2007, \href
  {http://dx.doi.org/10.1111/j.1365-2966.2007.12153.x} {\mnras}, 380, 877

\bibitem[{{Silk} \& {Rees}(1998)}]{Silk98}
{Silk}, J., \& {Rees}, M.~J. 1998, \href
  {http://arxiv.org/abs/arXiv:astro-ph/9801013} {\aap}, 331, L1

\bibitem[{{Silverman} {et~al.}(2011){Silverman}, {Kampczyk}, {Jahnke},
  {Andrae}, {Lilly}, {Elvis}, {Civano}, {Mainieri}, {Vignali}, {Zamorani},
  {Nair}, {Le F{\`e}vre}, {de Ravel}, {Bardelli}, {Bongiorno}, {Bolzonella},
  {Cappi}, {Caputi}, {Carollo}, {Contini}, {Coppa}, {Cucciati}, {de la Torre},
  {Franzetti}, {Garilli}, {Halliday}, {Hasinger}, {Iovino}, {Knobel},
  {Koekemoer}, {Kova{\v c}}, {Lamareille}, {Le Borgne}, {Le Brun}, {Maier},
  {Mignoli}, {Pello}, {P{\'e}rez-Montero}, {Ricciardelli}, {Peng}, {Scodeggio},
  {Tanaka}, {Tasca}, {Tresse}, {Vergani}, {Zucca}, {Brusa}, {Cappelluti},
  {Comastri}, {Finoguenov}, {Fu}, {Gilli}, {Hao}, {Ho}, \&
  {Salvato}}]{Silverman11}
{Silverman}, J.~D., {et~al.} 2011, \href
  {http://dx.doi.org/10.1088/0004-637X/743/1/2} {\apj}, 743, 2

\bibitem[{{Simpson} {et~al.}(2006){Simpson}, {Mart{\'{\i}}nez-Sansigre},
  {Rawlings}, {Ivison}, {Akiyama}, {Sekiguchi}, {Takata}, {Ueda}, \&
  {Watson}}]{Simpson06}
{Simpson}, C., {et~al.} 2006, \href
  {http://dx.doi.org/10.1111/j.1365-2966.2006.10907.x} {\mnras}, 372, 741

\bibitem[{{Skibba} {et~al.}(2014){Skibba}, {Smith}, {Coil}, {Moustakas},
  {Aird}, {Blanton}, {Bray}, {Cool}, {Eisenstein}, {Mendez}, {Wong}, \&
  {Zhu}}]{Skibba14}
{Skibba}, R.~A., {et~al.} 2014, \href
  {http://dx.doi.org/10.1088/0004-637X/784/2/128} {\apj}, 784, 128

\bibitem[{{Smith} {et~al.}(2003){Smith}, {Peacock}, {Jenkins}, {White},
  {Frenk}, {Pearce}, {Thomas}, {Efstathiou}, \& {Couchman}}]{Smith03}
{Smith}, R.~E., {et~al.} 2003, \href
  {http://dx.doi.org/10.1046/j.1365-8711.2003.06503.x} {\mnras}, 341, 1311

\bibitem[{{Smol{\v c}i{\'c}} {et~al.}(2008){Smol{\v c}i{\'c}}, {Schinnerer},
  {Scodeggio}, {Franzetti}, {Aussel}, {Bondi}, {Brusa}, {Carilli}, {Capak},
  {Charlot}, {Ciliegi}, {Ilbert}, {Ivezi{\'c}}, {Jahnke}, {McCracken},
  {Obri{\'c}}, {Salvato}, {Sanders}, {Scoville}, {Trump}, {Tremonti}, {Tasca},
  {Walcher}, \& {Zamorani}}]{Smolcic08}
{Smol{\v c}i{\'c}}, V., {et~al.} 2008, \href {http://dx.doi.org/10.1086/588028}
  {\apjs}, 177, 14

\bibitem[{{Soltan}(1982)}]{Soltan82}
{Soltan}, A. 1982, \mnras, 200, 115

\bibitem[{{Springel} {et~al.}(2005){Springel}, {Di Matteo}, \&
  {Hernquist}}]{Springel05a}
{Springel}, V., {Di Matteo}, T., \& {Hernquist}, L. 2005, \href
  {http://dx.doi.org/10.1086/428772} {\apjl}, 620, L79

\bibitem[{{Stern} {et~al.}(2005){Stern}, {Eisenhardt}, {Gorjian}, {Kochanek},
  {Caldwell}, {Eisenstein}, {Brodwin}, {Brown}, {Cool}, {Dey}, {Green},
  {Jannuzi}, {Murray}, {Pahre}, \& {Willner}}]{Stern05}
{Stern}, D., {et~al.} 2005, \href {http://dx.doi.org/10.1086/432523} {\apj},
  631, 163

\bibitem[{{Stern} {et~al.}(2012){Stern}, {Assef}, {Benford}, {Blain}, {Cutri},
  {Dey}, {Eisenhardt}, {Griffith}, {Jarrett}, {Lake}, {Masci}, {Petty},
  {Stanford}, {Tsai}, {Wright}, {Yan}, {Harrison}, \& {Madsen}}]{Stern12}
{Stern}, D., {et~al.} 2012, \href
  {http://dx.doi.org/10.1088/0004-637X/753/1/30} {\apj}, 753, 30

\bibitem[{{Sutherland} \& {Saunders}(1992)}]{Sutherland92}
{Sutherland}, W., \& {Saunders}, W. 1992, \mnras, 259, 413

\bibitem[{{Szokoly} {et~al.}(2004){Szokoly}, {Bergeron}, {Hasinger}, {Lehmann},
  {Kewley}, {Mainieri}, {Nonino}, {Rosati}, {Giacconi}, {Gilli}, {Gilmozzi},
  {Norman}, {Romaniello}, {Schreier}, {Tozzi}, {Wang}, {Zheng}, \&
  {Zirm}}]{Szokoly04}
{Szokoly}, G.~P., {et~al.} 2004, \href {http://dx.doi.org/10.1086/424707}
  {\apjs}, 155, 271

\bibitem[{{Tremaine} {et~al.}(2002){Tremaine}, {Gebhardt}, {Bender}, {Bower},
  {Dressler}, {Faber}, {Filippenko}, {Green}, {Grillmair}, {Ho}, {Kormendy},
  {Lauer}, {Magorrian}, {Pinkney}, \& {Richstone}}]{Tremaine02}
{Tremaine}, S., {et~al.} 2002, \href {http://dx.doi.org/10.1086/341002} {\apj},
  574, 740

\bibitem[{{Ueda} {et~al.}(2003){Ueda}, {Akiyama}, {Ohta}, \& {Miyaji}}]{Ueda03}
{Ueda}, Y., {et~al.} 2003, \href {http://dx.doi.org/10.1086/378940} {\apj},
  598, 886

\bibitem[{{Ueda} {et~al.}(2008){Ueda}, {Watson}, {Stewart}, {Akiyama},
  {Schwope}, {Lamer}, {Ebrero}, {Carrera}, {Sekiguchi}, {Yamada}, {Simpson},
  {Hasinger}, \& {Mateos}}]{Ueda08}
{Ueda}, Y., {et~al.} 2008, \href {http://dx.doi.org/10.1086/591083} {\apjs},
  179, 124

\bibitem[{{Urry} \& {Padovani}(1995)}]{Urry95}
{Urry}, C.~M., \& {Padovani}, P. 1995, \href {http://dx.doi.org/10.1086/133630}
  {\pasp}, 107, 803

\bibitem[{{van den Bosch}(2002)}]{Vandenbosch02}
{van den Bosch}, F.~C. 2002, \href
  {http://dx.doi.org/10.1046/j.1365-8711.2002.05171.x} {\mnras}, 331, 98

\bibitem[{{Wake} {et~al.}(2008){Wake}, {Croom}, {Sadler}, \&
  {Johnston}}]{Wake08}
{Wake}, D.~A., {et~al.} 2008, \href
  {http://dx.doi.org/10.1111/j.1365-2966.2008.14039.x} {\mnras}, 391, 1674

\bibitem[{{Willner} {et~al.}(2012){Willner}, {Ashby}, {Barmby}, {Chapman},
  {Coil}, {Cooper}, {Huang}, {Ivison}, \& {Koo}}]{Willner12}
{Willner}, S.~P., {et~al.} 2012, \href
  {http://dx.doi.org/10.1088/0004-637X/756/1/72} {\apj}, 756, 72

\bibitem[{{Wright} {et~al.}(2010){Wright}, {Eisenhardt}, {Mainzer}, {Ressler},
  {Cutri}, {Jarrett}, {Kirkpatrick}, {Padgett}, {McMillan}, {Skrutskie},
  {Stanford}, {Cohen}, {Walker}, {Mather}, {Leisawitz}, {Gautier}, {McLean},
  {Benford}, {Lonsdale}, {Blain}, {Mendez}, {Irace}, {Duval}, {Liu}, {Royer},
  {Heinrichsen}, {Howard}, {Shannon}, {Kendall}, {Walsh}, {Larsen}, {Cardon},
  {Schick}, {Schwalm}, {Abid}, {Fabinsky}, {Naes}, \& {Tsai}}]{Wright10}
{Wright}, E.~L., {et~al.} 2010, \href
  {http://dx.doi.org/10.1088/0004-6256/140/6/1868} {\aj}, 140, 1868

\bibitem[{{Yan} {et~al.}(2013){Yan}, {Donoso}, {Tsai}, {Stern}, {Assef},
  {Eisenhardt}, {Blain}, {Cutri}, {Jarrett}, {Stanford}, {Wright}, {Bridge}, \&
  {Riechers}}]{Yan13}
{Yan}, L., {et~al.} 2013, \href {http://dx.doi.org/10.1088/0004-6256/145/3/55}
  {\aj}, 145, 55

\bibitem[{{Yang} {et~al.}(2006){Yang}, {Mushotzky}, {Barger}, \&
  {Cowie}}]{Yang06}
{Yang}, Y., {et~al.} 2006, \href {http://dx.doi.org/10.1086/502706} {\apj},
  645, 68

\bibitem[{{Zehavi} {et~al.}(2005){Zehavi}, {Zheng}, {Weinberg}, {Frieman},
  {Berlind}, {Blanton}, {Scoccimarro}, {Sheth}, {Strauss}, {Kayo}, {Suto},
  {Fukugita}, {Nakamura}, {Bahcall}, {Brinkmann}, {Gunn}, {Hennessy},
  {Ivezi{\'c}}, {Knapp}, {Loveday}, {Meiksin}, {Schlegel}, {Schneider},
  {Szapudi}, {Tegmark}, {Vogeley}, {York}, \& {SDSS Collaboration}}]{Zehavi05}
{Zehavi}, I., {et~al.} 2005, \href {http://dx.doi.org/10.1086/431891} {\apj},
  630, 1

\bibitem[{{Zehavi} {et~al.}(2011){Zehavi}, {Zheng}, {Weinberg}, {Blanton},
  {Bahcall}, {Berlind}, {Brinkmann}, {Frieman}, {Gunn}, {Lupton}, {Nichol},
  {Percival}, {Schneider}, {Skibba}, {Strauss}, {Tegmark}, \&
  {York}}]{Zehavi11}
{Zehavi}, I., {et~al.} 2011, \href
  {http://dx.doi.org/10.1088/0004-637X/736/1/59} {\apj}, 736, 59

\bibitem[{{Zheng} {et~al.}(2009){Zheng}, {Zehavi}, {Eisenstein}, {Weinberg}, \&
  {Jing}}]{Zheng09}
{Zheng}, Z., {et~al.} 2009, \href
  {http://dx.doi.org/10.1088/0004-637X/707/1/554} {\apj}, 707, 554

\end{thebibliography}
